\documentclass[aps,prd,showpacs,floatfix,nofootinbib,twocolumn]{revtex4-2}
\usepackage{graphicx,xcolor,dcolumn,booktabs,bm}
\usepackage{amsmath,amssymb}
\usepackage{longtable}
\usepackage[pass]{geometry}
\usepackage{pifont}
\usepackage{txfonts}
\usepackage{threeparttable}
\usepackage{indentfirst}
\usepackage{epstopdf}   
\usepackage{cases}
\usepackage{multirow}
\usepackage{placeins}
\graphicspath{{}}

\usepackage[
  colorlinks=true,
  citecolor=black,
  linkcolor=black,
  urlcolor=black,
  breaklinks=true
]{hyperref}
\usepackage{comment}

\allowdisplaybreaks
\begin{document}
\title{Meson spectra in a screened relativized quark model}
\author{
Zhe-Qi Huang$^{1,2,3}$}
%\email{2024012028@bhu.edu.cn}
\author{
Lei Huang$^{4}$}
%\email{@bhu.edu.cn}
\author{
Xiao-Lei Zhang$^{3}$}
%\email{@bhu.edu.cn}
\author{
Xin-Lei Ge$^{1}$}\email{gexl\_bhu@163.com}

\author{
Hao Chen$^{2,5}$}\email{chenhao_qhnu@outlook.com
}
\author{Cheng-Qun Pang$^{3,6}$}\email{xuehua45@163.com}
\affiliation{
$^1$ 
College of Physical Science and Technology,
Bohai University, Jinzhou 121013, China
\\
$^2$ 
College of Physics and Electronic Information Engineering,
Qinghai Normal University, Xining 810000, China
\\
$^3$ School of Physics and Optoelectronic Engineering,
Ludong University, Yantai 264000, China\\
$^4$ Heze Medical College, Heze 274000, China\\
$^5$ 
Academy of Plateau Science and Sustainability,
Xining 810016, China
\\
$^6$ 
Lanzhou Center for Theoretical Physics,
Key Laboratory of Quantum Theory and Applications of MoE,
and Key Laboratory of Theoretical Physics of Gansu Province,
Lanzhou University, Lanzhou, Gansu 730000, China
}

\date{\today}

\begin{abstract}

In this work, we employed the modified Godfrey--Isgur  model to systematically investigate the mass spectra of several light, heavy-light, and heavy-heavy meson systems. In this model, a screened potential is employed to phenomenologically account for modifications to the long-range confining interaction associated with unquenched effects. We performed a global refit of the relevant model parameters to a combined set of 90 adopted meson masses spanning all three sectors, ensuring a consistent parameter determination across the different meson systems. 
Using the resulting parameter set, we calculated the low-lying and highly excited  spectra and the corresponding mixed states in different meson families, and compared our results with existing experimental data and other theoretical results. The assignment of several newly observed structures (such as $B_c^*$ and $D_{s1}(2933)$) has also been discussed based on their predicted masses. Overall, the model describes several main spectroscopic features across different flavors and provides mass predictions for highly excited states that have not yet been experimentally established.

\end{abstract}

\maketitle
\section{Introduction}\label{sec1}

Hadron spectroscopy provides a powerful means of probing the nonperturbative nature of the strong interaction.
As experimental precision improves and data samples grow, the meson listings compiled by the Particle Data Group (PDG) continue to expand.
Many ground states and low-lying excitations in the light, heavy-light, and heavy-heavy sectors are relatively well established experimentally, whereas  information on higher radial and orbital excitations continues to accumulate~\cite{ParticleDataGroup:2026mpi}.
In recent years, the LHCb, BESIII, Belle II, and ATLAS collaborations have reported a series of new hadronic structures, providing new experimental information for mapping out the meson spectrum and testing different theoretical models~\cite{LHCb:2024vfz,BESIII:2022sfx,Belle:2025syc,ATLAS:2026ubk}.
In particular, LHCb has reported the first observation of orbitally excited $B_c^+$ states consistent with the expected $B_c(1P)^+$ multiplet~\cite{LHCb:2025uce,LHCb:2025ubr}. The collaboration has also reported the observation of a new charm-strange meson, $D_{s1}(2933)^+$, in an amplitude analysis of  $B^0\to D^+D^-K^+\pi^-$ decays. 
The state was determined to have $J^P=1^+$, making it a candidate for the $D_s(2P_1^{(\prime)})^+$ assignment~\cite{LHCb:2026sup}.
In addition, LHCb has reported a new resonance in the $B_{s0}^*(5700)^0$  with a measured mass of $5698.9\pm1.5\pm0.6~\mathrm{MeV}$~\cite{LHCb:2026knl}. Under the $J^P=0^+$ hypothesis, this structure may be identified with the $B_{s0}^*(5700)^0$ state, although a $J^P=1^+$ interpretation is also compatible with the data. If the $J^P=0^+$ assignment is established, the state would provide new experimental constraints on the lowest scalar bottom-strange state.

These advances further highlight the need for systematic spectroscopic studies  across different meson flavor sectors, particularly of highly excited states, including those not yet experimentally established.

A variety of theoretical frameworks have been developed to describe the increasingly rich  meson spectrum revealed by experiments. 
Representative approaches include constituent quark models~\cite{Godfrey:1985xj,Barnes:2005pb}, covariant Dyson--Schwinger/Bethe--Salpeter  approaches~\cite{Fischer:2014cfa}, quasipotential approaches~\cite{Ebert:2009ua}, models based on two-body Dirac equations~\cite{Crater:2010fc}, basis light-front quantization~\cite{Li:2017mlw}, chiral models~\cite{Parganlija:2012fy}, unitarized models~\cite{vanBeveren:2020eis}, effective field theories~\cite{Brambilla:2004jw}, and lattice Quantum Chromodynamics (QCD) calculations~\cite{Dudek:2013yja,Lang:2014yfa}.

Among these approaches, the relativized quark model introduced in Ref.~\cite{Godfrey:1985xj} in 1985, commonly known as the Godfrey--Isgur (GI) model, has been widely used in meson spectroscopy. The GI model combines relativistic kinematics with short-range one-gluon exchange, long-range confinement, and spin-dependent interactions in a unified quark--antiquark Hamiltonian, providing a good description of the low-lying mass spectra of many conventional mesons~\cite{Godfrey:1985xj}. However, coupled-channel effects associated with unquenched dynamics can play an important role in hadron spectroscopy, particularly for states near hadronic thresholds and for highly excited states. Such effects have been invoked to account for the unexpectedly low masses or near-threshold properties of states such as $D_{s0}(2317)$, $D_{s1}(2460)$, $X(3872)$, and $\Lambda_c(2940)$~\cite{Lang:2014yfa,Tan:2019qwe,Luo:2019qkm}. Therefore, accounting for the influence of unquenched dynamics, at least phenomenologically, is important in conventional quark-model studies of highly excited meson spectra.

A commonly used phenomenological treatment is to replace the linear confining potential in the GI model with a screened form, leading to what is commonly referred to as the modified Godfrey--Isgur (MGI) model~\cite{Song:2015nia,Song:2015fha,Li:2023wgq}.
The screened potential phenomenologically accounts for some effects associated with light-quark pair creation and couplings to hadronic continua  by softening the long-range confining interaction. The resulting deviations from the unscreened model tend to become more pronounced for highly excited states~\cite{Bali:2005fu,Li:2009ad}.
Similar screening prescriptions have been used to study highly excited charmed mesons, charmed-strange mesons, charmonia, bottomonia, and $B_c$ mesons~\cite{Song:2015fha,Song:2015nia,Li:2009zu,Wang:2018rjg,Li:2023wgq}.
Previous studies have shown that screening effects tend to lower the predicted masses of highly excited states relative to corresponding unscreened potential-model results~\cite{Li:2009ad,Li:2023wgq}. The screened-potential prescription therefore offers a phenomenological means of estimating part of the impact of unquenched dynamics on highly excited meson spectra.

Based on these considerations, we employed the MGI model to systematically investigate the  spectra of  meson systems with different flavor contents.
We performed a global refit of the relevant model parameters using experimentally established  meson states from the light, heavy-light, and heavy-heavy sectors and used the resulting fitted parameters to calculate the low-lying and highly excited mass spectra, including the relevant mixed states, in different meson families.
We then compared our results with existing experimental data and other theoretical calculations to assess the applicability of the  MGI framework across meson systems with different flavors contents.

In light of recent experimental developments, we discuss the spectroscopic interpretation of the observed $B_c(1P)^+$ structures and the $B_c^{*+}$ state~\cite{LHCb:2025uce,LHCb:2025ubr,ATLAS:2026ubk}. We also examine possible assignments for the nonstrange charmed mesons $D_1^*(2600)$, $D_2(2740)$, and $D_3^*(2750)$~\cite{LHCb:2019juy}; the charmed-strange mesons $D_{s0}(2590)^+$ and $D_{s1}(2933)^+$~\cite{LHCb:2020gnv,LHCb:2026sup}; and the bottom and bottom-strange states $B_J(5970)^0$, $B_{sJ}(6063)^0$, $B_{sJ}(6114)^0$, and the $B_{s0}^*(5700)^0$ candidate structure~\cite{ParticleDataGroup:2026mpi,LHCb:2020pet,LHCb:2026knl}.
In the light-meson sector, we examine experimentally observed excited states such as $\rho_3(1690)$, $a_4(1970)$, $K_3^*(1780)$, and $K_5^*(2380)$, with particular attention  to the observed $\phi(2170)$ and its possible correspondence with the nearby $3{}^3S_1$ and $2{}^3D_1$ strangeonium states.
In the heavy-quarkonium sector, we further examine the possible correspondence of $\eta_c(3945)$, $h_c(4000)$, $\chi_{c1}(4010)$, $h_c(4300)$, $Y(4500)$, $Y(4710)$, and $\Upsilon(10753)$ to conventional quarkonium states and the limitations of these interpretations~\cite{LHCb:2024vfz,BESIII:2022joj,BESIII:2022kcv,BESIII:2023wqy,Belle:2025syc}. Motivated by the experimentally observed enhancement near the $t\bar t$ threshold, we also briefly explore a formal extrapolation of the model to the $t\bar t$ system~\cite{CMS:2025kzt,ATLAS:2026dbe,CMS:2026ani}. Such an extrapolation should be regarded as qualitative.

The remainder of this paper is organized as follows. 
In Sec.~\ref{sec2}, we briefly review the MGI model and present the variational procedure and global parameter determination. 
Section~\ref{sec3} presents the mass spectra of different meson systems, compares them with experimental data and other theoretical calculations, and discusses possible spectroscopic assignments for several recently observed states. The main results and conclusions are given in Sec.~\ref{sec4}.

\section{Screened Relativized Quark Model}\label{sec2}

We employ the MGI model to calculate the mass spectra of mesons with different flavor contents. As in the original GI model, each meson is treated as a relativistic quark--antiquark bound state. In the center-of-mass frame of the meson, the effective Hamiltonian of the MGI model is written as~\cite{Godfrey:1985xj,Song:2015nia,Li:2023wgq}
\begin{equation}
\label{eq:MGI_Hamiltonian}
\tilde{H}
=
\sqrt{\mathbf{p}^{2}+m_1^{2}}
+
\sqrt{\mathbf{p}^{2}+m_2^{2}}
+
\tilde{V}^{\mathrm{eff}}(\mathbf{p},\mathbf{r}),
\end{equation}
{where \(\tilde{V}^{\mathrm{eff}}(\mathbf{p},\mathbf{r}) = \tilde{V}^{\mathrm{conf}} + \tilde{V}^{\mathrm{hyp}} + \tilde{V}^{\mathrm{so}}\) denotes the effective quark--antiquark interaction. Quantities with a tilde denote the corresponding relativized interactions. In the nonrelativistic limit, it reduces to the standard nonrelativistic potential \(V^{\mathrm{eff}}(r)\),
\begin{equation}
V^{\mathrm{eff}}(r) = V^{\mathrm{conf}}(r) + V^{\mathrm{hyp}}(r) + V^{\mathrm{so}}(r). \label{eq:V_eff_nonrelativistic}
\end{equation}
The spin-independent potential \(V^{\mathrm{conf}}(r)\) is given by
\begin{equation}
V^{\mathrm{conf}}(r) = S(r) + G(r),
\end{equation}
where
\begin{equation}
    S(r) = \frac{b}{\mu}\left(1-e^{-\mu r}\right) + c
\end{equation}
is the screened scalar confining potential, and
\begin{equation}
    G(r) = -\frac{4\alpha_s(r)}{3r}
\end{equation}
is the Coulomb-like potential. The parameter \(b\) denotes the string tension, while \(c\) is a constant. The screening parameter \(\mu\) controls the distance scale and strength of the screening effect. In the limit \(\mu\to0\), \(S(r)\) reduces to the linear confining potential \(br+c\), whereas at large distances it approaches \(b/\mu+c\), reflecting the screening-induced flattening of the confining interaction. A finite value of \(\mu\) therefore parametrizes the screening modification relative to the original GI model. Here, \(\alpha_s(r)\) is the running coupling constant.
}

The hyperfine interaction \(V^{\mathrm{hyp}}(r)\) consists of a contact term and a tensor term.
{The contact interaction is given by
\begin{equation}
V^{\mathrm{cont}}(r)
=
\frac{4\alpha_s(r)}{3m_1m_2}
\frac{8\pi}{3}
\mathbf{S}_1\cdot\mathbf{S}_2
\delta^3(\mathbf{r}),
\end{equation}
where \(\mathbf{S}_1\) and \(\mathbf{S}_2\) denote the spins of the quark and antiquark, respectively, and \(\delta^3(\mathbf{r})\) is the three-dimensional Dirac delta function.

The tensor interaction is written as
\begin{equation}
V^{\mathrm{tens}}(r)
=
\frac{4\alpha_s(r)}{3m_1m_2r^3}
\left[
3\frac{
(\mathbf{S}_1\cdot\mathbf{r})
(\mathbf{S}_2\cdot\mathbf{r})
}{r^2}
-
\mathbf{S}_1\cdot\mathbf{S}_2
\right].
\end{equation}

The spin--orbit interaction \(V^{\mathrm{so}}(r)\) is separated into vector and scalar contributions. The vector spin--orbit term is given by
\begin{equation}
V^{\mathrm{so(v)}}(r)
=
\frac{1}{r}
\frac{\mathrm{d}G(r)}{\mathrm{d}r}
\Bigg[
\frac{\mathbf{S}_1}{2m_1^2}
+
\frac{\mathbf{S}_2}{2m_2^2}
+
\frac{\mathbf{S}_1+\mathbf{S}_2}{m_1m_2}
\Bigg]
\cdot\mathbf{L},
\end{equation}
whereas the scalar spin--orbit term is
\begin{equation}
V^{\mathrm{so(s)}}(r)
=
-\frac{1}{2r}
\frac{\mathrm{d}S(r)}{\mathrm{d}r}
\left(
\frac{\mathbf{S}_1}{m_1^2}
+
\frac{\mathbf{S}_2}{m_2^2}
\right)
\cdot\mathbf{L},
\end{equation}
where \(\mathbf{L}\) denotes the relative orbital angular momentum of the quark--antiquark pair.
}

Beyond the relativistic kinetic-energy terms, relativistic corrections to the quark--antiquark interaction are implemented in two steps. First, the nonlocality associated with the relativistic motion of the constituent quarks is modeled by smearing the quark--antiquark interaction with a mass-dependent Gaussian function $\rho_{12}\left(\mathbf{r}-\mathbf{r'}\right)$~\cite{Godfrey:1985xj,Song:2015nia,Li:2023wgq}:
\begin{equation}
    \rho_{12}\left(\mathbf{r}-\mathbf{r}^{\prime}\right)
    =
    \frac{\sigma_{12}^3}{\pi^{3/2}}
    \exp\left[-\sigma_{12}^2\left(\mathbf{r}-\mathbf{r}^{\prime}\right)^2\right],
    \label{smearing}
\end{equation}
with
\begin{align}
\label{smearing_sigma}
   \sigma_{12}^2=\sigma_0^2\Bigg[\frac{1}{2}+\frac{1}{2}\left(\frac{4m_1m_2}{(m_1+m_2)^2}\right)^4\Bigg]+
  s^2\left(\frac{2m_1m_2}{m_1+m_2}\right)^2,
\end{align}
where \(\sigma_0\) and \(s\) are universal smearing parameters. \(m_1\) and \(m_2\) are the masses of the constituent quark and antiquark, respectively.

The mass dependence of \(\sigma_{12}\) allows the same smearing prescription to interpolate continuously between different flavor sectors. For an equal-mass system with \(m_{1}=m_{2}=m\), Eq.~\eqref{smearing_sigma} reduces to
\begin{equation}
\sigma_{12}^{2}
=
\sigma_{0}^{2}
+
s^{2}m^{2}.
\end{equation}
For increasingly heavy equal-mass quarkonia, \(\sigma_{12}\) therefore becomes large, and the Gaussian kernel becomes progressively narrower. In the heavy-quark limit, it approaches the local form
\begin{equation}
\rho_{12}
\left(
\mathbf{r}-\mathbf{r}^{\prime}
\right)
\rightarrow
\delta^{3}
\left(
\mathbf{r}-\mathbf{r}^{\prime}
\right).
\end{equation}
This behavior is relevant to heavy-heavy systems such as charmonium and bottomonium, for which the spatial smearing becomes less pronounced as the constituent-quark masses increase. For the \(B_{c}\) system, although the two constituent masses are unequal, both are heavy, and the resulting smearing kernel is likewise narrower than in systems containing light constituent quarks.

For a strongly asymmetric heavy-light system with \(m_{2}\gg m_{1}\), Eq.~\eqref{smearing_sigma} approaches
\begin{equation}
\sigma_{12}^{2}
\rightarrow
\frac{\sigma_{0}^{2}}{2}
+
4s^{2}m_{1}^{2}.
\end{equation}
Thus, in the heavy-light limit, the smearing scale remains finite and is governed predominantly by the light constituent mass rather than increasing indefinitely with the heavy-quark mass. This mass dependence makes the same smearing prescription applicable to heavy-light mesons such as the \(D\), \(D_{s}\), \(B\), and \(B_{s}\) systems.

For light-light systems, both constituent masses are relatively small, so \(\sigma_{12}\) remains finite and the Gaussian smearing remains appreciable, reflecting the stronger role of relativistic motion in the light-quark sector. Consequently, the mass-dependent Gaussian smearing prescription provides a common phenomenological interpolation among light, heavy-light, and heavy-heavy meson systems rather than imposing the same spatial smearing scale on all flavor sectors. This feature provides a natural basis for employing the same relativized interaction framework across the different meson sectors considered in this work.

The smearing prescription is applied to the Coulomb-like potential \(G(r)\) and the screened scalar confining potential \(S(r)\) as follows:
\begin{equation}
\tilde{G}(r) = \int \mathrm{d}^3\mathbf{r}^{\prime} \rho_{12}({\bf r}-{\bf r}^\prime)G(r^\prime),
\end{equation}
and
\begin{equation}
\tilde{S}(r) = \int \mathrm{d}^3\mathbf{r}^{\prime}
\rho_{12} ({\bf r}-{\bf r}^\prime)S(r^\prime).
\end{equation}
The spin-dependent interaction terms are smeared in the same manner:
\begin{equation}
\tilde{V}^i(r) = \int \mathrm{d}^3\mathbf{r}^{\prime}
\rho_{12} ({\bf r}-{\bf r}^\prime)V^i(r^\prime),
\end{equation}
where the index $i=\{\mathrm{cont},\,\mathrm{tens},\,\mathrm{so(v)},\,\mathrm{so(s)}\}$ labels the contact, tensor, vector spin--orbit, and scalar spin--orbit interactions, respectively.

Second, the interactions are further relativized by introducing momentum-dependent factors that reduce to unity in the nonrelativistic limit~\cite{Godfrey:1985xj,Song:2015nia,Li:2023wgq}.
Accordingly, the smeared Coulomb-like term is modified according to
\begin{equation}
\tilde G(r)\ \longrightarrow\
\left(1+\frac{\mathbf p^2}{E_1E_2}\right)^{\!1/2}\!
\tilde G(r)\!
\left(1+\frac{\mathbf p^2}{E_1E_2}\right)^{\!1/2},
\end{equation}
where \(E_1\) and \(E_2\) are the energies of the constituent quark and antiquark, respectively, with \(E_{1(2)}=\sqrt{m_{1(2)}^2+\mathbf p^2}\).

Similarly, the smeared spin-dependent interactions are further relativized according to
\begin{equation}
\label{vsoij}
\tilde V^{\,i}(r)
\longrightarrow
\left(\frac{m_1 m_2}{E_1 E_2}\right)^{\!1/2+\epsilon_i}
\tilde V^{\,i}(r)
\left(\frac{m_1 m_2}{E_1 E_2}\right)^{\!1/2+\epsilon_i},
\end{equation}
where $\epsilon_i=\left\{ \epsilon_{\mathrm{cont}},\, \epsilon_{\mathrm{tens}},\, \epsilon_{\mathrm{so(v)}},\, \epsilon_{\mathrm{so(s)}} \right\}$ are phenomenological parameters associated with the relativistic corrections to the above mentioned four kinds of interactions.

After the above smearing and momentum-dependent corrections, the resulting interaction terms constitute the relativized effective potential \(\tilde{V}^{\mathrm{eff}}(\mathbf{p},\mathbf{r})\) entering the MGI Hamiltonian introduced in Eq. (\ref{eq:MGI_Hamiltonian}).

For equal-mass quarkonia, such as \(c\bar{c}\) and \(b\bar{b}\), the \(L\)--\(S\) coupling scheme is generally adopted, and the meson states are denoted by $n\,{}^{2S+1}L_J$.
For mesons composed of quarks with unequal masses, such as \(D\), \(D_s\), \(B\), \(B_s\), and \(B_c\), the antisymmetric spin--orbit interaction can mix the \({}^{1}L_J\) and \({}^{3}L_J\) states with \(J=L\), which have the same \(J^P\). The corresponding mass eigenstates are written as
\begin{align}
\left|L'_J\right\rangle
&=
\left|{}^1L_J\right\rangle\cos\theta
+
\left|{}^3L_J\right\rangle\sin\theta,
\\
\left|L_J\right\rangle
&=
-\left|{}^1L_J\right\rangle\sin\theta
+
\left|{}^3L_J\right\rangle\cos\theta,
\end{align}

To solve the Hamiltonian, we expand the meson wave functions in a simple harmonic oscillator (SHO) basis:
\begin{align}
\Psi_{nLM_L}(\mathbf{r})
&=
R_{nL}(r,\beta)
Y_{LM_L}(\Omega_r),
\\
\Psi_{nLM_L}(\mathbf{p})
&=
R_{nL}(p,\beta)
Y_{LM_L}(\Omega_p).
\end{align}
The radial wave function in coordinate space is written as
\begin{equation}
\begin{aligned}
R_{nL}(r,\beta)
={}&
\beta^{3/2}
\sqrt{
\frac{2n!}{\Gamma(n+L+3/2)}
}
(\beta r)^L
\\
&\times
\exp\left(
-\frac{\beta^2r^2}{2}
\right)
L_n^{L+1/2}(\beta^2r^2),
\end{aligned}
\end{equation}
whereas the radial wave function in momentum space is
\begin{equation}
\begin{aligned}
R_{nL}(p,\beta)
={}&
\frac{(-1)^n(-i)^L}{\beta^{3/2}}
\sqrt{
\frac{2n!}{\Gamma(n+L+3/2)}
}
\left(
\frac{p}{\beta}
\right)^L
\\
&\times
\exp\left(
-\frac{p^2}{2\beta^2}
\right)
L_n^{L+1/2}
\left(
\frac{p^2}{\beta^2}
\right).
\end{aligned}
\end{equation}
Here, \(L_n^{L+1/2}\) is the generalized Laguerre polynomial, \(Y_{LM_L}\) is the spherical harmonic, and \(\Omega_r\) and \(\Omega_p\) denote the angular coordinates of \(\mathbf{r}\) and \(\mathbf{p}\), respectively. The parameter \(\beta\) is the variational scale parameter of the SHO basis. Here, \(n=0,1,2,\ldots\) labels the radial SHO basis functions and is distinct from the radial quantum number \(n=1,2,\ldots\) used in the spectroscopic notation \(n\,{}^{2S+1}L_J\).

We determine a common set of model parameters through a weighted
least-squares fit to the masses of light mesons, heavy-light mesons,
and heavy quarkonia. The fit uses 90 experimental mass inputs selected
from the PDG 2026 compilation~\cite{ParticleDataGroup:2026mpi}. The experimental states used for fitting, along with their masses, are listed in Table~\ref{tab1}.

At each evaluation of the objective function in the global fit,
the SHO scale parameter $\beta$ was optimized independently
for each fitted state using the current model parameters. For a fixed set of model parameters and a specified physical
radial state, the optimal value of \(\beta\) was determined by minimizing
the corresponding mass eigenvalue within the prescribed range
\begin{equation}
\beta_{\min}
\leq
\beta
\leq
\beta_{\max}.
\end{equation}
In practice, we adopted
\begin{equation}
0.1~\mathrm{GeV}
\leq
\beta
\leq
5.0~\mathrm{GeV},
\end{equation}
as the search range.

The free model parameters were collected in the vector
\begin{equation}
\begin{aligned}
\mathbf{x}
=
\bigl(
&m_c,\,
m_b,\,
m_q,\,
m_s,\,
\mu_{\mathrm{L}},\,
\mu_{\mathrm{H}},\,
c,\,
b,\,
\sigma_0,\,
s,
\\
&\epsilon_{\mathrm{cont}},\,
\epsilon_{\mathrm{tens}},\,
\epsilon_{\mathrm{so(v)}},\,
\epsilon_{\mathrm{so(s)}}
\bigr).
\end{aligned}
\end{equation}
Here, $q=u,d$, and $m_q=m_u=m_d$ denotes the common constituent
mass of the light $u$ and $d$ quarks.

Previous studies using screened models have adopted different
effective screening parameters for different meson
systems~\cite{Pang:2017dlw,Song:2015nia,Song:2015fha,
Wang:2019mhs,Wang:2018rjg,Pang:2018gcn}.
Motivated by this system-dependent treatment, we introduce two
screening parameters, $\mu_{\mathrm{L}}$ and $\mu_{\mathrm{H}}$,
in the global fit.
They allow distinct effective screening scales for purely light
mesons and mesons containing heavy quarks, respectively:
\begin{equation}
\mu=
\begin{cases}
\mu_{\mathrm{L}},
& q\bar q,\ q\bar s,\ s\bar s,\\[2pt]
\mu_{\mathrm{H}},
& D,\ D_s,\ B,\ B_s,\ c\bar c,\ b\bar b,\ B_c.
\end{cases}
\end{equation}

Following the distinction between model errors and fitting
weights discussed in Ref.~\cite{Dobaczewski2014ErrorGuide},
we introduce, in addition to the experimental uncertainty,
an additional theoretical error scale determined by parameter
sensitivity for each flavor composition, so that a few states
with very small experimental uncertainties do not carry
excessive weight in the global fit.

Specifically, a change of one unit in the last digit of each
reference parameter (such as the parameter $b=0.2151~\mathrm{GeV}^2$, the change of one unit in the last digit of this parameter will be $0.0001~\mathrm{GeV}^2$) used for the error estimate defines its
variation step. For each flavor composition, the
$1{}^3P_0$, $2{}^3P_0$, and $3{}^3P_0$ levels are selected
as reference states. With $\beta=0.5~\mathrm{GeV}$ fixed,
the parameters relevant to that composition are varied
simultaneously by plus or minus one step in different
combinations. The largest absolute mass change among
the three reference states relative to the results at
the central parameter values is taken as the theoretical
error for that composition. The fitted states with the
same flavor composition share the same error scale,
denoted by $\sigma_{i,\mathrm{th}}$.

In the order $q\bar q$, $s\bar q$, $s\bar s$,
$c\bar q$, $c\bar s$, $c\bar c$, $b\bar q$,
$b\bar s$, $b\bar c$, and $b\bar b$, the adopted
theoretical errors are $2.0$, $1.5$, $1.2$, $1.8$,
$1.7$, $2.4$, $1.7$, $1.7$, $2.4$, and
$2.4~\mathrm{MeV}$, respectively.

The $\chi^2$ function used for the parameter fit is
defined as
\begin{equation}
\chi^2
=
\sum_{i=1}^{90}
\frac{
\left[M_i^{\mathrm{th}}-M_i^{\mathrm{exp}}\right]^2
}{
\sigma_{i,\mathrm{exp}}^2
+
\sigma_{i,\mathrm{th}}^2
}.
\end{equation}
Here, $M_i^{\mathrm{th}}$ and $M_i^{\mathrm{exp}}$
are the theoretical and experimental masses, respectively,
and $\sigma_{i,\mathrm{exp}}$ is the experimental uncertainty.
The nominal number of degrees of freedom is $90-14=76$.
The final fit gives $\chi^2=5250$ and
$\chi^2/\mathrm{d.o.f.}=69$. The model parameters determined by global fitting are listed in Table~\ref{tab2}.

\begin{table}[!t]
\caption{Parameters of the  MGI model in meson system.}
\label{tab2}\centering\footnotesize
\setlength{\tabcolsep}{3pt}\renewcommand{\arraystretch}{1.15}
\begin{tabular}{cccc}\toprule
Parameter & Value & Parameter & Value\\\midrule
$m_b$ & $5.100\,\mathrm{GeV}$ & $m_q$ & $0.3353\,\mathrm{GeV}$\\
$m_c$ & $1.754\,\mathrm{GeV}$ & $m_s$ & $0.5308\,\mathrm{GeV}$\\
$b$ & $0.2151\,\mathrm{GeV}^2$ & $c$ & $-0.5113\,\mathrm{GeV}$\\
$\mu_L$ & $0.02846\,\mathrm{GeV}$ & $\mu_H$ & $0.07838\,\mathrm{GeV}$\\
$s$ & $1.438$ & $\sigma_0$ & $1.302\,\mathrm{GeV}$\\
$\epsilon_{\rm cont}$ & $-0.2724$ & $\epsilon_{\rm tens}$ & $-0.3646$\\
$\epsilon_{\rm so(v)}$ & $-0.2349$ & $\epsilon_{\rm so(s)}$ & $0.004493$\\
\bottomrule\end{tabular}
\end{table}
% \clearpage
\begin{table*}[!t]
\centering
\caption{Experimentally
observed mass values of
mesons  which are used to
calculate the minimum $\chi^2$ and to
obtain the model parameters
listed in Table \ref{tab2}. Masses and uncertainties are in MeV; the optimized SHO scale $\beta$ is in GeV. Each theoretical entry is written as $M_{\rm th}(\beta,\sigma_{\rm th})$, where $\sigma_{\rm th}$ is the flavor-dependent theoretical error. The combined uncertainty is $\sigma_{\rm tot}=\sqrt{\sigma_{\rm exp}^2+\sigma_{\rm th}^2}$. Experimental inputs reproduce the fitted observation file.}
\label{tab1}
\begingroup
\fontsize{8}{9.5}\selectfont
\setlength{\tabcolsep}{2pt}
\renewcommand{\arraystretch}{1.06}
\resizebox{\textwidth}{!}{%
\begin{tabular}{@{}lccrc@{\hspace{10pt}}lccrc@{}}
\toprule
Meson & $n^{2S+1}L_J$ & $M_{\rm exp}\pm\sigma_{\rm exp}$ & $M_{\rm th}(\beta,\sigma_{\rm th})$ & $\sigma_{\rm tot}$ & Meson & $n^{2S+1}L_J$ & $M_{\rm exp}\pm\sigma_{\rm exp}$ & $M_{\rm th}(\beta,\sigma_{\rm th})$ & $\sigma_{\rm tot}$ \\
\midrule
$B_c^+$ & $1^{1}S_{0}$ & $6274.47\pm0.32$ & $6267\,(1.54,2.4)$ & $2.4$ & $D_{s1}^*(2860)^{\pm}$ & $1^{3}D_{1}$ & $2859\pm27$ & $2798\,(0.653,1.7)$ & $27$ \\ % observation_id=1,46
$B_c(2S)^{\pm}$ & $2^{1}S_{0}$ & $6871.2\pm1.0$ & $6856\,(1.16,2.4)$ & $2.6$ & $D_{s3}^*(2860)^{\pm}$ & $1^{3}D_{3}$ & $2860.5\pm7.0$ & $2856\,(0.499,1.7)$ & $7.2$ \\ % observation_id=2,47
$\eta_c(1S)$ & $1^{1}S_{0}$ & $2984.09\pm0.30$ & $2979\,(1.33,2.4)$ & $2.4$ & $D_{s0}(2590)^+$ & $2^{1}S_{0}$ & $2591\pm9.2$ & $2620\,(0.683,1.7)$ & $9.4$ \\ % observation_id=3,48
$\eta_c(2S)$ & $2^{1}S_{0}$ & $3637.8\pm0.60$ & $3621\,(0.979,2.4)$ & $2.5$ & $B^0$ & $1^{1}S_{0}$ & $5279.63\pm0.20$ & $5278\,(0.713,1.7)$ & $1.8$ \\ % observation_id=4,49
$J/\psi(1S)$ & $1^{3}S_{1}$ & $3096.9\pm0.0060$ & $3105\,(0.965,2.4)$ & $2.4$ & $B^*$ & $1^{3}S_{1}$ & $5324.75\pm0.20$ & $5334\,(0.599,1.7)$ & $1.8$ \\ % observation_id=5,50
$\psi(2S)$ & $2^{3}S_{1}$ & $3686.097\pm0.010$ & $3675\,(0.861,2.4)$ & $2.4$ & $B_2^*(5747)^+$ & $1^{3}P_{2}$ & $5737.3\pm0.70$ & $5754\,(0.411,1.7)$ & $1.9$ \\ % observation_id=6,51
$h_c(1P)$ & $1^{1}P_{1}$ & $3525.37\pm0.14$ & $3519\,(0.983,2.4)$ & $2.4$ & $B_s^0$ & $1^{1}S_{0}$ & $5366.91\pm0.11$ & $5362\,(0.825,1.7)$ & $1.7$ \\ % observation_id=7,52
$\chi_{c0}(1P)$ & $1^{3}P_{0}$ & $3415.5\pm0.19$ & $3424\,(1.03,2.4)$ & $2.4$ & $B_s^*$ & $1^{3}S_{1}$ & $5415.8\pm1.5$ & $5420\,(0.730,1.7)$ & $2.3$ \\ % observation_id=8,53
$\chi_{c1}(1P)$ & $1^{3}P_{1}$ & $3510.67\pm0.050$ & $3511\,(0.958,2.4)$ & $2.4$ & $B_{s2}^*(5840)^0$ & $1^{3}P_{2}$ & $5839.92\pm0.14$ & $5843\,(0.559,1.7)$ & $1.7$ \\ % observation_id=9,54
$\chi_{c2}(1P)$ & $1^{3}P_{2}$ & $3556.17\pm0.070$ & $3555\,(0.774,2.4)$ & $2.4$ & $\pi^{\pm}$ & $1^{1}S_{0}$ & $139.57039\pm0.00017$ & $173.1\,(0.759,2.0)$ & $2.0$ \\ % observation_id=10,55
$\chi_{c0}(3915)$ & $2^{3}P_{0}$ & $3922.1\pm1.8$ & $3866\,(0.847,2.4)$ & $3.0$ & $\pi(1300)$ & $2^{1}S_{0}$ & $1300\pm100$ & $1259\,(0.637,2.0)$ & $100$ \\ % observation_id=11,56
$\chi_{c2}(3930)$ & $2^{3}P_{2}$ & $3922.5\pm1.0$ & $3950\,(0.713,2.4)$ & $2.6$ & $\pi(1800)$ & $3^{1}S_{0}$ & $1810\pm10$ & $1831\,(0.577,2.0)$ & $10$ \\ % observation_id=12,57
$\psi(4040)$ & $3^{3}S_{1}$ & $4039.6\pm4.3$ & $4049\,(0.713,2.4)$ & $4.9$ & $b_1(1235)$ & $1^{1}P_{1}$ & $1229.5\pm3.2$ & $1166\,(0.650,2.0)$ & $3.8$ \\ % observation_id=13,58
$\psi(3770)$ & $1^{3}D_{1}$ & $3778.1\pm0.70$ & $3801\,(0.848,2.4)$ & $2.5$ & $\rho(770)^0$ & $1^{3}S_{1}$ & $775.26\pm0.23$ & $759.8\,(0.587,2.0)$ & $2.0$ \\ % observation_id=14,59
$\psi_2(3823)$ & $1^{3}D_{2}$ & $3823.51\pm0.34$ & $3826\,(0.713,2.4)$ & $2.4$ & $\rho(1450)$ & $2^{3}S_{1}$ & $1465\pm25$ & $1451\,(0.527,2.0)$ & $25$ \\ % observation_id=15,60
$\psi(4160)$ & $2^{3}D_{1}$ & $4191\pm5.0$ & $4125\,(0.711,2.4)$ & $5.5$ & $\rho(1900)$ & $3^{3}S_{1}$ & $1909\pm30$ & $1967\,(0.406,2.0)$ & $30$ \\ % observation_id=16,61
$\eta_b(1S)$ & $1^{1}S_{0}$ & $9398.7\pm2.0$ & $9397.7\,(2.33,2.4)$ & $3.1$ & $\rho(2150)$ & $4^{3}S_{1}$ & $2227\pm13$ & $2377\,(0.439,2.0)$ & $13$ \\ % observation_id=17,62
$\eta_b(2S)$ & $2^{1}S_{0}$ & $9999\pm4.2$ & $9988.5\,(1.63,2.4)$ & $4.8$ & $a_1(1260)$ & $1^{3}P_{1}$ & $1202\pm15$ & $1183\,(0.644,2.0)$ & $15$ \\ % observation_id=18,63
$\Upsilon(1S)$ & $1^{3}S_{1}$ & $9460.4\pm0.10$ & $9466.5\,(1.89,2.4)$ & $2.4$ & $a_1(1640)$ & $2^{3}P_{1}$ & $1655\pm16$ & $1756\,(0.545,2.0)$ & $16$ \\ % observation_id=19,64
$\Upsilon(2S)$ & $2^{3}S_{1}$ & $10023.4\pm0.50$ & $10019\,(1.45,2.4)$ & $2.4$ & $a_2(1320)$ & $1^{3}P_{2}$ & $1318.2\pm0.60$ & $1319\,(0.504,2.0)$ & $2.1$ \\ % observation_id=20,65
$h_b(1P)$ & $1^{1}P_{1}$ & $9899.3\pm0.80$ & $9895.0\,(1.55,2.4)$ & $2.5$ & $a_2(1700)$ & $2^{3}P_{2}$ & $1706\pm14$ & $1836\,(0.483,2.0)$ & $14$ \\ % observation_id=21,66
$\chi_{b0}(1P)$ & $1^{3}P_{0}$ & $9859.44\pm0.52$ & $9854.9\,(1.69,2.4)$ & $2.4$ & $\pi_2(1670)$ & $1^{1}D_{2}$ & $1670.6\pm2.1$ & $1631\,(0.556,2.0)$ & $2.9$ \\ % observation_id=22,67
$\chi_{b1}(1P)$ & $1^{3}P_{1}$ & $9892.78\pm0.40$ & $9889.1\,(1.55,2.4)$ & $2.4$ & $\pi_2(2005)$ & $2^{1}D_{2}$ & $1963\pm22$ & $2071\,(0.481,2.0)$ & $22$ \\ % observation_id=23,68
$\chi_{b2}(1P)$ & $1^{3}P_{2}$ & $9912.21\pm0.40$ & $9910.9\,(1.23,2.4)$ & $2.4$ & $\rho(1700)$ & $1^{3}D_{1}$ & $1720\pm20$ & $1421\,(0.679,2.0)$ & $20$ \\ % observation_id=24,69
$\Upsilon(3S)$ & $3^{3}S_{1}$ & $10355.1\pm0.50$ & $10356\,(1.19,2.4)$ & $2.4$ & $\rho_3(1690)$ & $1^{3}D_{3}$ & $1688.8\pm2.1$ & $1678\,(0.443,2.0)$ & $2.9$ \\ % observation_id=25,70
$h_b(2P)$ & $2^{1}P_{1}$ & $10259.8\pm1.2$ & $10258\,(1.21,2.4)$ & $2.7$ & $\rho_3(1990)$ & $2^{3}D_{3}$ & $1982\pm14$ & $2121\,(0.406,2.0)$ & $14$ \\ % observation_id=26,71
$\chi_{b0}(2P)$ & $2^{3}P_{0}$ & $10232.5\pm0.64$ & $10230\,(1.33,2.4)$ & $2.5$ & $a_4(1970)$ & $1^{3}F_{4}$ & $1967\pm16$ & $1983\,(0.399,2.0)$ & $16$ \\ % observation_id=27,72
$\chi_{b1}(2P)$ & $2^{3}P_{1}$ & $10255.46\pm0.55$ & $10254\,(1.23,2.4)$ & $2.4$ & $\rho_3(2250)$ & $3^{3}D_{3}$ & $2248\pm36$ & $2498\,(0.381,2.0)$ & $36$ \\ % observation_id=28,73
$\chi_{b2}(2P)$ & $2^{3}P_{2}$ & $10268.65\pm0.55$ & $10269\,(1.11,2.4)$ & $2.4$ & $\rho_5(2350)$ & $1^{3}G_{5}$ & $2330\pm35$ & $2254\,(0.367,2.0)$ & $35$ \\ % observation_id=29,74
$\Upsilon(4S)$ & $4^{3}S_{1}$ & $10579.4\pm1.2$ & $10610\,(1.01,2.4)$ & $2.7$ & $a_0(980)$ & $1^{3}P_{0}$ & $980\pm20$ & $568.3\,(0.843,2.0)$ & $20$ \\ % observation_id=30,75
$\Upsilon(10860)$ & $5^{3}S_{1}$ & $10885.2\pm2.1$ & $10817\,(0.884,2.4)$ & $3.2$ & $a_0(1450)$ & $2^{3}P_{0}$ & $1439\pm34$ & $1495\,(0.664,2.0)$ & $34$ \\ % observation_id=31,76
$\Upsilon(11020)$ & $6^{3}S_{1}$ & $11000\pm4.0$ & $10994\,(0.713,2.4)$ & $4.6$ & $K^0$ & $1^{1}S_{0}$ & $497.611\pm0.013$ & $471.5\,(0.764,1.5)$ & $1.5$ \\ % observation_id=32,77
$\chi_{b1}(3P)$ & $3^{3}P_{1}$ & $10513.42\pm0.67$ & $10525\,(1.02,2.4)$ & $2.5$ & $K(1460)$ & $2^{1}S_{0}$ & $1482.4\pm16$ & $1431\,(0.641,1.5)$ & $16$ \\ % observation_id=33,78
$\chi_{b2}(3P)$ & $3^{3}P_{2}$ & $10524.02\pm0.78$ & $10537\,(0.951,2.4)$ & $2.5$ & $K^*(892)^0$ & $1^{3}S_{1}$ & $895.56\pm0.20$ & $898.6\,(0.603,1.5)$ & $1.5$ \\ % observation_id=34,79
$D^0$ & $1^{1}S_{0}$ & $1864.85\pm0.040$ & $1862\,(0.713,1.8)$ & $1.8$ & $K^*(1410)$ & $2^{3}S_{1}$ & $1414\pm15$ & $1580\,(0.538,1.5)$ & $15$ \\ % observation_id=35,80
$D_0(2550)^0$ & $2^{1}S_{0}$ & $2549\pm19$ & $2513\,(0.591,1.8)$ & $19$ & $K_2^*(1430)^0$ & $1^{3}P_{2}$ & $1432.4\pm1.3$ & $1440\,(0.528,1.5)$ & $2.0$ \\ % observation_id=36,81
$D^*(2007)^0$ & $1^{3}S_{1}$ & $2006.86\pm0.050$ & $2013\,(0.559,1.8)$ & $1.8$ & $K^*(1680)$ & $1^{3}D_{1}$ & $1718\pm18$ & $1614\,(0.674,1.5)$ & $18$ \\ % observation_id=37,82
$D_2^*(2460)^0$ & $1^{3}P_{2}$ & $2461.1\pm0.70$ & $2462\,(0.508,1.8)$ & $1.9$ & $K_3^*(1780)$ & $1^{3}D_{3}$ & $1779\pm8.0$ & $1795\,(0.468,1.5)$ & $8.1$ \\ % observation_id=38,83
$D_0^*(2300)^0$ & $1^{3}P_{0}$ & $2343\pm10$ & $2235\,(0.713,1.8)$ & $10$ & $K_4^*(2045)$ & $1^{3}F_{4}$ & $2048\pm8.5$ & $2094\,(0.406,1.5)$ & $8.6$ \\ % observation_id=39,84
$D_3^*(2750)$ & $1^{3}D_{3}$ & $2763.1\pm3.2$ & $2756\,(0.443,1.8)$ & $3.7$ & $K_5^*(2380)^0$ & $1^{3}G_{5}$ & $2382\pm24$ & $2359\,(0.386,1.5)$ & $24$ \\ % observation_id=40,85
$D_s^{\pm}$ & $1^{1}S_{0}$ & $1969\pm1.4$ & $1963\,(0.831,1.7)$ & $2.2$ & $K_0^*(700)$ & $1^{3}P_{0}$ & $838\pm11$ & $900.8\,(0.789,1.5)$ & $11$ \\ % observation_id=41,86
$D_s^{*\pm}$ & $1^{3}S_{1}$ & $2112.2\pm0.40$ & $2110\,(0.713,1.7)$ & $1.8$ & $K_0^*(1430)$ & $2^{3}P_{0}$ & $1425\pm50$ & $1692\,(0.625,1.5)$ & $50$ \\ % observation_id=42,87
$D_{s1}^*(2700)$ & $2^{3}S_{1}$ & $2714\pm5.0$ & $2681\,(0.606,1.7)$ & $5.3$ & $\phi(1020)$ & $1^{3}S_{1}$ & $1019.46\pm0.016$ & $1023\,(0.600,1.2)$ & $1.2$ \\ % observation_id=43,88
$D_{s0}^*(2317)^{\pm}$ & $1^{3}P_{0}$ & $2318\pm0.70$ & $2373\,(0.755,1.7)$ & $1.9$ & $\phi(1680)$ & $2^{3}S_{1}$ & $1680\pm20$ & $1700\,(0.590,1.2)$ & $20$ \\ % observation_id=44,89
$D_{s2}^*(2573)$ & $1^{3}P_{2}$ & $2569.1\pm0.80$ & $2561\,(0.574,1.7)$ & $1.9$ & $\phi_3(1850)$ & $1^{3}D_{3}$ & $1854\pm7.0$ & $1905\,(0.505,1.2)$ & $7.1$ \\ % observation_id=45,90
\bottomrule
\end{tabular}%
}
\endgroup
\end{table*}

\section{The Meson Spectrum}\label{sec3}

{Using the model parameters determined in Sec.~\ref{sec2}, we present the mass spectra of light, heavy-light, and heavy-heavy mesons obtained from the final 90-state fit. In this section, the results will be compared with available experimental data and other theoretical calculations, with emphasis on low-lying levels, spin-dependent splittings, and higher radial and orbital excitations. We will discuss possible spectroscopic assignments for selected observed states using mass comparisons and available quantum-number information, and provide mass predictions for states not yet experimentally established. Both the agreement and the discrepancies across different flavor sectors are examined, together with the limitations of assignments based on masses alone. Finally, the \(t\bar t\) system is considered as a qualitative extrapolation of the potential model, and the spectrum of this meson system is obtained.
}

\subsection{The light meson spectrum}

In this subsection, we discuss the spectra in the order of
Tables~\ref{tab:merged_light_negative},
\ref{tab:merged_strange_negative}, and~\ref{tab:merged_strange_positive}:
unflavored light mesons, negative-parity strange mesons, and
positive-parity strange mesons, respectively. The experimental data \cite{ParticleDataGroup:2026mpi}
, the GI model~\cite{Godfrey:1985xj},
relativistic quasipotential model (RQM)~\cite{Ebert:2009ub}, and BS (model A)~\cite{Koll:2000ke}
results provide the comparisons shown in these
tables.

\subsubsection{Unflavored light mesons}

For the negative-parity states in Table~\ref{tab:merged_light_negative},
the fitted nonstrange $1{}^1S_0$ and $1{}^3S_1$ masses are
$173.1$ and $759.8~\mathrm{MeV}$, differing from $\pi^\pm$ and
$\rho(770)$ by $+33.5$ and $-15.5~\mathrm{MeV}$.
The $2{}^1S_0$ and $3{}^1S_0$ states, fitted using $\pi(1300)$
and $\pi(1800)$, occur at $1259$ and $1831~\mathrm{MeV}$.
The $2S$ and $3S$ assignments in the $\pi$ family follow the radial classification used in Ref.~\cite{Wang:2017iai}. The calculated masses differ from the tabulated experimental central values by $-41$ and $+21~\mathrm{MeV}$, respectively.
The fitted $\rho(1450)$ and $\rho(1900)$ assignments to
$2{}^3S_1$ and $3{}^3S_1$ likewise agree with the classification
studied in Refs.~\cite{He:2013ttg,Li:2021qgz}.

The isoscalar columns compare physical candidates with the unmixed
$n\bar n$ and $s\bar s$ basis spectra.
The nonstrange $2{}^1S_0$ and $3{}^1S_0$ states lie $35~\mathrm{MeV}$
below $\eta(1295)$ and $80~\mathrm{MeV}$ above $\eta(1760)$,
respectively. These additional comparisons are compatible with the
$2S$ and $3S$ radial assignments considered in Ref.~\cite{Yu:2011ta}. 
For the vector states, the same nonstrange $2{}^3S_1$ and $3{}^3S_1$
states differ from $\omega(1420)$ and $\omega(1960)$ by
$+41$ and $+7~\mathrm{MeV}$.
Within the approximation of a common nonstrange basis, these comparisons provide mass-based support for the corresponding radial candidate assignments discussed in Refs.~\cite{Wang:2019jch,Wang:2022xxi}.

For the strangeonium vectors, the $1{}^3S_1$ state was fitted
using $\phi(1020)$ and gives $1023~\mathrm{MeV}$.
The $2{}^3S_1$ mass of $1700~\mathrm{MeV}$ is a fitted result, $20~\mathrm{MeV}$ above $\phi(1680)$.
It is compatible with the first radial vector excitation assigned
to this resonance in Ref.~\cite{Pang:2019ttv}. 
We predict that the mass of $3{}^3S_1$ is $2197~\mathrm{MeV}$, which supports  the $3{}^3S_1$ assignment  to $\phi(2170)$ \cite{Pang:2019ttv}.

For the $D$ waves, the fitted $1{}^3D_3$ mass of
$1678~\mathrm{MeV}$ is $10.8~\mathrm{MeV}$ below
$\rho_3(1690)$ and reproduces the state under the assignment used
in Ref.~\cite{He:2013ttg}.
The fitted masses of $1{}^1D_2$ and $2{}^1D_2$ states are
$1631$ and $2071~\mathrm{MeV}$, with differences of
$-39.6$ and $+108~\mathrm{MeV}$ from $\pi_2(1670)$ and
$\pi_2(2005)$. The latter is one of the first-radial-excitation
candidates examined in Ref.~\cite{Wang:2014sea}.
The same basis levels lie $14$ and $41~\mathrm{MeV}$ above
$\eta_2(1645)$ and $\eta_2(2030)$, respectively.
These isoscalar masses were not fit inputs. Their masses provide comparisons for the nonstrange $1{}^1D_2$ assignment of
$\eta_2(1645)$ and the $2{}^1D_2$ candidate interpretation of
$\eta_2(2030)$ studied in Ref.~\cite{Wang:2014sea}.

The positive-parity continuation of Table~\ref{tab:merged_light_negative}
contains the $P$-, $F$-, and $H$-wave states.
The fitted $a_1(1260)$ and $a_2(1320)$ states occur at
$1183$ and $1319~\mathrm{MeV}$, with differences of
$-19$ and $+0.8~\mathrm{MeV}$ from the inputs in Table~\ref{tab1}.
For the unfitted nonstrange $2{}^1P_1$ state, the prediction
$1717~\mathrm{MeV}$ lies $123~\mathrm{MeV}$ above
$h_1(1595)$ and can be compared with its proposed first-radial-excitation
classification in Ref.~\cite{Chen:2015iqa}.
In the strangeonium sector, $h_1(1415)$ is an unfitted comparison for the
$1{}^1P_1$ state, giving $1453~\mathrm{MeV}$. The $1{}^3P_1$ and $1{}^3P_2$ masses
of $1457$ and $1554~\mathrm{MeV}$ are predictions for
states not used in the fit. They lie $29.3$ and
$36.7~\mathrm{MeV}$ above $f_1(1420)$ and $f_2'(1525)$,
respectively, supporting their placement among the predominantly
strange $1P$ states considered in
Refs.~\cite{Chen:2015iqa,Ye:2012gu}.

In the fitting programe, we assign $a_0(980)$ and $a_0(1450)$ to $1{}^3P_0$ and $2{}^3P_0$, respectively. Their masses are $568$ and $1495~\mathrm{MeV}$, giving residuals of $-412$ and $+56~\mathrm{MeV}$;  the low scalar assignment therefore remains poorly reproduced.

For the higher-spin positive-parity states, the $1{}^3F_4$ state was constrained by the $a_4(1970)$ input, while $2{}^3F_4$ is a prediction.
Their masses of $1983$ and $2372~\mathrm{MeV}$ also lie
$35.0~\mathrm{MeV}$ below $f_4(2050)$ and
$52~\mathrm{MeV}$ above $f_4(2300)$, neither of which was
included in the fit. These additional comparisons support the
$1{}^3F_4$ and $2{}^3F_4$ isoscalar assignments of
Ref.~\cite{Pang:2015eha}.
The masses of unfitted $1{}^1F_3$ and $2{}^1F_3$ states are predicted at
$1977$ and $2355~\mathrm{MeV}$, giving differences of
$-48$ and $+110~\mathrm{MeV}$ from $h_3(2025)$ and
$b_3(2245)$. These masses are compatible with the respective
$1F$ and $2F$ classifications considered in the same study.
The $1{}^3H_6$ prediction, $2499~\mathrm{MeV}$, is only
$29~\mathrm{MeV}$ above the  $f_6(2510)$ mass and
provides further mass-based support for its lowest $H$-wave
assignment~\cite{Pang:2015eha}.

The lowest nonstrange and strangeonium $1{}^3H_5$ masses are $2534$ and $2730~\mathrm{MeV}$, respectively, providing independent model-based mass references for the missing $5^{++}$ family discussed in Ref.~\cite{Pang:2018gcn}.

The remaining predictions for higher unflavored light-meson excitations are listed in Table~\ref{tab:merged_light_negative}.

\subsubsection{Negative-parity strange mesons}
\par
Table~\ref{tab:merged_strange_negative} presents the $S$-, $D$-,
and $G$-wave negative-parity $q\bar s$ states, with $q=u,d$.
Their calculated masses, $471.5$ and $898.6~\mathrm{MeV}$,
are compared here with the charged references, giving differences
of $-22.2$ and $+6.72~\mathrm{MeV}$.
The corresponding NR masses are $493.0$ and
$898.0~\mathrm{MeV}$~\cite{Bhavyashri:2005rb}. 
The fitted $2{}^1S_0$ state at $1431~\mathrm{MeV}$ lies
$51.4~\mathrm{MeV}$ below the experimental value of $K(1460)$.
The radial assignment adopted in this fit is the same as that in
Ref.~\cite{Pang:2017dlw}.
The $3S$ and higher states were not used as inputs and are predictions:
for example, the $3{}^1S_0$ and $3{}^3S_1$ masses are
$1982$ and $2085~\mathrm{MeV}$, while the $6S$ pair is
predicted at $3087$ and $3137~\mathrm{MeV}$.
Their decreasing hyperfine separation, from $103$ to
$50.0~\mathrm{MeV}$ between these two radial states, supplies
a test for future measurements.

The $1{}^3D_3$ state was fitted using $K_3^*(1780)$.
Its calculated mass of $1795~\mathrm{MeV}$ is
$16~\mathrm{MeV}$ above the adopted input in Table~\ref{tab1} under the
same assignment as in Ref.~\cite{Pang:2017dlw}.
The $J^P=2^-$ partners, however, were not fit inputs:
the predicted $1D$ and $1D^\prime$ masses of $1756$ and
$1768~\mathrm{MeV}$ lie $17$ and $51.0~\mathrm{MeV}$ below
$K_2(1770)$ and $K_2(1820)$.
Their placement as the mixed $1D$ pair is compatible with the
classification of Ref.~\cite{Pang:2017dlw}, although the calculated
$12~\mathrm{MeV}$ separation is smaller than the observed
$46~\mathrm{MeV}$ separation.
For this pair, diagonalization gives $\theta_{1D}=-79.4^\circ$
 in our convention, in which the primed state is the heavier member. 
The corresponding computed angle from GI model
is $12.8^\circ$ in its lower-mass-state convention.

For the first radially excited $J^P=2^-$ doublet, we predict masses
of $2191$ and $2204~\mathrm{MeV}$, with
$\theta_{2D}=88.6^\circ$; both the masses and the mixing angle
are predictions.
The higher predicted state is $43.0~\mathrm{MeV}$ below
$K_2(2250)$, supporting its assignment to the first radially
excited $D$-wave family.
Ref.~\cite{Pang:2017dlw} likewise discusses this state as a
$2D^\prime$ candidate and infers a partner mass of approximately
$1994~\mathrm{MeV}$ from its mass and mixing relations;
the present $2D$ prediction of $2191~\mathrm{MeV}$ is approximately
$197~\mathrm{MeV}$ higher.
With its original convention retained, GI model gives a mixing
angle of $20.2^\circ$ for this pair.

The higher $3D$ and $4D$ doublets have masses of
$2562$ and $2575~\mathrm{MeV}$, and
$2887$ and $2899~\mathrm{MeV}$, respectively,
with corresponding mixing angles of $88.9^\circ$ and $89.9^\circ$.
Ref.~\cite{Pang:2017dlw} gives unmixed masses of
$2380$ and $2388~\mathrm{MeV}$ for
$3{}^1D_2$ and $3{}^3D_2$, respectively,
and $2570$ and $2575~\mathrm{MeV}$ for
$4{}^1D_2$ and $4{}^3D_2$, respectively.
Although these unmixed states cannot be equated with the
mixed-state eigenvalues obtained here, the corresponding states
in the present calculation have a higher overall mass scale.
This higher mass scale may be related to the different parameter
set obtained from the fit across flavor sectors.
Compared with Ref.~\cite{Pang:2017dlw}, the present calculation
uses different confinement and screening parameters for light mesons.
These choices modify the slope of the confining term
as the separation increases and may therefore contribute to
shifts of highly excited states.
The constituent-quark masses and the constant term obtained in the global fit may also jointly
contribute to the state positions.
Furthermore, in comparing highly excited strange-meson spectra,
Ref.~\cite{Taboada-Nieto:2022igy} also relates the differences
between calculations primarily to how the screened confining
interaction is determined and to the distinction between
a global fit across flavor sectors and a fit restricted to
the strange-meson sector.

For comparison, a Regge-phenomenology study gives masses of
$2504\pm190~\mathrm{MeV}$ and
$2804\pm210~\mathrm{MeV}$ for
$3{}^3D_2$ and $4{}^3D_2$, respectively;
their central values are close to the mass scale of the higher
excitations obtained here~\cite{Oudichhya:2023lva}.

For the $G$ waves, the $K_5^*(2380)$ input was assigned to $1{}^3G_5$ in the fit. The calculated mass is $2359~\mathrm{MeV}$, with a residual of $-23~\mathrm{MeV}$.  This assignment is also adopted in the literature~\cite{Pang:2017dlw}. 
The neighboring $J^P=4^-$ partners are predicted at $2374$
and $2386~\mathrm{MeV}$ with $\theta_{1G}=76.8^\circ$.
Their masses and mixing are predictions for missing states.
The corresponding GI angle is $10.8^\circ$ in its own convention.
For $K_4(2500)$, Refs.~\cite{Pang:2017dlw,Wang:2025lin}
both discuss a $2G$ interpretation.
However, the corresponding mass of $2708~\mathrm{MeV}$
listed in the present table exceeds the experimental central value
by $218~\mathrm{MeV}$.
A recent study of high-spin strange mesons gives an unmixed
$2G$ basis-state mass of approximately $2538~\mathrm{MeV}$,
which is closer to the experimental central value than the present
result~\cite{Wang:2025lin}.
This upward deviation resembles that found for the higher radial
$D$-wave excitations and more likely reflects differences in
extrapolating the highly excited spectrum with different parameter sets.

\onecolumngrid
\clearpage

\setlength{\paperwidth}{11in}
\setlength{\paperheight}{8.5in}
\newgeometry{left=0.5in,right=0.5in,top=0.5in,bottom=0.5in}
\pdfpagewidth=11in
\pdfpageheight=8.5in

\setlength{\textwidth}{10in}
\setlength{\hsize}{10in}
\setlength{\linewidth}{10in}
\setlength{\columnwidth}{10in}

\begingroup
\fontsize{8}{9.5}\selectfont
\setlength{\tabcolsep}{1.75pt}
\setlength{\LTcapwidth}{\textwidth}
\renewcommand{\arraystretch}{1.12}
\begin{longtable}{@{}cccccccccccccccc@{}}
\caption{
Mass spectra of unflavored light mesons: negative-parity states,
with the positive-parity states continued on the next page.
All masses are in MeV.
The nonstrange results in ``This work'' apply to both the isovector
and the $n\bar n$ isoscalar columns;
$n\bar n=(u\bar u+d\bar d)/\sqrt{2}$ and $s\bar s$ are unmixed
flavor bases.
For experimental states included in the fit, the masses and uncertainties
are adopted from Table~\ref{tab1}; the other experimental entries follow the PDG compilation~\cite{ParticleDataGroup:2026mpi}.
Comparison models: GI~\cite{Godfrey:1985xj},
relativistic quasipotential model (RQM)~\cite{Ebert:2009ub},
and instantaneous Bethe--Salpeter model A
(BS)~\cite{Koll:2000ke}. 
}
\label{tab:merged_light_negative}\\

\toprule
 & &
\multicolumn{6}{c}{$I=1$} &
\multicolumn{2}{c}{$I=0$, $n\bar n$} &
\multicolumn{6}{c}{$I=0$, $s\bar s$}\\
\cmidrule(lr){3-8}
\cmidrule(lr){9-10}
\cmidrule(lr){11-16}

$n^{2S+1}L_J$ & $J^{PC}$ &
This work & Expt. state & Mass & GI & RQM & BS &
Expt. state & Mass &
This work & Expt. state & Mass & GI & RQM & BS\\
\midrule
\endfirsthead

\caption[]{Negative-parity unflavored light mesons (MeV), continued.}\\
\toprule
 & &
\multicolumn{6}{c}{$I=1$} &
\multicolumn{2}{c}{$I=0$, $n\bar n$} &
\multicolumn{6}{c}{$I=0$, $s\bar s$}\\
\cmidrule(lr){3-8}
\cmidrule(lr){9-10}
\cmidrule(lr){11-16}

$n^{2S+1}L_J$ & $J^{PC}$ &
This work & Expt. state & Mass & GI & RQM & BS &
Expt. state & Mass &
This work & Expt. state & Mass & GI & RQM & BS\\
\midrule
\endhead

\midrule
\multicolumn{16}{r}{Continued on next page}\\
\endfoot

\bottomrule
\endlastfoot

$1{}^1S_0$ & $0^{-+}$ & $173.1$ & $\pi^\pm$ & $139.57039\pm0.00017$ &
$150.0$ & $154.0$ & $138.0$ &
$\eta(548)$ & $547.862\pm0.017$ & $672.1$ & $\eta'(958)$ & $957.78\pm0.06$ &
$960.0$ & $743.0$ & $\cdots$\\

$1{}^3S_1$ & $1^{--}$ & $759.8$ & $\rho(770)$ & $775.26\pm0.23$ &
$770.0$ & $776.0$ & $778.0$ &
$\omega(782)$ & $782.66\pm0.13$ & $1023$ & $\phi(1020)$ & $1019.46\pm0.016$ &
$1020$ & $1038$ & $954.0$\\

$2{}^1S_0$ & $0^{-+}$ & $1259$ & $\pi(1300)$~\cite{Wang:2017iai} & $1300\pm100$ &
$1300$ & $1292$ & $1357$ &
$\eta(1295)$~\cite{Yu:2011ta} & $1294\pm4$ & $1571$ & $\eta(1475)$ & $1476\pm4$ &
$1630$ & $1536$ & $1533$\\

$2{}^3S_1$ & $1^{--}$ & $1451$ & $\rho(1450)$~\cite{He:2013ttg} & $1465\pm25$ &
$1450$ & $1486$ & $1553$ &
$\omega(1420)$~\cite{Wang:2019jch} & $1410\pm60$ & $1700$ & $\phi(1680)$~\cite{Pang:2019ttv} & $1680\pm20$ &
$1690$ & $1698$ & $1605$\\

$3{}^1S_0$ & $0^{-+}$ & $1831$ & $\pi(1800)$~\cite{Wang:2017iai} & $1810\pm10$ &
$1880$ & $1788$ & $2012$ &
$\eta(1760)$~\cite{Yu:2011ta} & $1751\pm15$ & $2108$ & $\eta(2100)$ & $2103\pm50$ &
$\cdots$ & $2085$ & $\cdots$\\

$3{}^3S_1$ & $1^{--}$ & $1967$ & $\rho(1900)$~\cite{Li:2021qgz,He:2013ttg} & $1909\pm30.232$ &
$2000$ & $1921$ & $2118$ &
$\omega(1960)$~\cite{Wang:2022xxi} & $1960\pm25$ & $2197$ & $\phi(2170)$ & $2164\pm5$ &
$\cdots$ & $2119$ & $\cdots$\\

$4{}^1S_0$ & $0^{-+}$ & $2277$ & $\pi(2070)$~\cite{Wang:2017iai} & $2070\pm35$ &
$\cdots$ & $2073$ & $2498$ &
$\eta(2100)$ ~\cite{Wang:2017iai}& $2050^{+30+75}_{-24-26}$ & $2530$ & & &
$\cdots$ & $2439$ & $2177$\\

$4{}^3S_1$ & $1^{--}$ & $2377$ & $\rho(2150)$~\cite{He:2013ttg} & $2227\pm12.728$ &
$\cdots$ & $2195$ & $2567$ &
$\omega(2330)$~\cite{Wang:2022xxi}  & $2300\pm30$ & $2597$ & & &
$\cdots$ & $2472$ & $\cdots$\\

$5{}^1S_0$ & $0^{-+}$ & $2645$ & $\pi(2360)$~\cite{Wang:2017iai} & $2360\pm25$ &
$\cdots$ & $2385$ & $2898$ &
$\eta(2320)$ & $2320\pm15$ & $2884$ & & &
$\cdots$ & $2749$ & $\cdots$\\

$5{}^3S_1$ & $1^{--}$ & $2726$ & & &
$\cdots$ & $2491$ & $\cdots$ &
& & $2938$ & & &
$\cdots$ & $2782$ & $\cdots$\\

$6{}^1S_0$ & $0^{-+}$ & $2968$ & & &
$\cdots$ & $\cdots$ & $\cdots$ &
& & $3194$ & & &
$\cdots$ & $\cdots$ & $\cdots$\\

$6{}^3S_1$ & $1^{--}$ & $3032$ & & &
$\cdots$ & $\cdots$ & $\cdots$ &
& & $3238$ & & &
$\cdots$ & $\cdots$ & $\cdots$\\

$1{}^1D_2$ & $2^{-+}$ & $1631$ & $\pi_2(1670)$~\cite{Wang:2014sea} & $1670.6\pm2.05$ &
$1680$ & $1643$ & $1633$ &
$\eta_2(1645)$~\cite{Wang:2014sea} & $1617\pm5$ & $1875$ & $\eta_2(1870)$~\cite{Wang:2014sea}  & $1842\pm8$ &
$1890$ & $1909$ & $\cdots$\\

$1{}^3D_1$ & $1^{--}$ & $1421$ & $\rho(1700)$~\cite{He:2013ttg} & $1720\pm20$ &
$1660$ & $1557$ & $1605$ &
$\omega(1650)$~\cite{Wang:2022xxi}  & $1670\pm30$ & $1776$ & & &
$1880$ & $1845$ & $\cdots$\\

$1{}^3D_2$ & $2^{--}$ & $1645$ & & &
$1700$ & $1661$ & $\cdots$ &
& & $1884$ & & &
$1910$ & $1908$ & $\cdots$\\

$1{}^3D_3$ & $3^{--}$ & $1678$ & $\rho_3(1690)$~\cite{He:2013ttg} & $1688.8\pm2.1$ &
$1680$ & $1714$ & $1698$ &
$\omega_3(1670)$~\cite{Wang:2022xxi}  & $1667\pm4$ & $1905$ & $\phi_3(1850)$ & $1854\pm7$ &
$1900$ & $1950$ & $1899$\\

$2{}^1D_2$ & $2^{-+}$ & $2071$ & $\pi_2(2005)$~\cite{Wang:2014sea} & $1963\pm22$ &
$2130$ & $1960$ & $2156$ &
$\eta_2(2030)$~\cite{Wang:2014sea} & $2030\pm5\pm15$ & $2306$ & & &
$\cdots$ & $2321$ & $\cdots$\\

$2{}^3D_1$ & $1^{--}$ & $1936$ & $\rho(2000)$~\cite{He:2013ttg} & $2000\pm30$ &
$2150$ & $1895$ & $\cdots$ &
& & $2233$&&&
$\cdots$ & $2258$ & $\cdots$\\

$2{}^3D_2$ & $2^{--}$ & $2089$ & $\rho_2(1940)$~\cite{Guo:2019wpx} & $1940\pm40$ &
$2150$ & $1983$ & $\cdots$ &
$\omega_2(1975)$~\cite{Guo:2019wpx} & $1975\pm20$ & $2315$ & & &
$\cdots$ & $2323$ & $\cdots$\\

$2{}^3D_3$ & $3^{--}$ & $2121$ & $\rho_3(1990)$ ~\cite{He:2013ttg}& $1982\pm14$ &
$2130$ & $2066$ & $2157$ &
$\omega_3(1945)$~\cite{Wang:2022xxi}  & $1945\pm20$ & $2337$ & & &
$\cdots$ & $2338$ & $\cdots$\\

$3{}^1D_2$ & $2^{-+}$ & $2446$ & $\pi_2(2285)$~\cite{Wang:2014sea}  & $2285\pm32$ &
$\cdots$ & $2216$ & $2592$ &
$\eta_2(2250)$~\cite{Wang:2014sea}  & $2248\pm20$ & $2673$ & & &
$\cdots$ & $2662$ & $\cdots$\\

$3{}^3D_1$ & $1^{--}$ & $2339$ & $\rho(2270)$~\cite{He:2013ttg} & $2265\pm40$ &
$\cdots$ & $2168$ & $\cdots$ &
$\omega(2220)$~\cite{Wang:2022xxi}   & $2188\pm21$ & $2610$ & & &
$\cdots$ & $2607$ & $\cdots$\\

$3{}^3D_2$ & $2^{--}$ & $2464$ & $\rho_2(2225)$~\cite{Guo:2019wpx} & $2225\pm35$ &
$\cdots$ & $2241$ & $\cdots$ &
$\omega_2(2195)$~\cite{Guo:2019wpx} & $2195\pm30$ & $2681$ & & &
$\cdots$ & $2667$ & $\cdots$\\

$3{}^3D_3$ & $3^{--}$ & $2498$ & $\rho_3(2250)$~\cite{He:2013ttg} & $2248\pm36.235$ &
$\cdots$ & $2309$ & $2208$ &
$\omega_3(2285)$~\cite{Wang:2022xxi}  & $2285\pm60$ & $2705$ & & &
$\cdots$ & $2727$ & $\cdots$\\

$4{}^1D_2$ & $2^{-+}$ & $2775$ & & &
$\cdots$ & $\cdots$ & $\cdots$ &
& & $2994$ & & &
$\cdots$ & $\cdots$ & $\cdots$\\

$4{}^3D_1$ & $1^{--}$ & $2686$ & & &
$\cdots$ & $\cdots$ & $\cdots$ &
& & $2940$ & & &
$\cdots$ & $\cdots$ & $\cdots$\\

$4{}^3D_2$ & $2^{--}$ & $2792$ & & &
$\cdots$ & $\cdots$ & $\cdots$ &
& & $3002$ & & &
$\cdots$ & $\cdots$ & $\cdots$\\

$4{}^3D_3$ & $3^{--}$ & $2826$ & & &
$\cdots$ & $\cdots$ & $\cdots$ &
& & $3027$ & & &
$\cdots$ & $\cdots$ & $\cdots$\\

$1{}^1G_4$ & $4^{-+}$ & $2268$ & $\pi_4(2250)$~\cite{Pang:2015eha} & $2250\pm15$ &
$2330$ & $2092$ & $\cdots$ &
$\eta_4(2330)$ ~\cite{Pang:2015eha} & $2328\pm28$ & $2477$ & & &
$2510$ & $2469$ & $\cdots$\\

$1{}^3G_3$ & $3^{--}$ & $2249$ & & &
$2370$ & $2002$ & $\cdots$ &
$\omega_3(2255)$~\cite{Wang:2022xxi}  & $2255\pm15$ & $2483$ & & &
$2540$ & $2403$ & $\cdots$\\

$1{}^3G_4$ & $4^{--}$ & $2280$ & $\rho_4(2230)$~\cite{Pang:2015eha}  & $2230\pm25$ &
$2340$ & $2122$ & $\cdots$ &
$\omega_4(2250)$~\cite{Pang:2015eha}  & $2250\pm30$ & $2486$ & & &
$2520$ & $2481$ & $\cdots$\\

$1{}^3G_5$ & $5^{--}$ & $2254$ & $\rho_5(2350)$~\cite{Pang:2015eha}  & $2330\pm35$ &
$2300$ & $2264$ & $2279$ &
$\omega_5(2250)$~\cite{Pang:2015eha}  & $2250\pm70$ & $2460$ & & &
$2470$ & $2559$ & $\cdots$\\

\end{longtable}

\clearpage
\addtocounter{table}{-1}
\renewcommand{\theHtable}{\arabic{table}.continued}
\renewcommand{\arraystretch}{1.0}

\begin{longtable}{@{}cccccccccccccccc@{}}
\caption[]{
Mass spectra of unflavored light mesons (continued):
positive-parity states.
Masses, flavor bases, assignment references, and comparison models
follow the definitions in the first part of
Table~\ref{tab:merged_light_negative}.
}
\label{tab:merged_light_positive}\\

\toprule
 & &
\multicolumn{6}{c}{$I=1$} &
\multicolumn{2}{c}{$I=0$, $n\bar n$} &
\multicolumn{6}{c}{$I=0$, $s\bar s$}\\
\cmidrule(lr){3-8}
\cmidrule(lr){9-10}
\cmidrule(lr){11-16}

$n^{2S+1}L_J$ & $J^{PC}$ &
This work & Expt. state & Mass & GI & RQM & BS &
Expt. state & Mass &
This work & Expt. state & Mass & GI & RQM & BS\\
\midrule
\endfirsthead

\caption[]{Positive-parity unflavored light mesons (MeV), continued.}\\
\toprule
 & &
\multicolumn{6}{c}{$I=1$} &
\multicolumn{2}{c}{$I=0$, $n\bar n$} &
\multicolumn{6}{c}{$I=0$, $s\bar s$}\\
\cmidrule(lr){3-8}
\cmidrule(lr){9-10}
\cmidrule(lr){11-16}

$n^{2S+1}L_J$ & $J^{PC}$ &
This work & Expt. state & Mass & GI & RQM & BS &
Expt. state & Mass &
This work & Expt. state & Mass & GI & RQM & BS\\
\midrule
\endhead

\midrule
\multicolumn{16}{r}{Continued on next page}\\
\endfoot

\bottomrule
\endlastfoot

$1{}^1P_1$ & $1^{+-}$ & $1166$ & $b_1(1235)$ & $1229.5\pm3.2$ &
$1220$ & $1258$ & $1240$ &
$h_1(1170)$ & $1166\pm6$ & $1453$ & $h_1(1415)$ & $1416.5\pm8$ &
$1470$ & $1485$ & $1454$\\

$1{}^3P_0$ & $0^{++}$ & $568.3$ & $a_0(980)$~\cite{Liu:2008zze}  & $980\pm20$ &
$1090$ & $1176$ & $1321$ &
$f_0(500)$~\cite{Liu:2008zze} & $400--800$ & $1133$ & $f_0(980)$ & $990\pm20$ &
$1360$ & $1420$ & $\cdots$\\

$1{}^3P_1$ & $1^{++}$ & $1183$ & $a_1(1260)$~\cite{Chen:2015iqa} & $1202\pm15$ &
$1240$ & $1254$ & $1240$ &
$f_1(1285)$~\cite{Chen:2015iqa} & $1281.8\pm0.5$ & $1457$ & $f_1(1420)$~\cite{Chen:2015iqa} &
$1427.7^{+1.7}_{-1.4}$ &
$1480$ & $1464$ & $1454$\\

$1{}^3P_2$ & $2^{++}$ & $1319$ & $a_2(1320)$~\cite{Ye:2012gu} & $1318.2\pm0.6$ &
$1310$ & $1317$ & $1312$ &
$f_2(1270)$~\cite{Ye:2012gu} & $1275.4\pm0.8$ & $1554$ & $f_2'(1525)$~\cite{Ye:2012gu} & $1517.3\pm2.4$ &
$1530$ & $1529$ & $1495$\\

$2{}^1P_1$ & $1^{+-}$ & $1717$ & & &
$1780$ & $1721$ & $1876$ &
$h_1(1595)$~\cite{Chen:2015iqa} & $1594\pm15{}^{+10}_{-60}$ & $1983$ & $h_1(1900)$~\cite{Wang:2024lba} & $1908\pm6{}^{+8}_{-4}$ &
$2010$ & $2024$ & $\cdots$\\

$2{}^3P_0$ & $0^{++}$ & $1495$ & $a_0(1450)$ ~\cite{Liu:2008zze} & $1439\pm34$ &
$1780$ & $1679$ & $1931$ &
$f_0(1370)$ ~\cite{Liu:2008zze} & $1200--1500$ & $1845$ & $f_0(1710)$ & $1723^{+7}_{-6}$ &
$1990$ & $1969$ & $\cdots$\\

$2{}^3P_1$ & $1^{++}$ & $1756$ & $a_1(1640)$~\cite{Chen:2015iqa} & $1655\pm16$ &
$1820$ & $1742$ & $1876$ &
& & $2000$ & & &
$2030$ & $2016$ & $\cdots$\\

$2{}^3P_2$ & $2^{++}$ & $1836$ & $a_2(1700)$~\cite{Pang:2014laa} & $1706\pm14$ &
$1820$ & $1779$ & $1931$ &
& & $2061$ & & &
$2040$ & $2030$ & $\cdots$\\

$3{}^1P_1$ & $1^{+-}$ & $2158$ & $b_1(1960)$ \cite{Chen:2015iqa} & $1960\pm30$ &
$\cdots$ & $2007$ & $2373$ &
$h_1(1965)$~\cite{Wang:2024lba}& $1965\pm45$ & $2409$ & & &
$\cdots$ & $\cdots$ & $\cdots$\\

$3{}^3P_0$ & $0^{++}$ & $2021$ & $a_0(1710)$ & $1713\pm19$ &
$\cdots$ & $1993$ & $2423$ &
& & $2317$ & & &
$\cdots$ & $2364$ & $\cdots$\\

 & & & $a_0(2020)$ & $2025\pm30$ &
 & & & & & & & & & & \\

 & & & $a_0(1950)$ & $1931\pm14\pm22$ &
 & & & & & & & & & & \\

$3{}^3P_1$ & $1^{++}$ & $2196$ & $a_1(2095)$~\cite{Chen:2015iqa} & $2096\pm17\pm121$ &
$\cdots$ & $2039$ & $2373$ &
$f_1(1970)$~\cite{Chen:2015iqa} & $1971\pm15$ & $2426$ & & &
$\cdots$ & $2403$ & $\cdots$\\

$3{}^3P_2$ & $2^{++}$ & $2263$ & $a_2(2175)$~\cite{Pang:2014laa} & $2175\pm40$ &
$\cdots$ & $2048$ & $\cdots$ &
$f_2(2240)$~\cite{Ye:2012gu} & $2240\pm15$ & $2476$ & & &
$\cdots$ & $2412$ & $\cdots$\\

$4{}^1P_1$ & $1^{+-}$ & $2532$ & $b_1(2240)$~\cite{Chen:2015iqa} & $2224\pm35$ &
$\cdots$ & $2264$ & $\cdots$ &
$h_1(2215)$~\cite{Wang:2024lba} & $2215\pm40$ & $2770$ & & &
$\cdots$ & $\cdots$ & $\cdots$\\

$4{}^3P_0$ & $0^{++}$ & $2431$ & & &
$\cdots$ & $2250$ & $\cdots$ &
& & $2700$ & & &
$\cdots$ & $2699$ & $\cdots$\\

$4{}^3P_1$ & $1^{++}$ & $2567$ & & &
$\cdots$ & $2286$ & $\cdots$ &
& & $2786$ & & &
$\cdots$ & $2729$ & $\cdots$\\

$4{}^3P_2$ & $2^{++}$ & $2623$ & & &
$\cdots$ & $2297$ & $\cdots$ &
& & $2829$ & & &
$\cdots$ & $2734$ & $\cdots$\\

$5{}^1P_1$ & $1^{+-}$ & $2858$ & & &
$\cdots$ & $\cdots$ & $\cdots$ &
& & $3085$ & & &
$\cdots$ & $\cdots$ & $\cdots$\\

$5{}^3P_0$ & $0^{++}$ & $2776$ & & &
$\cdots$ & $\cdots$ & $\cdots$ &
& & $3029$ & & &
$\cdots$ & $\cdots$ & $\cdots$\\

$5{}^3P_1$ & $1^{++}$ & $2889$ & & &
$\cdots$ & $\cdots$ & $\cdots$ &
& & $3100$ & & &
$\cdots$ & $\cdots$ & $\cdots$\\

$5{}^3P_2$ & $2^{++}$ & $2939$ & & &
$\cdots$ & $\cdots$ & $\cdots$ &
& & $3138$ & & &
$\cdots$ & $\cdots$ & $\cdots$\\

$6{}^3P_0$ & $0^{++}$ & $3083$ & & &
$\cdots$ & $\cdots$ & $\cdots$ &
& & $3322$ & & &
$\cdots$ & $\cdots$ & $\cdots$\\

$1{}^1F_3$ & $3^{+-}$ & $1977$ & $b_3(2030)$~\cite{Pang:2015eha}& $2032\pm12$ &
$2030$ & $1884$ & $\cdots$ &
$h_3(2025)$~\cite{Pang:2015eha} & $2025\pm20$ & $2200$ & & &
$\cdots$ & $\cdots$ & $\cdots$\\

$1{}^3F_2$ & $2^{++}$ & $1903$ & $a_2(2030)$ ~\cite{Pang:2014laa}& $2030\pm20$ &
$2050$ & $1797$ & $1879$ &
$f_2(2010)$ ~\cite{Ye:2012gu}& $2011^{+62}_{-76}$ & $2175$ & & &
$2240$ & $2143$ & $\cdots$\\

$1{}^3F_3$ & $3^{++}$ & $1990$ & $a_3(2030)$~\cite{Chen:2026mdo}& $2031\pm12$ &
$2050$ & $1910$ & $1951$ &
$f_3(2050)$ ~\cite{Chen:2026mdo}& $2048\pm8$ & $2210$ & & &
$2230$ & $2215$ & $2192$\\

$1{}^3F_4$ & $4^{++}$ & $1983$ & $a_4(1970)$~\cite{Wang:2025hve} & $1967\pm16$ &
$2010$ & $2018$ & $2011$ &
$f_4(2050)$~\cite{Pang:2015eha} & $2018\pm11$ & $2200$ & & &
$2200$ & $2286$ & $\cdots$\\

$2{}^1F_3$ & $3^{+-}$ & $2355$ & $b_3(2245)$~\cite{Pang:2015eha} & $2245\pm50$ &
$\cdots$ & $2164$ & $\cdots$ &
$h_3(2275)$~\cite{Pang:2015eha} & $2275\pm25$ & $2573$ & & &
$\cdots$ & $\cdots$ & $\cdots$\\

$2{}^3F_2$ & $2^{++}$ & $2291$ & $a_2(2255)$ ~\cite{Pang:2014laa}& $2255\pm20$ &
$\cdots$ & $2091$ & $\cdots$ &
$f_2(2300)$ ~\cite{Ye:2012gu}& $2297\pm28$ & $2546$ & & &
$\cdots$ & $2514$ & $\cdots$\\

$2{}^3F_3$ & $3^{++}$ & $2368$ & $a_3(2275)$~\cite{Chen:2026mdo} & $2275\pm35$ &
$\cdots$ & $2191$ & $2401$ &
$f_3(2300)$ ~\cite{Chen:2026mdo}& $2334\pm25$ & $2582$ & & &
$\cdots$ & $2585$ & $\cdots$\\

$2{}^3F_4$ & $4^{++}$ & $2372$ & $a_4(2255)$~\cite{Wang:2025hve} & $2237\pm5$ &
$\cdots$ & $2284$ & $2451$ &
$f_4(2300)$~\cite{Pang:2015eha} & $2320\pm60$ & $2580$ & & &
$\cdots$ & $2657$ & $\cdots$\\

$3{}^1F_3$ & $3^{+-}$ & $2688$ & & &
$\cdots$ & $\cdots$ & $\cdots$ &
& & $2901$ & & &
$\cdots$ & $\cdots$ & $\cdots$\\

$3{}^3F_2$ & $2^{++}$ & $2627$ & & &
$\cdots$ & $\cdots$ & $\cdots$ &
& & $2871$ & & &
$\cdots$ & $\cdots$ & $\cdots$\\

$3{}^3F_3$ & $3^{++}$ & $2701$ & & &
$\cdots$ & $\cdots$ & $2792$ &
& & $2908$ & & &
$\cdots$ & $\cdots$ & $\cdots$\\

$3{}^3F_4$ & $4^{++}$ & $2710$ & & &
$\cdots$ & $\cdots$ & $\cdots$ &
& & $2913$ & & &
$\cdots$ & $\cdots$ & $\cdots$\\

$1{}^1H_5$ & $5^{+-}$ & $2524$ & & &
$\cdots$ & $2328$ & $\cdots$ &
& & $2722$ & & &
$\cdots$ & $\cdots$ & $\cdots$\\

$1{}^3H_4$ & $4^{++}$ & $2532$ & & &
$\cdots$ & $2234$ & $2402$ &
& & $2744$ & & &
$\cdots$ & $2634$ & $\cdots$\\

$1{}^3H_5$ & $5^{++}$ & $2534$ & & &
$\cdots$ & $2359$ & $2463$ &
& & $2730$ & & &
$\cdots$ & $2720$ & $2730$\\

$1{}^3H_6$ & $6^{++}$ & $2499$ & $a_6(2450)$ ~\cite{Pang:2015eha}& $2450\pm130$ &
$\cdots$ & $2475$ & $2517$ &
$f_6(2510)$~\cite{Pang:2015eha} & $2470\pm50$ & $2696$ & & &
$\cdots$ & $2809$ & $\cdots$\\

$2{}^1H_5$ & $5^{+-}$ & $2828$ & & &
$\cdots$ & $\cdots$ & $\cdots$ &
& & $3024$ & & &
$\cdots$ & $\cdots$ & $\cdots$\\

$2{}^3H_4$ & $4^{++}$ & $2829$ & & &
$\cdots$ & $\cdots$ & $\cdots$ &
& & $3037$ & & &
$\cdots$ & $\cdots$ & $\cdots$\\

$2{}^3H_5$ & $5^{++}$ & $2837$ & & &
$\cdots$ & $\cdots$ & $2825$ &
& & $3031$ & & &
$\cdots$ & $\cdots$ & $\cdots$\\

$2{}^3H_6$ & $6^{++}$ & $2813$ & & &
$\cdots$ & $\cdots$ & $2826$ &
& & $3007$ & & &
$\cdots$ & $\cdots$ & $\cdots$\\

\end{longtable}

\clearpage
\endgroup

\setlength{\paperwidth}{8.5in}
\setlength{\paperheight}{11in}
\restoregeometry
\pdfpagewidth=8.5in
\pdfpageheight=11in

\noindent\begin{minipage}{\textwidth}
\makeatletter\def\@captype{table}\makeatother
\centering
\footnotesize
\caption{Negative-parity strange-meson spectrum ($I=\frac12$, $q\bar s$ with $q=u,d$). Masses are in MeV; a parenthesis after a mass gives the pair mixing angle $\theta_{nL}$ in degrees. Comparison models: GI~\cite{Godfrey:1985xj}, RQM~\cite{Ebert:2009ub}, and BS (model A)~\cite{Koll:2000ke}. Each model retains its own mixing convention. GI angles are the computed values in the lower-mass-state convention of \href{https://mmikhasenko.github.io/GIModel.jl/dev/paper/reports/mixing_angles}{GIModel.jl}, while GI masses are from Ref.~\cite{Godfrey:1985xj}. RQM $J=L$ entries are mixed-state eigenvalues.}
\label{tab:merged_strange_negative}
\setlength{\tabcolsep}{3pt}
\renewcommand{\arraystretch}{1.02}
\begin{tabular*}{\textwidth}{@{\extracolsep{\fill}}cccccccc@{}}
\toprule
$n^{2S+1}L_J$ & $J^P$ & This work & Expt. state & Mass & GI & RQM & BS \\
\midrule
$1{}^1S_0$ & $0^{-}$ & $471.5$ & $K^\pm$ & $493.677\pm0.015$ & $470.0$ & $482.0$ & $499.0$ \\
$1{}^3S_1$ & $1^{-}$ & $898.6$ & $K^*(892)^\pm$ & $891.88\pm0.23$ & $900.0$ & $897.0$ & $870.0$ \\
$2{}^1S_0$ & $0^{-}$ & $1431$ & $K(1460)$~\cite{Pang:2017dlw} & $1482.4\pm15.635$ & $1450$ & $1538$ & $1508$ \\
$2{}^3S_1$ & $1^{-}$ & $1580$ & $K^*(1410)$ & $1414\pm15$ & $1580$ & $1675$ & $1687$ \\
$3{}^1S_0$ & $0^{-}$ & $1982$ & $K(1830)$ & $1874\pm43{}^{+59}_{-115}$ & $2020$ & $2065$ & $2159$ \\
$3{}^3S_1$ & $1^{-}$ & $2085$ &  &  & $2110$ & $2156$ & $\cdots$ \\
$4{}^1S_0$ & $0^{-}$ & $2413$ & $\cdots$ & $\cdots$ & $\cdots$ & $\cdots$ & $2652$ \\
$4{}^3S_1$ & $1^{-}$ & $2490$ & $\cdots$ & $\cdots$ & $\cdots$ & $\cdots$ & $\cdots$ \\
$5{}^1S_0$ & $0^{-}$ & $2773$ & $\cdots$ & $\cdots$ & $\cdots$ & $\cdots$ & $3062$ \\
$5{}^3S_1$ & $1^{-}$ & $2835$ & $\cdots$ & $\cdots$ & $\cdots$ & $\cdots$ & $\cdots$ \\
$6{}^1S_0$ & $0^{-}$ & $3087$ & $\cdots$ & $\cdots$ & $\cdots$ & $\cdots$ & $3420$ \\
$6{}^3S_1$ & $1^{-}$ & $3137$ & $\cdots$ & $\cdots$ & $\cdots$ & $\cdots$ & $\cdots$ \\
$1D$ & $2^{-}$ & $1756(-79.4)$ & $K_2(1770)$~\cite{Pang:2017dlw} & $1773\pm8$ & $1780(12.8)$ & $1709(44.0)$ & $1750$ \\
$1{}^3D_1$ & $1^{-}$ & $1614$ & $K^*(1680)$ & $1718\pm18$ & $1780$ & $1699$ & $\cdots$ \\
$1D^\prime$ & $2^{-}$ & $1768(-79.4)$ & $K_2(1820)$~\cite{Pang:2017dlw} & $1819\pm12$ & $1810(12.8)$ & $1824(44.0)$ & $1750$ \\
$1{}^3D_3$ & $3^{-}$ & $1795$ & $K_3^*(1780)$~\cite{Pang:2017dlw} & $1779\pm8$ & $1790$ & $1789$ & $1800$ \\
{$2D$ }& $2^{-}$ & $2191(88.6)$ & $\cdots$ & $\cdots$ & $2230(20.2)$ & $2066(45.0)$ & $\cdots$ \\
{$2{}^3D_1$} & $1^{-}$ & $2096$ & $\cdots$ & $\cdots$ & $2250$ & $2063$ & $\cdots$ \\
{$2D^\prime$} & $2^{-}$ & $2204(88.6)$ & $K_2(2250)$~\cite{Pang:2017dlw} & $2247\pm17$ & $2260(20.2)$ & $2163(45.0)$ & $2292$ \\
$2{}^3D_3$ & $3^{-}$ & $2232$ & $\cdots$ & $\cdots$ & $2240$ & $2182$ & $\cdots$ \\
{$3D^\prime$ }& $2^{-}$ & $2575(88.9)$ & $\cdots$ & $\cdots$ & $\cdots$ & $\cdots$ & $\cdots$ \\
$3{}^3D_1$ & $1^{-}$ & $2484$ & $\cdots$ & $\cdots$ & $\cdots$ & $\cdots$ & $\cdots$ \\
{$3D$} & $2^{-}$ & $2562(88.9)$ & $\cdots$ & $\cdots$ & $\cdots$ & $\cdots$ & $\cdots$ \\
$3{}^3D_3$ & $3^{-}$ & $2603$ & $\cdots$ & $\cdots$ & $\cdots$ & $\cdots$ & $\cdots$ \\
{$4D^\prime$} & $2^{-}$ & $2899(89.9)$ & $\cdots$ & $\cdots$ & $\cdots$ & $\cdots$ & $\cdots$ \\
$4{}^3D_1$ & $1^{-}$ & $2821$ & $\cdots$ & $\cdots$ & $\cdots$ & $\cdots$ & $\cdots$ \\
{$4D$} & $2^{-}$ & $2887(89.9)$ & $\cdots$ & $\cdots$ & $\cdots$ & $\cdots$ & $\cdots$ \\
$4{}^3D_3$ & $3^{-}$ & $2928$ & $\cdots$ & $\cdots$ & $\cdots$ & $\cdots$ & $\cdots$ \\
$1G$ & $4^{-}$ & $2374(76.8)$ & $\cdots$ & $\cdots$ & $2410(10.8)$ & $2255(44.0)$ & $2352$ \\
$1{}^3G_3$ & $3^{-}$ & $2370$ & $\cdots$ & $\cdots$ & $2460$ & $2207$ & $\cdots$ \\
$1G^\prime$ & $4^{-}$ & $2386(76.8)$ & $\cdots$ & $\cdots$ & $2440(10.8)$ & $2285(44.0)$ & $2352$ \\
$1{}^3G_5$ & $5^{-}$ & $2359$ & $K_5^*(2380)$~\cite{Pang:2017dlw} & $2382\pm24$ & $2390$ & $2356$ & $2397$ \\
{$2G$}& $4^{-}$ & $2708(74.3)$ & $K_4(2500)$ \cite{Wang:2025lin}& $2490\pm20$ & $\cdots$ & $2575$ & $2771$ \\
\bottomrule
\end{tabular*}
\end{minipage}
\par\vspace{\dbltextfloatsep}
\twocolumngrid

\begin{table*}[!t]
\centering
\footnotesize
\caption{Positive-parity strange-meson spectrum ($I=\frac12$, $q\bar s$ with $q=u,d$). Masses are in MeV; parentheses give pair mixing angles $\theta_{nL}$ in degrees. The fitted inputs are specified in Table~\ref{tab1}; other states and all angles are predictions. Experimental masses shared with the fit are adopted from Table~\ref{tab1}; other entries follow the PDG compilation~\cite{ParticleDataGroup:2026mpi}. Comparison models: GI~\cite{Godfrey:1985xj}, RQM~\cite{Ebert:2009ub}, and BS (model A)~\cite{Koll:2000ke}. Mixing conventions remain model dependent. GI angles are the computed values in the lower-mass-state convention of \href{https://mmikhasenko.github.io/GIModel.jl/dev/paper/reports/mixing_angles}{GIModel.jl}; GI masses are from Ref.~\cite{Godfrey:1985xj}. RQM $J=L$ masses are mixed-state eigenvalues.}
\label{tab:merged_strange_positive}
\setlength{\tabcolsep}{3pt}
\renewcommand{\arraystretch}{1.02}
\begin{tabular*}{\textwidth}{@{\extracolsep{\fill}}cccccccc@{}}
\toprule
$n^{2S+1}L_J$ & $J^P$ & This work & Expt. state & Mass & GI & RQM & BS \\
\midrule
$1P^\prime$ & $1^{+}$ & $1332(-55.4)$ & $K_1(1400)$~\cite{Pang:2017dlw} & $1403\pm7$ & $1380(4.30)$ & $1412(44.0)$ & $1353$ \\
$1{}^3P_0$ & $0^{+}$ & $900.8$ & $K_0^*(700)$ & $838\pm11$ & $1240$ & $1362$ & $1426$ \\
$1P$ & $1^{+}$ & $1309(-55.4)$ & $K_1(1270)$~\cite{Pang:2017dlw} & $1256\pm6$ & $1340(4.30)$ & $1294(44.0)$ & $1353$ \\
$1{}^3P_2$ & $2^{+}$ & $1440$ & $K_2^*(1430)^\pm$~\cite{Pang:2017dlw} & $1427.3\pm1.5$ & $1430$ & $1424$ & $1406$ \\
$2P$ & $1^{+}$ & $1856(-86.7)$ & $K_1(1861)$~\cite{Li:2024hrf} & $1861\pm10^{+16}_{-46}$ & $1900(14.2)$ & $1757(45.0)$ & $2005$ \\
$2{}^3P_0$ & $0^{+}$ & $1692$ & $K_0^*(1430)$ & $1425\pm50$ & $1890$ & $1791$ & $2058$ \\
$2P^\prime$ & $1^{+}$ & $1881(-86.7)$ & $K_1(1911)$~\cite{Li:2024hrf} & $1911\pm37^{+124}_{-48}$ & $1930(14.2)$ & $1893(45.0)$ & $2005$ \\
$2{}^3P_2$ & $2^{+}$ & $1951$ & $K_2^*(1980)$~\cite{Pang:2017dlw} & $1990^{+60}_{-50}$ & $1940$ & $1896$ & $2019$ \\
$3P^\prime$ & $1^{+}$ & $2314(-87.9)$ & $\cdots$ & $\cdots$ & $\cdots$ & $2200(45.0)$ & $\cdots$ \\
$3{}^3P_0$ & $0^{+}$ & $2184$ & $\cdots$ & $\cdots$ & $\cdots$ & $2160$ & $2561$ \\
$3P$ & $1^{+}$ & $2289(-87.9)$ & $\cdots$ & $\cdots$ & $\cdots$ & $2164(45.0)$ & $\cdots$ \\
$3{}^3P_2$ & $2^{+}$ & $2371$ & $\cdots$ & $\cdots$ & $\cdots$ & $2206$ & $\cdots$ \\
$4P^\prime$ & $1^{+}$ & $2679(-88.3)$ & $\cdots$ & $\cdots$ & $\cdots$ & $\cdots$ & $\cdots$ \\
$4{}^3P_0$ & $0^{+}$ & $2578$ & $K_0^*(2600)$~\cite{Pang:2025esm} & $2662\pm59\pm201$ & $\cdots$ & $\cdots$ & $\cdots$ \\
$4P$ & $1^{+}$ & $2656(-88.3)$ & $\cdots$ & $\cdots$ & $\cdots$ & $\cdots$ & $\cdots$ \\
$4{}^3P_2$ & $2^{+}$ & $2727$ & $\cdots$ & $\cdots$ & $\cdots$ & $\cdots$ & $\cdots$ \\
$5P^\prime$ & $1^{+}$ & $2997(-88.1)$ & $\cdots$ & $\cdots$ & $\cdots$ & $\cdots$ & $\cdots$ \\
$5{}^3P_0$ & $0^{+}$ & $2913$ & $\cdots$ & $\cdots$ & $\cdots$ & $\cdots$ & $\cdots$ \\
$5P$ & $1^{+}$ & $2976(-88.1)$ & $\cdots$ & $\cdots$ & $\cdots$ & $\cdots$ & $\cdots$ \\
$5{}^3P_2$ & $2^{+}$ & $3040$ & $\cdots$ & $\cdots$ & $\cdots$ & $\cdots$ & $\cdots$ \\
$1F^\prime$ & $3^{+}$ & $2103(83.5)$ & $\cdots$ & $\cdots$ & $2120(12.6)$ & $2080(44.0)$ & $2074$ \\
$1{}^3F_2$ & $2^{+}$ & $2045$ & $\cdots$ & $\cdots$ & $2150$ & $1964$ & $\cdots$ \\
$1F$ & $3^{+}$ & $2091(83.5)$ & $\cdots$ & $\cdots$ & $2120(12.6)$ & $2009(44.0)$ & $2074$ \\
$1{}^3F_4$ & $4^{+}$ & $2094$ & $K_4^*(2045)$~\cite{Pang:2017dlw} & $2048\pm8.5$ & $2110$ & $2096$ & $2121$ \\
$2F^\prime$ & $3^{+}$ & $2477(79.7)$ & $\cdots$ & $\cdots$ & $\cdots$ & $2348$ & $\cdots$ \\
$2{}^3F_2$ & $2^{+}$ & $2424$ & $\cdots$ & $\cdots$ & $\cdots$ & $\cdots$ & $\cdots$ \\
$2F$ & $3^{+}$ & $2466(79.7)$ & $K_3(2320)$~\cite{Wang:2025lin} & $2324\pm24$ & $\cdots$ & $2348$ & $2544$ \\
$2{}^3F_4$ & $4^{+}$ & $2478$ & $\cdots$ & $\cdots$ & $\cdots$ & $2436$ & $\cdots$ \\
$3F^\prime$ & $3^{+}$ & $2806(80.1)$ & $\cdots$ & $\cdots$ & $\cdots$ & $\cdots$ & $\cdots$ \\
$3{}^3F_2$ & $2^{+}$ & $2755$ & $\cdots$ & $\cdots$ & $\cdots$ & $\cdots$ & $\cdots$ \\
$3F$ & $3^{+}$ & $2796(80.1)$ & $\cdots$ & $\cdots$ & $\cdots$ & $\cdots$ & $\cdots$ \\
$3{}^3F_4$ & $4^{+}$ & $2813$ & $\cdots$ & $\cdots$ & $\cdots$ & $\cdots$ & $\cdots$ \\
$1H^\prime$ & $5^{+}$ & $2635(73.4)$ & $\cdots$ & $\cdots$ & $\cdots$ & $\cdots$ & $\cdots$ \\
$1{}^3H_4$ & $4^{+}$ & $2641$ & $\cdots$ & $\cdots$ & $\cdots$ & $\cdots$ & $\cdots$ \\
$1H$ & $5^{+}$ & $2624(73.4)$ & $\cdots$ & $\cdots$ & $\cdots$ & $\cdots$ & $\cdots$ \\
$1{}^3H_6$ & $6^{+}$ & $2599$ & $\cdots$ & $\cdots$ & $\cdots$ & $\cdots$ & $\cdots$ \\
$2H^\prime$ & $5^{+}$ & $2937(71.1)$ & $\cdots$ & $\cdots$ & $\cdots$ & $\cdots$ & $\cdots$ \\
$2{}^3H_4$ & $4^{+}$ & $2936$ & $\cdots$ & $\cdots$ & $\cdots$ & $\cdots$ & $\cdots$ \\
$2H$ & $5^{+}$ & $2927(71.1)$ & $\cdots$ & $\cdots$ & $\cdots$ & $\cdots$ & $\cdots$ \\
$2{}^3H_6$ & $6^{+}$ & $2912$ & $\cdots$ & $\cdots$ & $\cdots$ & $\cdots$ & $\cdots$ \\
\bottomrule
\end{tabular*}
\end{table*}
\subsubsection{Positive-parity strange mesons}
Table~\ref{tab:merged_strange_positive} contains the $P$-, $F$-,
and $H$-wave strange mesons.
The $1{}^3P_2$ mass of $1440~\mathrm{MeV}$ was constrained
by the $K_2^*(1430)$ input and lies $7.6~\mathrm{MeV}$ above
the $1432.4~\mathrm{MeV}$ input in Table~\ref{tab1}.  This fit uses the same lowest
tensor assignment as Ref.~\cite{Pang:2017dlw}.
The $2{}^3P_2$ state, in contrast, is a prediction:
$1951~\mathrm{MeV}$ is $39.0~\mathrm{MeV}$ below the
experimental value of $K_2^*(1980)$, within its quoted mass interval.
This result supports the first-radial-excitation interpretation
favored in Ref.~\cite{Pang:2017dlw}.

We treat $K_0^*(700)$ and $K_0^*(1430)$ as $1{}^3P_0$ and $2{}^3P_0$, respectively. Their predicted masses are $900.8$ and $1692~\mathrm{MeV}$, with residuals of $+62.8$ and $+267~\mathrm{MeV}$ relative to the adopted $838$ and $1425~\mathrm{MeV}$ inputs.

The lowest axial-vector masses were not included in the fit.
The predicted $1P$ and $1P^\prime$ states at $1309$ and
$1332~\mathrm{MeV}$ can be associated with $K_1(1270)$ and
$K_1(1400)$, as in Ref.~\cite{Pang:2017dlw}, with differences
of $+53$ and $-71~\mathrm{MeV}$ from the experimental masses.
This supports their placement in the lowest axial-vector pair,
although the predicted splitting is $23.0~\mathrm{MeV}$ compared
with the experimental $147~\mathrm{MeV}$.
The associated mixing angle, $\theta_{1P}=-55.4^\circ$, is a
model output, not an angle fitted to decay data and can be compared with  the result of Refs.  \cite{Li:2005eq,Li:2006we,Shi:2023kiy}.
The GI value of $4.30^\circ$ uses the convention stated in the
caption.

For the next axial-vector pair, the predicted $2P$ and
$2P^\prime$ masses of $1856$ and $1881~\mathrm{MeV}$ lie
$5$ and $30~\mathrm{MeV}$ below $K_1(1861)$ and
$K_1(1911)$, respectively. Both resonances are outside the fit,
and these results support the same pair classification proposed
in the strong-decay study of Ref.~\cite{Li:2024hrf}.
Our predicted mixing angle for this pair is $\theta_{2P}=-86.7^\circ$;
the GI column gives $14.2^\circ$.
The mass agreement supports the radial assignment without
determining the experimental mixing angle.
At higher excitation, the $3P$ pair is predicted at $2289$
and $2314~\mathrm{MeV}$ with $\theta_{3P}=-87.9^\circ$.
The $4P$ and $5P$ pairs occur at $(2656,2679)$ and
$(2976,2997)~\mathrm{MeV}$, respectively, with
$\theta=-88.3^\circ$ and $\theta=-88.1^\circ$, respectively. These are predictions for the missing
axial-vector states listed in the table.

For the $F$ waves, $K_4^*(2045)$ was included in the fit as
$1{}^3F_4$, the classification used in Ref.~\cite{Pang:2017dlw}.
The fitted result is $2094~\mathrm{MeV}$, or
$46~\mathrm{MeV}$ above the present reference.
The masses of neighboring $J^P=3^+$ states are unfitted predictions:
the $1F$ pair occurs at $2091$ and $2103~\mathrm{MeV}$
with $\theta_{1F}=83.5^\circ$; its GI comparison angle is
$12.6^\circ$ in the convention stated in the caption.
The predicted $2F$ pair lies at $2466$ and
$2477~\mathrm{MeV}$ with $\theta_{2F}=79.7^\circ$.
For the higher $F$-wave excitations, the two $J^P=3^+$ states
obtained here have masses of $2796$ and $2806~\mathrm{MeV}$,
with a splitting of $10~\mathrm{MeV}$.
The corresponding mixing angle is $80.1^\circ$.
Ref.~\cite{Pang:2017dlw} gives unmixed masses of
$2550$ and $2546~\mathrm{MeV}$ for
$3{}^1F_3$ and $3{}^3F_3$, respectively.
These two masses are also close to each other but are lower overall
than the present results.
This difference may arise from the different parameter fits and
long-range confining interactions.
These predictions provide both mass estimates and spin-mixing
information for searches for higher positive-parity states.

Finally, the $H$-wave states are predictions beyond the fitted
spectrum. The $1H$ multiplet spans $2599$--$2641~\mathrm{MeV}$,
with its $J^P=5^+$ partners at $2624$ and
$2635~\mathrm{MeV}$ and $\theta_{1H}=73.4^\circ$.
The $2H$ multiplet lies in the range
$2912$--$2937~\mathrm{MeV}$; its $J^P=5^+$ pair is at
$2927$ and $2937~\mathrm{MeV}$ with
$\theta_{2H}=71.1^\circ$.
Experimental masses and decay information for these states
would test both the radial spectrum and the predicted mixing.

The remaining predictions for higher strange-meson excitations are listed  in Tables~\ref{tab:merged_strange_negative} and \ref{tab:merged_strange_positive}.
\subsection{The $D$ and $D_s$ meson spectra}

Using the fitted parameters in Table~\ref{tab2}, we calculated the $D$ and $D_s$ meson mass spectra. The results are listed in Tables~\ref{tab:D} and~\ref{tab:Ds}, respectively, together with available experimental data and other theoretical predictions.

\par
\onecolumngrid
\begingroup
\footnotesize
\setlength{\tabcolsep}{2.2pt}
\renewcommand{\arraystretch}{1.0}
\setlength{\LTleft}{0pt}
\setlength{\LTright}{0pt}
\setlength{\LTcapwidth}{\textwidth}
\setlength{\LTpre}{10pt}
\setlength{\LTpost}{4pt}
\setlength{\parindent}{0pt}
\begin{longtable}{@{\extracolsep{\fill}}l*{8}{c}@{}}
\caption[Mass spectrum of the $D$ meson family. Masses are given in $\mathrm{MeV}$; mixing angles follow the conventions of the corresponding references.] {Mass spectrum of the $D$ meson family. Masses are given in $\mathrm{MeV}$; mixing angles follow the conventions of the corresponding references. 
Experimental entries adopted in this work; see the PDG compilation~\cite{ParticleDataGroup:2026mpi}. For states included in the fit, the experimental masses and uncertainties reproduce Table~\ref{tab1}; other entries retain their cited experimental references.  NR: nonrelativistic constituent quark model~\cite{Li:2010vx}; GI: Godfrey--Isgur model~\cite{Godfrey:2015dva}; RQM: relativistic quasipotential model~\cite{Ebert:2009ua}; PM: semirelativistic potential model~\cite{Ni:2021pce}; BS: instantaneous Bethe--Salpeter model~\cite{Li:2024dyi}.  The $1P_1$ and $1P_1^\prime$ rows use the masses of $D_1(2430)^0$ and $D_1(2420)^0$, respectively, as comparison values. In the ``This work'' column, primed states denote the higher-mass members of the mixed pairs.  
An ellipsis indicates that no value is quoted in the present table.
}
\label{tab:D}\\
\toprule[1pt]
\toprule[1pt]
States & Meson & Expt. & This work & NR & GI & RQM & PM & BS \\
\midrule[0.6pt]
\endfirsthead
\caption[]{Mass spectrum of the $D$ meson family (continued).}\\
\toprule[1pt]
\toprule[1pt]
States & Meson & Expt. & This work & NR & GI & RQM & PM & BS \\
\midrule[0.6pt]
\endhead
\midrule
\multicolumn{9}{r}{\footnotesize Continued on next page}\\
\endfoot
\bottomrule[1pt]
\bottomrule[1pt]
\endlastfoot
$1^1S_0$ & $D^0$ & $1864.85\pm0.04$ & $1862$ & $1867$ & $1877$ & $1871$ & $1865$ & $1865$ \\
$2^1S_0$ & $D_0(2550)^0$ & $2549\pm19$ & $2513$ & $2555$ & $2581$ & $2581$ & $2547$ & $2564$ \\
$3^1S_0$ &  & $\cdots$ & $2920$ & \ensuremath{\cdots} & $3068$ & $3062$ & $3029$ & $3042$ \\
$4^1S_0$ &  & \ensuremath{\cdots} & $3221$ & \ensuremath{\cdots} & $3468$ & $3452$ & \ensuremath{\cdots} & $\cdots$ \\
$5^1S_0$ &  & \ensuremath{\cdots} & $3458$ & \ensuremath{\cdots} & $3814$ & $3793$ & \ensuremath{\cdots} & $\cdots$ \\
\addlinespace[3pt]

$1^3S_1$ & $D^{*0}$ & $2006.86\pm0.05$ & $2013$ & $2010$ & $2041$ & $2010$ & $2008$ & $2007$ \\
$2^3S_1$ & {$D_1^*(2600)^0$} & {$2641.9\pm1.8\pm4.5$} & $2575$ & $2636$ & $2643$ & $2632$ & $2636$ & $2627$ \\
$3^3S_1$ &  & \ensuremath{\cdots} & $2958$ & \ensuremath{\cdots} & $3110$ & $3096$ & $3093$ & $3116$ \\
$4^3S_1$ &  & \ensuremath{\cdots} & $3248$ & \ensuremath{\cdots} & $3497$ & $3482$ & \ensuremath{\cdots} & $\cdots$ \\
$5^3S_1$ &  & \ensuremath{\cdots} & $3478$ & \ensuremath{\cdots} & $3837$ & $3822$ & \ensuremath{\cdots} & $\cdots$ \\
\addlinespace[3pt]
$1^3P_0$ & $D_0^*(2300)^0$ & $2343\pm10$ & $2235$ & $2252$ & $2399$ & $2406$ & $2313$ & $2300$ \\
$1^3P_2$ & $D_2^*(2460)^0$ & $2461.1\pm0.7$ & $2462$ & $2466$ & $2502$ & $2460$ & $2475$ & $2461$ \\
$1P_1$ & $D_1(2430)^0$ & $2412\pm9$ & $2373$ & $2402$ & $2456$ & $2426$ & $2424$ & $2412$ \\
$1P'_1$ & $D_1(2420)^0$ & $2422.1\pm0.6$ & $2414$ & $2417$ & $2467$ & $2469$ & $2453$ & $2422$ \\
$\theta_{1P}$ &  & \ensuremath{\cdots} & $-38.3^{\circ}$ & \ensuremath{\cdots} & $-25.7^{\circ}$ & $35.5^{\circ}$ & $-34.0^{\circ}$ & $\cdots$ \\
\addlinespace[3pt]

$2^3P_0$ &  & \ensuremath{\cdots} & $2744$ & $2752$ & $2931$ & $2919$ & $2849$ & $2915$ \\
$2^3P_2$ &  & \ensuremath{\cdots} & $2857$ & $2971$ & $2957$ & $3012$ & $2955$ & $2960$ \\
$2P_1$ &  & \ensuremath{\cdots} & $2808$ & $2886$ & $2924$ & $2932$ & $2900$ & $2919$ \\
$2P'_1$ &  & \ensuremath{\cdots} & $2819$ & $2926$ & $2961$ & $3021$ & $2936$ & $2949$ \\
$\theta_{2P}$ &  & \ensuremath{\cdots} & $-51.9^{\circ}$ & \ensuremath{\cdots} & $-29.4^{\circ}$ & $37.5^{\circ}$ & $-23.5^{\circ}$ & $\cdots$ \\
\addlinespace[3pt]

$3^3P_0$ &  & \ensuremath{\cdots} & $3084$ & \ensuremath{\cdots} & $3343$ & $3346$ & \ensuremath{\cdots} & $3266$ \\
$3^3P_2$ &  & \ensuremath{\cdots} & $3163$ & \ensuremath{\cdots} & $3353$ & $3407$ & \ensuremath{\cdots} & $3329$ \\
$3P_1$ &  & \ensuremath{\cdots} & $3125$ & \ensuremath{\cdots} & $3328$ & $3365$ & \ensuremath{\cdots} & $3260$ \\
$3P'_1$ &  & \ensuremath{\cdots} & $3133$ & \ensuremath{\cdots} & $3360$ & $3461$ & \ensuremath{\cdots} & $3279$ \\
$\theta_{3P}$ &  & \ensuremath{\cdots} & $-57.8^{\circ}$ & \ensuremath{\cdots} & $-28.1^{\circ}$ & $38.4^{\circ}$ & \ensuremath{\cdots} & $\cdots$ \\
\addlinespace[3pt]

$4^3P_0$ &  & \ensuremath{\cdots} & $3347$ & \ensuremath{\cdots} & $3697$ & \ensuremath{\cdots} & \ensuremath{\cdots} & $\cdots$ \\
$4^3P_2$ &  & \ensuremath{\cdots} & $3407$ & \ensuremath{\cdots} & $3701$ & \ensuremath{\cdots} & \ensuremath{\cdots} & $\cdots$ \\
$4P_1$ &  & \ensuremath{\cdots} & $3377$ & \ensuremath{\cdots} & $3681$ & \ensuremath{\cdots} & \ensuremath{\cdots} & $\cdots$ \\
$4P'_1$ &  & \ensuremath{\cdots} & $3383$ & \ensuremath{\cdots} & $3709$ & \ensuremath{\cdots} & \ensuremath{\cdots} & $\cdots$ \\
$\theta_{4P}$ &  & \ensuremath{\cdots} & $-59.6^{\circ}$ & \ensuremath{\cdots} & $-26.9^{\circ}$ & \ensuremath{\cdots} & \ensuremath{\cdots} & $\cdots$ \\
\addlinespace[3pt]
$1^3D_1$ & {$D_1^*(2760)^0$} & {$2781\pm18\pm13$} & $2676$ & $2740$ & $2817$ & $2788$ & $2754$ & $\cdots$ \\
$1^3D_3$ & {$D_3^*(2750)^0$} & $2763.1\pm3.2$ & $2756$ & $2719$ & $2833$ & $2863$ & $2782$ & $\cdots$ \\
$1D_2$ &  & \ensuremath{\cdots} & $2734$ & $2693$ & $2816$ & $2806$ & $2755$ & $\cdots$ \\
$1D'_2$ & {$D_2(2740)^0$} & {$2751\pm3\pm7$} & $2735$ & $2789$ & $2845$ & $2850$ & $2827$ & $\cdots$ \\
$\theta_{1D}$ &  & \ensuremath{\cdots} & $-88.9^{\circ}$ & \ensuremath{\cdots} & $-38.2^{\circ}$ & $40.7^{\circ}$ & $-40.2^{\circ}$ & $\cdots$ \\
\addlinespace[3pt]
$2^3D_1$ &  & \ensuremath{\cdots} & $3017$ & $3168$ & $3231$ & $3228$ & $3143$ & $\cdots$ \\
$2^3D_3$ &  & \ensuremath{\cdots} & $3075$ & $3170$ & $3226$ & $3335$ & $3202$ & $\cdots$ \\
$2D_2$ &  & \ensuremath{\cdots} & $3055$ & $3145$ & $3212$ & $3259$ & $3168$ & $\cdots$ \\
$2D'_2$ &  & \ensuremath{\cdots} & $3058$ & $3215$ & $3248$ & $3307$ & $3221$ & $\cdots$ \\
$\theta_{2D}$ &  & \ensuremath{\cdots} & $71.3^{\circ}$ & \ensuremath{\cdots} & $-37.4^{\circ}$ & $39.0^{\circ}$ & $-40.2^{\circ}$ & $\cdots$ \\
\addlinespace[3pt]

$3^3D_1$ &  & \ensuremath{\cdots} & $3284$ & \ensuremath{\cdots} & $3588$ & \ensuremath{\cdots} & \ensuremath{\cdots} & $\cdots$ \\
$3^3D_3$ &  & \ensuremath{\cdots} & $3333$ & \ensuremath{\cdots} & $3579$ & \ensuremath{\cdots} & \ensuremath{\cdots} & $\cdots$ \\
$3D_2$ &  & \ensuremath{\cdots} & $3314$ & \ensuremath{\cdots} & $3566$ & \ensuremath{\cdots} & \ensuremath{\cdots} & $\cdots$ \\
$3D'_2$ &  & \ensuremath{\cdots} & $3316$ & \ensuremath{\cdots} & $3600$ & \ensuremath{\cdots} & \ensuremath{\cdots} & $\cdots$ \\
$\theta_{3D}$ &  & \ensuremath{\cdots} & $-86.2^{\circ}$ & \ensuremath{\cdots} & $-36.9^{\circ}$ & \ensuremath{\cdots} & \ensuremath{\cdots} & $\cdots$ \\
\addlinespace[3pt]
$4^3D_1$ &  & $\cdots$ & $3502$ & $\cdots$ & $\cdots$ & $\cdots$ & $\cdots$ & $\cdots$ \\
$4^3D_3$ &  & $\cdots$ & $3543$ & $\cdots$ & $\cdots$ & $\cdots$ & $\cdots$ & $\cdots$ \\
$4D_2$ &  & $\cdots$ & $3526$ & $\cdots$ & $\cdots$ & $\cdots$ & $\cdots$ & $\cdots$ \\
$4D'_2$ &  & $\cdots$ & $3529$ & $\cdots$ & $\cdots$ & $\cdots$ & $\cdots$ & $\cdots$ \\
$\theta_{4D}$ &  & \ensuremath{\cdots} & $-68.2^{\circ}$ & \ensuremath{\cdots} & \ensuremath{\cdots} & \ensuremath{\cdots} & \ensuremath{\cdots} & $\cdots$ \\
\addlinespace[3pt]
$1^3F_2$ &  & \ensuremath{\cdots} & $2973$ & \ensuremath{\cdots} & $3132$ & $3090$ & $3096$ & $\cdots$ \\
$1^3F_4$ &  & \ensuremath{\cdots} & $2993$ & \ensuremath{\cdots} & $3113$ & $3187$ & $3034$ & $\cdots$ \\
$1F_3$ &  & \ensuremath{\cdots} & $2983$ & \ensuremath{\cdots} & $3108$ & $3129$ & $3022$ & $\cdots$ \\
$1F'_3$ &  & \ensuremath{\cdots} & $3001$ & \ensuremath{\cdots} & $3143$ & $3145$ & $3129$ & $\cdots$ \\
$\theta_{1F}$ &  & \ensuremath{\cdots} & $50.8^{\circ}$ & \ensuremath{\cdots} & $-39.5^{\circ}$ & $39.6^{\circ}$ & $-41.0^{\circ}$ & $\cdots$ \\
\addlinespace[3pt]

$2^3F_2$ &  & \ensuremath{\cdots} & $3240$ & \ensuremath{\cdots} & $3490$ & \ensuremath{\cdots} & \ensuremath{\cdots} & $\cdots$ \\
$2^3F_4$ &  & \ensuremath{\cdots} & $3261$ & \ensuremath{\cdots} & $3466$ & $3610$ & \ensuremath{\cdots} & $\cdots$ \\
$2F_3$ &  & \ensuremath{\cdots} & $3250$ & \ensuremath{\cdots} & $3461$ & $3551$ & \ensuremath{\cdots} & $\cdots$ \\
$2F'_3$ &  & \ensuremath{\cdots} & $3263$ & \ensuremath{\cdots} & $3498$ & \ensuremath{\cdots} & \ensuremath{\cdots} & $\cdots$ \\
$\theta_{2F}$ &  & \ensuremath{\cdots} & $51.8^{\circ}$ & \ensuremath{\cdots} & $-39.4^{\circ}$ & \ensuremath{\cdots} & \ensuremath{\cdots} & $\cdots$ \\
\end{longtable}
\par
\endgroup
\twocolumngrid

\par
\onecolumngrid
\begingroup
\footnotesize
\setlength{\tabcolsep}{2.2pt}
\renewcommand{\arraystretch}{1.0}
\setlength{\LTleft}{0pt}
\setlength{\LTright}{0pt}
\setlength{\LTcapwidth}{\textwidth}
\setlength{\LTpre}{10pt}
\setlength{\LTpost}{4pt}
\setlength{\parindent}{0pt}
\begin{longtable}{@{\extracolsep{\fill}}l*{8}{c}@{}}
\caption[Mass spectrum of the $D_s$ meson family. Masses are given in $\mathrm{MeV}$; mixing angles follow the conventions of the corresponding references.]{Mass spectrum of the $D_s$ meson family. Masses are given in $\mathrm{MeV}$; mixing angles follow the conventions of the corresponding references. 
Experimental entries adopted in this work; see the PDG compilation~\cite{ParticleDataGroup:2026mpi}. For states included in the fit, the experimental masses and uncertainties reproduce Table~\ref{tab1}; other entries retain their cited experimental references.  NR: nonrelativistic constituent quark model~\cite{Li:2010vx};  GI: Godfrey--Isgur model~\cite{Godfrey:2015dva};  RQM: relativistic quasipotential model~\cite{Ebert:2009ua}; BS: instantaneous Bethe--Salpeter model~\cite{Li:2024dyi} PM: semirelativistic potential model~\cite{Ni:2021pce}.  The $1P_1$ and $1P_1^\prime$ rows list the masses of $D_{s1}(2460)^+$ and $D_{s1}(2536)^+$, respectively, for comparison. In the ``This work'' column, primed states denote the higher-mass members of the mixed pairs. An ellipsis indicates that no value is quoted in the present table. The central mass $2933~\mathrm{MeV}$ is quoted as in the discussion~\cite{LHCb:2026sup}. Its repetition in the $2P_1$ and $2P_1^\prime$ rows denotes alternative assignments of the same $D_{s1}(2933)^+$ resonance.
}
\label{tab:Ds}\\
\toprule[1pt]\toprule[1pt]
States & Meson & Expt. & This work
& NR
& GI
& RQM
& PM
& BS \\
\midrule[0.6pt]
\endfirsthead
\caption[]{Mass spectrum of the $D_s$ meson family (continued).}\\
\toprule[1pt]\toprule[1pt]
States & Meson & Expt. & This work
& NR
& GI
& RQM
& PM
& BS \\
\midrule[0.6pt]
\endhead
\midrule
\multicolumn{9}{r}{\footnotesize Continued on next page}\\
\endfoot
\bottomrule[1pt]
\bottomrule[1pt]
\endlastfoot
$1^1S_0$ & $D_s^+$ & $1969\pm1.4$ & $1963$ & $1969$ & $1979$ & $1969$ & $1969$ & $1968$ \\
$2^1S_0$ & $D_{s0}(2590)^+$ & $2591\pm9.2195$ & $2620$ & $2640$ & $2673$ & $2688$ & $2649$ & $2681$ \\
$3^1S_0$ &  & \ensuremath{\cdots} & $3028$ & \ensuremath{\cdots} & $3154$ & $3219$ & $3126$ & $3131$ \\
$4^1S_0$ &  & \ensuremath{\cdots} & $3330$ & \ensuremath{\cdots} & $3547$ & $3652$ & \ensuremath{\cdots} & $\cdots$ \\
$5^1S_0$ &  & \ensuremath{\cdots} & $3569$ & \ensuremath{\cdots} & $3894$ & $4033$ & \ensuremath{\cdots} & $\cdots$ \\
\addlinespace[3pt]

$1^3S_1$ & $D_s^{*+}$ & $2112.2\pm0.4$ & $2110$ & $2107$ & $2129$ & $2111$ & $2112$ & $2112$ \\
$2^3S_1$ & $D_{s1}^*(2700)^+$ & $2714\pm5$ & $2681$ & $2714$ & $2732$ & $2731$ & $2737$ & $2720$ \\
$3^3S_1$ &  & \ensuremath{\cdots} & $3066$ & \ensuremath{\cdots} & $3193$ & $3242$ & $3191$ & $3200$ \\
$4^3S_1$ &  & \ensuremath{\cdots} & $3357$ & \ensuremath{\cdots} & $3575$ & $3669$ & \ensuremath{\cdots} & $\cdots$ \\
$5^3S_1$ &  & \ensuremath{\cdots} & $3590$ & \ensuremath{\cdots} & $3912$ & $4048$ & \ensuremath{\cdots} & $\cdots$ \\
\addlinespace[3pt]
$1^3P_0$ & $D_{s0}^*(2317)^+$ & $2318\pm0.7$ & $2373$ & $2344$ & $2484$ & $2509$ & $2409$ & $2318$ \\
$1^3P_2$ & $D_{s2}^*(2573)^+$ & $2569.1\pm0.8$ & $2561$ & $2559$ & $2592$ & $2571$ & $2575$ & $2569$ \\
$1P_1$ & $D_{s1}(2460)^+$ & $2459.5\pm0.6$ & $2496$ & $2488$ & $2549$ & $2536$ & $2528$ & $2459$ \\
$1P'_1$ & $D_{s1}(2536)^+$ &  {$2535.12\pm0.06$} & $2515$ & $2510$ & $2556$ & $2574$ & $2545$ & $2535$ \\
$\theta_{1P}$ &  & \ensuremath{\cdots} & $-34.9^{\circ}$ & \ensuremath{\cdots} & $-37.5^{\circ}$ & $34.5^{\circ}$ & $-36.8^{\circ}$ & $\cdots$ \\
\addlinespace[3pt]

$2^3P_0$ &  & \ensuremath{\cdots} & $2859$ & $2830$ & $3005$ & $3054$ & $2940$ & $2995$ \\
$2^3P_2$ &  & \ensuremath{\cdots} & $2964$ & $3040$ & $3048$ & $3142$ & $3053$ & $3050$ \\
$2P_1$ & {$D_{s1}(2933)^+$} & {$2933^{+6+4}_{-5-3}$} & $2923$ & $2958$ & $3018$ & $3067$ & $3002$ & $3016$ \\
$2P'_1$ & {$D_{s1}(2933)^+$} & {$2933^{+6+4}_{-5-3}$} & $2928$ & $2995$ & $3038$ & $3154$ & $3026$ & $3034$ \\
$\theta_{2P}$ &  & \ensuremath{\cdots} & $-49.2^{\circ}$ & \ensuremath{\cdots} & $-30.4^{\circ}$ & $37.6^{\circ}$ & $-21.0^{\circ}$ & $\cdots$ \\
\addlinespace[3pt]

$3^3P_0$ &  & \ensuremath{\cdots} & $3197$ & \ensuremath{\cdots} & $3412$ & $3513$ & \ensuremath{\cdots} & $3339$ \\
$3^3P_2$ &  & \ensuremath{\cdots} & $3271$ & \ensuremath{\cdots} & $3439$ & $3580$ & \ensuremath{\cdots} & $3418$ \\
$3P_1$ &  & \ensuremath{\cdots} & $3240$ & \ensuremath{\cdots} & $3416$ & $3519$ & \ensuremath{\cdots} & $3356$ \\
$3P'_1$ &  & \ensuremath{\cdots} & $3244$ & \ensuremath{\cdots} & $3433$ & $3618$ & \ensuremath{\cdots} & $3368$ \\
$\theta_{3P}$ &  & \ensuremath{\cdots} & $-58.7^{\circ}$ & \ensuremath{\cdots} & $-27.7^{\circ}$ & $38.2^{\circ}$ & \ensuremath{\cdots} & $\cdots$ \\
\addlinespace[3pt]

$4^3P_0$ &  & \ensuremath{\cdots} & $3460$ & \ensuremath{\cdots} & $3764$ & \ensuremath{\cdots} & \ensuremath{\cdots} & $\cdots$ \\
$4^3P_2$ &  & \ensuremath{\cdots} & $3517$ & \ensuremath{\cdots} & $3783$ & \ensuremath{\cdots} & \ensuremath{\cdots} & $\cdots$ \\
$4P_1$ &  & \ensuremath{\cdots} & $3491$ & \ensuremath{\cdots} & $3764$ & \ensuremath{\cdots} & \ensuremath{\cdots} & $\cdots$ \\
$4P'_1$ &  & \ensuremath{\cdots} & $3495$ & \ensuremath{\cdots} & $3778$ & \ensuremath{\cdots} & \ensuremath{\cdots} & $\cdots$ \\
$\theta_{4P}$ &  & \ensuremath{\cdots} & $-62.2^{\circ}$ & \ensuremath{\cdots} & $-25.4^{\circ}$ & \ensuremath{\cdots} & \ensuremath{\cdots} & $\cdots$ \\
\addlinespace[3pt]
$1^3D_1$ & $D_{s1}^*(2860)^+$ & $2859\pm26.833$ & $2798$ & $2804$ & $2899$ & $2913$ & $2843$ & $\cdots$ \\
$1^3D_3$ & $D_{s3}^*(2860)^+$ & $2860.5\pm7.0007$ & $2856$ & $2811$ & $2917$ & $2971$ & $2882$ & $\cdots$ \\
$1D_2$ &  & \ensuremath{\cdots} & $2837$ & $2788$ & $2900$ & $2931$ & $2857$ & $\cdots$ \\
$1D'_2$ &  & \ensuremath{\cdots} & $2844$ & $2849$ & $2926$ & $2961$ & $2911$ & $\cdots$ \\
$\theta_{1D}$ &  & \ensuremath{\cdots} & $52.0^{\circ}$ & \ensuremath{\cdots} & $-38.5^{\circ}$ & $39.2^{\circ}$ & $-40.7^{\circ}$ & $\cdots$ \\
\addlinespace[3pt]

$2^3D_1$ &  & \ensuremath{\cdots} & $3134$ & $3217$ & $3306$ & $3383$ & $3233$ & $\cdots$ \\
$2^3D_3$ &  & \ensuremath{\cdots} & $3181$ & $3240$ & $3311$ & $3469$ & $3299$ & $\cdots$ \\
$2D_2$ &  & \ensuremath{\cdots} & $3163$ & $3217$ & $3298$ & $3403$ & $3267$ & $\cdots$ \\
$2D'_2$ &  & \ensuremath{\cdots} & $3168$ & $3260$ & $3323$ & $3456$ & $3306$ & $\cdots$ \\
$\theta_{2D}$ &  & \ensuremath{\cdots} & $56.3^{\circ}$ & \ensuremath{\cdots} & $-37.7^{\circ}$ & $41.2^{\circ}$ & $-41.3^{\circ}$ & $\cdots$ \\
\addlinespace[3pt]

$3^3D_1$ &  & \ensuremath{\cdots} & $3400$ & \ensuremath{\cdots} & $3658$ & \ensuremath{\cdots} & \ensuremath{\cdots} & $\cdots$ \\
$3^3D_3$ &  & \ensuremath{\cdots} & $3441$ & \ensuremath{\cdots} & $3661$ & \ensuremath{\cdots} & \ensuremath{\cdots} & $\cdots$ \\
$3D_2$ &  & \ensuremath{\cdots} & $3425$ & \ensuremath{\cdots} & $3650$ & \ensuremath{\cdots} & \ensuremath{\cdots} & $\cdots$ \\
$3D'_2$ &  & \ensuremath{\cdots} & $3427$ & \ensuremath{\cdots} & $3672$ & \ensuremath{\cdots} & \ensuremath{\cdots} & $\cdots$ \\
$\theta_{3D}$ &  & \ensuremath{\cdots} & $65.3^{\circ}$ & \ensuremath{\cdots} & $-37.2^{\circ}$ & \ensuremath{\cdots} & \ensuremath{\cdots} & $\cdots$ \\

\addlinespace[3pt]
$4^3D_1$ &  & $\cdots$ & $3618$ & $\cdots$ & $\cdots$ & $\cdots$ & $\cdots$ & $\cdots$ \\
$4^3D_3$ &  & $\cdots$ & $3654$ & $\cdots$ & $\cdots$ & $\cdots$ & $\cdots$ & $\cdots$ \\
$4D_2$ &  & $\cdots$ & $3640$ & $\cdots$ & $\cdots$ & $\cdots$ & $\cdots$ & $\cdots$ \\
$4D'_2$ &  & $\cdots$ & $3641$ & $\cdots$ & $\cdots$ & $\cdots$ & $\cdots$ & $\cdots$ \\
$\theta_{4D}$ &  & \ensuremath{\cdots} & $89.2^{\circ}$ & \ensuremath{\cdots} & \ensuremath{\cdots} & \ensuremath{\cdots} & \ensuremath{\cdots} & $\cdots$ \\
\addlinespace[3pt]
$1^3F_2$ &  & \ensuremath{\cdots} & $3083$ & \ensuremath{\cdots} & $3208$ & $3230$ & $3176$ & $\cdots$ \\
$1^3F_4$ &  & \ensuremath{\cdots} & $3093$ & \ensuremath{\cdots} & $3190$ & $3300$ & $3134$ & $\cdots$ \\
$1F_3$ &  & \ensuremath{\cdots} & $3084$ & \ensuremath{\cdots} & $3186$ & $3254$ & $3123$ & $\cdots$ \\
$1F'_3$ &  & \ensuremath{\cdots} & $3103$ & \ensuremath{\cdots} & $3218$ & $3266$ & $3205$ & $\cdots$ \\
$\theta_{1F}$ &  & \ensuremath{\cdots} & $50.7^{\circ}$ & \ensuremath{\cdots} & $-39.3^{\circ}$ & $40.5^{\circ}$ & $-40.7^{\circ}$ & $\cdots$ \\
\addlinespace[3pt]

$2^3F_2$ &  & \ensuremath{\cdots} & $3351$ & \ensuremath{\cdots} & $3562$ & \ensuremath{\cdots} & \ensuremath{\cdots} & $\cdots$ \\
$2^3F_4$ &  & \ensuremath{\cdots} & $3365$ & \ensuremath{\cdots} & $3544$ & $3754$ & \ensuremath{\cdots} & $\cdots$ \\
$2F_3$ &  & \ensuremath{\cdots} & $3356$ & \ensuremath{\cdots} & $3540$ & $3710$ & \ensuremath{\cdots} & $\cdots$ \\
$2F'_3$ &  & \ensuremath{\cdots} & $3368$ & \ensuremath{\cdots} & $3569$ & \ensuremath{\cdots} & \ensuremath{\cdots} & $\cdots$ \\
$\theta_{2F}$ &  & \ensuremath{\cdots} & $51.5^{\circ}$ & \ensuremath{\cdots} & $-39.1^{\circ}$ & \ensuremath{\cdots} & \ensuremath{\cdots} & $\cdots$ \\
\end{longtable}
\par
\endgroup
\twocolumngrid

For the ground states \mbox{$D(1{}^1S_0)$} and \mbox{$D^*(1{}^3S_1)$}, we obtained masses of \mbox{$1862$} and \mbox{$2013~\mathrm{MeV}$}, respectively, differing from the experimental values of \mbox{$D^0$} and \mbox{$D^{*0}$}, \mbox{$1864.85$} and \mbox{$2006.86~\mathrm{MeV}$}, by \mbox{$-2.85$} and \mbox{$+6.14~\mathrm{MeV}$}, respectively~\mbox{\cite{ParticleDataGroup:2026mpi}}. The fitted masses of \mbox{$D_s(1{}^1S_0)$} and \mbox{$D_s^*(1{}^3S_1)$} are \mbox{$1963$} and \mbox{$2110~\mathrm{MeV}$}, differing from the experimental values by \mbox{$-6.00$} and \mbox{$-2.2~\mathrm{MeV}$}, respectively. The calculated \mbox{$1S$} hyperfine splitting of the \mbox{$D$} and $D_s$ are \mbox{$151~\mathrm{MeV}$} and $147~\mathrm{MeV}$, compared with \mbox{$142~\mathrm{MeV}$} and  $143~\mathrm{MeV}$ obtained from the experimental masses. 
This indicates that the current parameters give ground-state masses within several MeV of the adopted inputs, while overestimating the hyperfine splittings.

For the \mbox{$1P$} nonstrange charmed mesons, we obtained \mbox{$M[D(1{}^3P_2)]=2462~\mathrm{MeV}$}, a mass that exceeds the tabulated \mbox{$D_2^*(2460)$} value by \mbox{$0.9~\mathrm{MeV}$}. Under the state correspondence adopted in Table~\mbox{\ref{tab:D}}, 
the masses of the axial-vector mixed states $M[D(1P_1)]=2373~\mathrm{MeV}$ and $M[D(1P_1^\prime)]=2414~\mathrm{MeV}$ are $39$ and $8.1~\mathrm{MeV}$ lower than the tabulated masses of $D_1(2430)$ and $D_1(2420)$, respectively.
Among these \mbox{$1P$} states, the largest absolute mass discrepancy occurs for \mbox{$M[D(1{}^3P_0)]=2235~\mathrm{MeV}$}, which lies \mbox{$108~\mathrm{MeV}$} below the quoted \mbox{$D_0^*(2300)$} mass of \mbox{$2343\pm10~\mathrm{MeV}$}~\mbox{\cite{ParticleDataGroup:2026mpi}}. These discrepancies may reflect the different uncertainty weights of the experimental data sets in the joint fit, together with the requirement that the central potential, tensor term, and spin--orbit coupling terms simultaneously satisfy constraints from hadron systems across multiple flavor sectors. These discrepancies indicate that the present global parameterization reproduces the nonstrange \mbox{$1P$} spectrum less accurately than the \mbox{$1S$} ground-state masses.

For the $1P$ charmed-strange meson $D_{s0}^*(2317)$, we obtained $M[D_s(1{}^3P_0)]=2373~\mathrm{MeV}$, compared with the tabulated GI and RQM results that exceed this value by $111$ and $136~\mathrm{MeV}$, respectively. This reduces the mass discrepancy between conventional potential-model predictions and experiment.
This indicates that the screened relativized quark model we employ partially alleviates the low-mass puzzle of $D_{s0}^*(2317)$~\cite{vanBeveren:2003kd,Lang:2014yfa}.

However, the calculated mass remains approximately $55~\mathrm{MeV}$ above the measured value. This is consistent with the finding of the screened GI study in Ref.~\cite{Song:2015nia} that the screened potential alone cannot adequately reproduce the low mass of $D_{s0}^*(2317)$.

For the $D_s$ axial-vector pair, the mixing angle is $\theta_{1P}=-34.9^\circ$. The predicted masses of the two $1^+$ mixed states, $M[D_s(1P_1)]=2496~\mathrm{MeV}$ and $M[D_s(1P_1^\prime)]=2515~\mathrm{MeV}$, are $36.5~\mathrm{MeV}$ above and $20.1~\mathrm{MeV}$ lower than the experimental masses of $D_{s1}(2460)^+$ and $D_{s1}(2536)^+$, respectively. 
The fitted $D_s(1{}^3P_2)$ mass of $2561~\mathrm{MeV}$ is $8.1~\mathrm{MeV}$ below the measured mass of $D_{s2}^*(2573)^+$~\cite{ParticleDataGroup:2026mpi}.

For higher excitations, the fitted result $M[D(2{}^1S_0)]=2513~\mathrm{MeV}$ lies $36~\mathrm{MeV}$ lower than the adopted mass of $2549\pm19~\mathrm{MeV}$, a candidate for the $2{}^1S_0$ assignment of $D_0{(2550)^0}$.

For the $J^P=0^-$ state $D_{s0}(2590)^+$ reported by LHCb,
the present fitted result,
$M[D_s(2{}^1S_0)]=2620~\mathrm{MeV}$,
is $29~\mathrm{MeV}$ above its central mass value of
$2591~\mathrm{MeV}$~\cite{LHCb:2020gnv}.
At the quoted precision, this result agrees with the value of
$2620~\mathrm{MeV}$ obtained within the screened-potential
framework of Ref.~\cite{Gao:2022bsb}; that study, which also
includes strong-decay calculations, supports a conventional
$2{}^1S_0$ interpretation of this state.
The present mass result is compatible with this interpretation.
On the other hand, together with the
present fitted result for
$M[D_s(2{}^3S_1)]=2681~\mathrm{MeV}$,
the calculated $2S$ hyperfine splitting is only
$61~\mathrm{MeV}$;
if $D_{s0}(2590)$ and $D_{s1}^*(2700)$ are identified as the
respective members of this doublet, the difference between
the adopted experimental central values is approximately
$123~\mathrm{MeV}$.

The differences from other model predictions generally increase with radial excitation. The present predictions, $M[D(3{}^1S_0)]=2920~\mathrm{MeV}$ and $M[D_s(3{}^1S_0)]=3028~\mathrm{MeV}$, are approximately $109$--$148$ and $98$--$191~\mathrm{MeV}$ below the tabulated GI, RQM, PM, and BS results, respectively; for $5{}^1S_0$, the differences from the GI results increase to $356$ and $325~\mathrm{MeV}$, respectively.
This trend is qualitatively consistent with the lowering of higher radial excitations by a screened confining potential~\cite{Li:2009ad,Li:2023wgq}. 

The $D_s(1{}^3D_1)$ and $D_s(1{}^3D_3)$ masses are $2798$ and $2856~\mathrm{MeV}$, respectively, lying $61$ and $4.5~\mathrm{MeV}$ lower than the experimental masses of $D_{s1}^*(2860)^+$ and $D_{s3}^*(2860)^+$, respectively~\cite{ParticleDataGroup:2026mpi}.

For nonstrange charmed mesons, the LHCb amplitude analysis of
$B^-\to D^{*+}\pi^-\pi^-$ determined the spin-parity quantum
numbers of $D_2(2740)$ and $D_3^*(2750)$ to be
$J^P=2^-$ and $3^-$, respectively~\cite{LHCb:2019juy}.
The two $J^P=2^-$ mixed $1D$ states obtained in this work,
$D(1D_2)$ and $D(1D_2^\prime)$, have masses of
$2734$ and $2735~\mathrm{MeV}$, respectively,
with a corresponding mixing angle of
$\theta_{1D}=-88.9^\circ$.
These masses are approximately $17$ and $16~\mathrm{MeV}$,
respectively, below the central mass value of
$2751~\mathrm{MeV}$ reported for $D_2(2740)^0$ in that analysis,
providing mass-based references for its $1D$ interpretation;
because the two calculated levels are nearly degenerate,
their specific correspondence to the observed state cannot
yet be distinguished on the basis of masses alone.
For $D_3^*(2750)^0$, which was included in the fit under the
$1{}^3D_3$ assignment, the mass obtained in this work is
$2756~\mathrm{MeV}$, approximately $7~\mathrm{MeV}$ below
the adopted experimental input of
$2763.1\pm3.2~\mathrm{MeV}$.

For $D_1^*(2600)$, the same analysis reports a mass of $2641.9\pm1.8\pm4.5~\mathrm{MeV}$; the present result, $M[D(2{}^3S_1)]=2575~\mathrm{MeV}$, is $66.9~\mathrm{MeV}$ lower than this central value. The calculated $D(1{}^3D_1)$ mass of $2676~\mathrm{MeV}$ lies $105~\mathrm{MeV}$ below the mass of $D_1^*(2760)^0$, $2781\pm18\pm13~\mathrm{MeV}$~\cite{LHCb:2015eqv}, which motivates an examination of possible mixing between \(2^{3}S_{1}\) and \(1^{3}D_{1}\). 
Ref.~\cite{Chen:2011rr} examined a mixed-state interpretation
of $D^*(2600)$ using both mass and strong-decay information
and proposed a mixing angle of approximately $19^\circ$
in its convention. 
However, the studies in Refs.~\cite{Song:2015fha,Godfrey:2015dva} support a predominantly $2{}^3S_1$ interpretation on the basis of strong-decay information. The corresponding decay widths and branching fractions have not yet been calculated in the present study; further comparisons incorporating the decay properties of the $1{}^3D_1$ state can be made in future work.

In a 2026 amplitude analysis of $B^0\to D^+D^-K^+\pi^-$ decays, LHCb reported $D_{s1}(2933)^+$ with a statistical significance exceeding $10\sigma$ and determined its spin-parity to be $J^P=1^+$~\cite{LHCb:2026sup}. The $D_s(2P_1)$ and $D_s(2P_1^\prime)$ masses are $2923$ and $2928~\mathrm{MeV}$, respectively, which lie $10.0$ and $5~\mathrm{MeV}$ lower than the measured central mass of $2933~\mathrm{MeV}$. The calculated mixing angle for this $D_s$ pair is $\theta_{2P}=-49.2^\circ$. Both states provide possible $2P$ axial-vector interpretations of $D_{s1}(2933)^+$, with the higher-mass member lying closer to the measured central value.

In addition, the present study provides mass predictions for higher excited states: the masses of $D(3{}^3S_1)$ and $D_s(3{}^3S_1)$ are $2958$ and $3066~\mathrm{MeV}$, respectively, while the corresponding $3P$ multiplets are in the ranges $3084$--$3163$ and $3197$--$3271~\mathrm{MeV}$, respectively. For higher orbital excitations, the predicted masses of $D(2{}^3D_3)$ and $D_s(2{}^3D_3)$ are $3075$ and $3181~\mathrm{MeV}$, respectively, whereas those of $D(2{}^3F_4)$ and $D_s(2{}^3F_4)$ are $3261$ and $3365~\mathrm{MeV}$, respectively. These results provide reference mass ranges for future experiments.

\subsection{The $B$ and $B_s$ meson spectra}

Tables~\ref{tab:B} and~\ref{tab:Bs} present our $B$ and $B_s$ meson mass spectra, together with comparisons with available experimental data and other theoretical predictions.

\par
\onecolumngrid
\begingroup
\footnotesize
\setlength{\tabcolsep}{2.2pt}
\renewcommand{\arraystretch}{1.0}
\setlength{\LTleft}{0pt}
\setlength{\LTright}{0pt}
\setlength{\LTcapwidth}{\textwidth}
\setlength{\LTpre}{10pt}
\setlength{\LTpost}{4pt}
\setlength{\parindent}{0pt}
\begin{longtable}{@{\extracolsep{\fill}}l*{8}{c}@{}}
\caption[Mass spectrum of the $B$ meson family. Masses are given in $\mathrm{MeV}$; mixing angles follow the conventions of the corresponding references.]{Mass spectrum of the $B$ meson family. Masses are given in $\mathrm{MeV}$; mixing angles follow the conventions of the corresponding references. 
Experimental entries adopted in this work; see the PDG compilation~\cite{ParticleDataGroup:2026mpi}. For states included in the fit, the experimental masses and uncertainties reproduce Table~\ref{tab1}; other entries retain their cited experimental references.  NR: nonrelativistic quark model~\cite{Lu:2016bbk}; GI: Godfrey--Isgur model~\cite{PhysRevD.94.054025}; RQM: relativistic quasipotential model~\cite{Ebert:2009ua}; PM: nonrelativistic potential model~\cite{Asghar:2018tha}; BS: instantaneous Bethe--Salpeter model~\cite{Li:2024dyi}. 
In the ``This work'' column, primed states denote the higher-mass members of the mixed pairs. An ellipsis indicates that no value is quoted in the present table.
The repeated $B_J(5970)^0$ entry in the $2^3S_1$ and $1^3D_1$ rows denotes alternative candidate assignments of the same observed structure.
}
\label{tab:B}\\
\toprule[1pt]\toprule[1pt]
States & Meson & Expt. & This work
& NR
& GI
& RQM
& PM
& BS \\
\toprule[1pt]
\endfirsthead
\caption[]{Mass spectrum of the $B$ meson family (continued).}\\
\toprule[1pt]\toprule[1pt]
States & Meson & Expt. & This work
& NR
& GI
& RQM
& PM
& BS \\
\toprule[1pt]
\endhead
\midrule
\multicolumn{9}{r}{\footnotesize Continued on next page}\\
\endfoot
\toprule[1pt]\toprule[1pt]
\endlastfoot
$1^1S_0$ & $B^0$ & $5279.63\pm0.2$ & $5278$ & $5280$ & $5312$ & $5280$ & $5268$ & $5280$ \\
$2^1S_0$ &  & \ensuremath{\cdots} & $5837$ & $5910$ & $5904$ & $5890$ & $5877$ & $5915$ \\
$3^1S_0$ &  & \ensuremath{\cdots} & $6206$ & $6369$ & $6335$ & $6379$ & $6288$ & $6316$ \\
$4^1S_0$ &  & \ensuremath{\cdots} & $6485$ & \ensuremath{\cdots} & $6689$ & $6781$ & $6631$ & $\cdots$ \\
$5^1S_0$ &  & \ensuremath{\cdots} & $6709$ & \ensuremath{\cdots} & $6997$ & $7129$ & \ensuremath{\cdots} & $\cdots$ \\
\addlinespace[3pt]
$1^3S_1$ & $B^{*0}$ & $5324.75\pm0.2$ & $5334$ & $5329$ & $5371$ & $5326$ & $5329$ & $5325$ \\
$2^3S_1$ & {$B_J(5970)^0$} & {$5971\pm5$} & $5866$ & $5939$ & $5933$ & $5906$ & $5905$ & $5952$ \\
$3^3S_1$ &  & \ensuremath{\cdots} & $6225$ & $6391$ & $6355$ & $6387$ & $6308$ & $6363$ \\
$4^3S_1$ &  & \ensuremath{\cdots} & $6499$ & \ensuremath{\cdots} & $6703$ & $6786$ & $6647$ & $\cdots$ \\
$5^3S_1$ &  & \ensuremath{\cdots} & $6719$ & \ensuremath{\cdots} & $7008$ & $7133$ & \ensuremath{\cdots} & $\cdots$ \\
\addlinespace[3pt]
$1^3P_0$ &  & \ensuremath{\cdots} & $5629$ & $5683$ & $5756$ & $5749$ & $5704$ & $5699$ \\
$1^3P_2$ & $B_2^*(5747)^+$ & $5737.3\pm0.7$ & $5754$ & $5768$ & $5797$ & $5741$ & $5769$ & $5737$ \\
$1P_1$ &  & \ensuremath{\cdots} & $5684$ & $5729$ & $5777$ & $5723$ & $5755$ & $5725$ \\
$1P'_1$ & {$B_1(5721)^0$} & {$5726.1\pm1.2$} & $5736$ & $5754$ & $5784$ & $5774$ & $5739$ & $5763$ \\
$\theta_{1P}$ &  & \ensuremath{\cdots} & $-36.2^{\circ}$ & $-34.6^{\circ}$ & $30.3^{\circ}$ & $35.0^{\circ}$ & $35.3^{\circ}$ & $\cdots$ \\
\addlinespace[3pt]
$2^3P_0$ &  & \ensuremath{\cdots} & $6059$ & $6145$ & $6213$ & $6221$ & $6129$ & $6199$ \\
$2^3P_2$ &  & \ensuremath{\cdots} & $6126$ & $6253$ & $6213$ & $6260$ & $6190$ & $6215$ \\
$2P_1$ &  & \ensuremath{\cdots} & $6088$ & $6185$ & $6197$ & $6209$ & $6175$ & $6195$ \\
$2P'_1$ &  & \ensuremath{\cdots} & $6108$ & $6241$ & $6228$ & $6281$ & $6161$ & $6224$ \\
$\theta_{2P}$ &  & \ensuremath{\cdots} & $-39.5^{\circ}$ & $-36.1^{\circ}$ & $32.3^{\circ}$ & $37.3^{\circ}$ & $35.3^{\circ}$ & $\cdots$ \\
\addlinespace[3pt]
$3^3P_0$ &  & \ensuremath{\cdots} & $6366$ & \ensuremath{\cdots} & $6576$ & $6629$ & $6480$ & $6510$ \\
$3^3P_2$ &  & \ensuremath{\cdots} & $6414$ & \ensuremath{\cdots} & $6570$ & $6678$ & $6539$ & $6500$ \\
$3P_1$ &  & \ensuremath{\cdots} & $6385$ & \ensuremath{\cdots} & $6557$ & $6650$ & $6525$ & $6499$ \\
$3P'_1$ &  & \ensuremath{\cdots} & $6399$ & \ensuremath{\cdots} & $6585$ & $6685$ & $6512$ & $6519$ \\
$\theta_{3P}$ &  & \ensuremath{\cdots} & $-41.4^{\circ}$ & \ensuremath{\cdots} & $31.6^{\circ}$ & $34.1^{\circ}$ & $35.3^{\circ}$ & $\cdots$ \\
\addlinespace[3pt]
$4^3P_0$ &  & \ensuremath{\cdots} & $6609$ & \ensuremath{\cdots} & $6890$ & \ensuremath{\cdots} & \ensuremath{\cdots} & $\cdots$ \\
$4^3P_2$ &  & \ensuremath{\cdots} & $6646$ & \ensuremath{\cdots} & $6883$ & \ensuremath{\cdots} & \ensuremath{\cdots} & $\cdots$ \\
$4P_1$ &  & \ensuremath{\cdots} & $6623$ & \ensuremath{\cdots} & $6872$ & \ensuremath{\cdots} & \ensuremath{\cdots} & $\cdots$ \\
$4P'_1$ &  & \ensuremath{\cdots} & $6634$ & \ensuremath{\cdots} & $6897$ & \ensuremath{\cdots} & \ensuremath{\cdots} & $\cdots$ \\
$\theta_{4P}$ &  & \ensuremath{\cdots} & $-42.5^{\circ}$ & \ensuremath{\cdots} & $31.0^{\circ}$ & \ensuremath{\cdots} & \ensuremath{\cdots} & $\cdots$ \\
\addlinespace[3pt]
$1^3D_1$ & {$B_J(5970)^0$} & {$5971\pm5$} & $5990$ & $6095$ & $6110$ & $6119$ & $6022$ & $\cdots$ \\
$1^3D_3$ &  & \ensuremath{\cdots} & $6035$ & $6014$ & $6106$ & $6091$ & $6031$ & $\cdots$ \\
$1D_2$ &  & \ensuremath{\cdots} & $6018$ & $6004$ & $6095$ & $6103$ & $6031$ & $\cdots$ \\
$1D'_2$ &  & \ensuremath{\cdots} & $6025$ & $6113$ & $6124$ & $6121$ & $6026$ & $\cdots$ \\
$\theta_{1D}$ &  & \ensuremath{\cdots} & $-38.4^{\circ}$ & $-39.6^{\circ}$ & $39.7^{\circ}$ & $38.0^{\circ}$ & $39.2^{\circ}$ & $\cdots$ \\
\addlinespace[3pt]
$2^3D_1$ &  & \ensuremath{\cdots} & $6299$ & $6497$ & $6475$ & $6534$ & $6384$ & $\cdots$ \\
$2^3D_3$ &  & \ensuremath{\cdots} & $6333$ & $6444$ & $6460$ & $6542$ & $6395$ & $\cdots$ \\
$2D_2$ &  & \ensuremath{\cdots} & $6319$ & $6435$ & $6450$ & $6528$ & $6393$ & $\cdots$ \\
$2D'_2$ &  & \ensuremath{\cdots} & $6323$ & $6513$ & $6486$ & $6554$ & $6391$ & $\cdots$ \\
$\theta_{2D}$ &  & \ensuremath{\cdots} & $-43.5^{\circ}$ & $-39.7^{\circ}$ & $39.0^{\circ}$ & $41.6^{\circ}$ & $39.2^{\circ}$ & $\cdots$ \\
\addlinespace[3pt]
$3^3D_1$ &  & \ensuremath{\cdots} & $6546$ & \ensuremath{\cdots} & $6792$ & \ensuremath{\cdots} & \ensuremath{\cdots} & $\cdots$ \\
$3^3D_3$ &  & \ensuremath{\cdots} & $6576$ & \ensuremath{\cdots} & $6775$ & \ensuremath{\cdots} & \ensuremath{\cdots} & $\cdots$ \\
$3D_2$ &  & \ensuremath{\cdots} & $6562$ & \ensuremath{\cdots} & $6767$ & \ensuremath{\cdots} & \ensuremath{\cdots} & $\cdots$ \\
$3D'_2$ &  & \ensuremath{\cdots} & $6567$ & \ensuremath{\cdots} & $6800$ & \ensuremath{\cdots} & \ensuremath{\cdots} & $\cdots$ \\
$\theta_{3D}$ &  & \ensuremath{\cdots} & $-44.1^{\circ}$ & \ensuremath{\cdots} & $38.6^{\circ}$ & \ensuremath{\cdots} & \ensuremath{\cdots} & $\cdots$ \\
\addlinespace[3pt]
$4^3D_1$ &  & $\cdots$ & $6752$ & $\cdots$ & $\cdots$ & $\cdots$ & $\cdots$ & $\cdots$ \\
$4^3D_3$ &  & $\cdots$ & $6779$ & $\cdots$ & $\cdots$ & $\cdots$ & $\cdots$ & $\cdots$ \\
$4D_2$ &  & $\cdots$ & $6765$ & $\cdots$ & $\cdots$ & $\cdots$ & $\cdots$ & $\cdots$ \\
$4D'_2$ &  & $\cdots$ & $6770$ & $\cdots$ & $\cdots$ & $\cdots$ & $\cdots$ & $\cdots$ \\
$\theta_{4D}$ &  & \ensuremath{\cdots} & $-44.0^{\circ}$ & \ensuremath{\cdots} & \ensuremath{\cdots} & \ensuremath{\cdots} & \ensuremath{\cdots} & $\cdots$ \\
\addlinespace[3pt]
$1^3F_2$ &  & \ensuremath{\cdots} & $6253$ & $6383$ & $6387$ & $6412$ & $6259$ & $\cdots$ \\
$1^3F_4$ &  & \ensuremath{\cdots} & $6260$ & $6243$ & $6364$ & $6380$ & $6252$ & $\cdots$ \\
$1F_3$ &  & \ensuremath{\cdots} & $6254$ & $6236$ & $6358$ & $6391$ & $6264$ & $\cdots$ \\
$1F'_3$ &  & \ensuremath{\cdots} & $6269$ & $6393$ & $6396$ & $6420$ & $6249$ & $\cdots$ \\
$\theta_{1F}$ &  & \ensuremath{\cdots} & $48.7^{\circ}$ & $-41.0^{\circ}$ & $41.1^{\circ}$ & $39.5^{\circ}$ & $40.9^{\circ}$ & $\cdots$ \\
\addlinespace[3pt]
$2^3F_2$ &  & \ensuremath{\cdots} & $6501$ & \ensuremath{\cdots} & $6704$ & \ensuremath{\cdots} & \ensuremath{\cdots} & $\cdots$ \\
$2^3F_4$ &  & \ensuremath{\cdots} & $6511$ & \ensuremath{\cdots} & $6679$ & $6784$ & \ensuremath{\cdots} & $\cdots$ \\
$2F_3$ &  & \ensuremath{\cdots} & $6504$ & \ensuremath{\cdots} & $6673$ & $6786$ & \ensuremath{\cdots} & $\cdots$ \\
$2F'_3$ &  & \ensuremath{\cdots} & $6514$ & \ensuremath{\cdots} & $6711$ & \ensuremath{\cdots} & \ensuremath{\cdots} & $\cdots$ \\
$\theta_{2F}$ &  & \ensuremath{\cdots} & $49.4^{\circ}$ & \ensuremath{\cdots} & $40.9^{\circ}$ & \ensuremath{\cdots} & \ensuremath{\cdots} & $\cdots$ \\
\end{longtable}
\par
\endgroup
\twocolumngrid

\par
\onecolumngrid
\begingroup
\footnotesize
\setlength{\tabcolsep}{2.2pt}
\renewcommand{\arraystretch}{1.0}
\setlength{\LTleft}{0pt}
\setlength{\LTright}{0pt}
\setlength{\LTcapwidth}{\textwidth}
\setlength{\LTpre}{10pt}
\setlength{\LTpost}{4pt}
\setlength{\parindent}{0pt}
\begin{longtable}{@{\extracolsep{\fill}}l*{8}{c}@{}}
\caption[Mass spectrum of the $B_s$ meson family. Masses are given in $\mathrm{MeV}$; mixing angles follow the conventions of the corresponding references.]{Mass spectrum of the $B_s$ meson family. Masses are given in $\mathrm{MeV}$; mixing angles follow the conventions of the corresponding references. Experimental entries adopted in this work; see the PDG compilation~\cite{ParticleDataGroup:2026mpi}. For states included in the fit, the experimental masses and uncertainties reproduce Table~\ref{tab1}; other entries retain their cited experimental references.  NR: nonrelativistic quark model~\cite{Lu:2016bbk}; GI: Godfrey--Isgur model~\cite{PhysRevD.94.054025}; RQM: relativistic quasipotential model~\cite{Ebert:2009ua}; PM: nonrelativistic potential model~\cite{Asghar:2018tha}; BS: instantaneous Bethe--Salpeter model~\cite{Li:2024dyi}. In the ``This work'' column, primed states denote the higher-mass members of the mixed pairs.  An ellipsis indicates that no value is quoted in the present table. 
}
\label{tab:Bs}\\
\toprule[1pt]
\toprule[1pt]
States & Meson & Expt. & This work
& NR
& GI
& RQM
& PM
& BS \\
\toprule[1pt]
\endfirsthead
\caption[]{Mass spectrum of the $B_s$ meson family (continued).}\\
\toprule[1pt]
\toprule[1pt]
States & Meson & Expt. & This work
& NR
& GI
& RQM
& PM
& BS \\
\toprule[1pt]
\endhead
\midrule
\multicolumn{9}{r}{\footnotesize Continued on next page}\\
\endfoot
\toprule[1pt]\toprule[1pt]
\endlastfoot
$1^1S_0$ & $B_s^0$ & $5366.91\pm0.11$ & $5362$ & $5362$ & $5394$ & $5372$ & $5377$ & $5367$ \\
$2^1S_0$ &  & \ensuremath{\cdots} & $5931$ & $5977$ & $5984$ & $5976$ & $5929$ & $5995$ \\
$3^1S_0$ &  & \ensuremath{\cdots} & $6302$ & $6415$ & $6410$ & $6467$ & $6305$ & $6385$ \\
$4^1S_0$ &  & \ensuremath{\cdots} & $6582$ & \ensuremath{\cdots} & $6759$ & $6874$ & $6619$ & $\cdots$ \\
$5^1S_0$ &  & \ensuremath{\cdots} & $6808$ & \ensuremath{\cdots} & $7063$ & $7231$ & \ensuremath{\cdots} & $\cdots$ \\
\addlinespace[3pt]
$1^3S_1$ & $B_s^{*0}$ & $5415.8\pm1.5$ & $5420$ & $5413$ & $5450$ & $5414$ & $5422$ & $5415$ \\
$2^3S_1$ & $B_{sJ}(6063)^0$~\cite{LHCb:2020pet} & {$6063.5\pm1.2\pm0.8$} & $5961$ & $6003$ & $6012$ & $5992$ & $5949$ & $6035$ \\
$3^3S_1$ &  & \ensuremath{\cdots} & $6322$ & $6435$ & $6429$ & $6475$ & $6319$ & $6444$ \\
$4^3S_1$ &  & \ensuremath{\cdots} & $6597$ & \ensuremath{\cdots} & $6773$ & $6879$ & $6631$ & $\cdots$ \\
$5^3S_1$ &  & \ensuremath{\cdots} & $6819$ & \ensuremath{\cdots} & $7076$ & $7235$ & \ensuremath{\cdots} & $\cdots$ \\
\addlinespace[3pt]
$1^3P_0$ & $B_{s0}^*(5700)^0$~\cite{LHCb:2026knl} & {$5698.9\pm1.5\pm0.6$} & $5746$ & $5756$ & $5831$ & $5833$ & $5770$ & $5829$ \\
$1^3P_2$ & $B_{s2}^*(5840)^0$ & $5839.92\pm0.14$ & $5843$ & $5851$ & $5876$ & $5842$ & $5822$ & $5840$ \\
$1P_1$ &  & \ensuremath{\cdots} & $5796$ & $5801$ & $5857$ & $5831$ & $5803$ & $5828$ \\
$1P'_1$ & {$B_{s1}(5830)^0$} & {$5828.73\pm0.20$} & $5823$ & $5836$ & $5861$ & $5865$ & $5801$ & $5854$ \\
$\theta_{1P}$ &  & \ensuremath{\cdots} & $-35.1^{\circ}$ & $-34.9^{\circ}$ & $39.1^{\circ}$ & $36.0^{\circ}$ & $35.3^{\circ}$ & $\cdots$ \\
\addlinespace[3pt]
$2^3P_0$ &  & \ensuremath{\cdots} & $6162$ & $6203$ & $6279$ & $6318$ & $6160$ & $\cdots$ \\
$2^3P_2$ &  & \ensuremath{\cdots} & $6221$ & $6309$ & $6295$ & $6359$ & $6208$ & $6301$ \\
$2P_1$ &  & \ensuremath{\cdots} & $6192$ & $6241$ & $6279$ & $6321$ & $6196$ & $6275$ \\
$2P'_1$ &  & \ensuremath{\cdots} & $6204$ & $6297$ & $6296$ & $6345$ & $6186$ & $6293$ \\
$\theta_{2P}$ &  & \ensuremath{\cdots} & $-38.3^{\circ}$ & $-36.1^{\circ}$ & $33.1^{\circ}$ & $34.0^{\circ}$ & $35.3^{\circ}$ & $\cdots$ \\
\addlinespace[3pt]
$3^3P_0$ &  & \ensuremath{\cdots} & $6468$ & \ensuremath{\cdots} & $6639$ & $6731$ & $6483$ & $6570$ \\
$3^3P_2$ &  & \ensuremath{\cdots} & $6512$ & \ensuremath{\cdots} & $6648$ & $6780$ & $6530$ & $6590$ \\
$3P_1$ &  & \ensuremath{\cdots} & $6488$ & \ensuremath{\cdots} & $6635$ & $6761$ & $6518$ & $6674$ \\
$3P'_1$ &  & \ensuremath{\cdots} & $6497$ & \ensuremath{\cdots} & $6650$ & $6768$ & $6508$ & $6586$ \\
$\theta_{3P}$ &  & \ensuremath{\cdots} & $-40.7^{\circ}$ & \ensuremath{\cdots} & $31.6^{\circ}$ & $36.7^{\circ}$ & $35.3^{\circ}$ & $\cdots$ \\
\addlinespace[3pt]
$4^3P_0$ &  & \ensuremath{\cdots} & $6711$ & \ensuremath{\cdots} & $6950$ & \ensuremath{\cdots} & \ensuremath{\cdots} & $\cdots$ \\
$4^3P_2$ &  & \ensuremath{\cdots} & $6746$ & \ensuremath{\cdots} & $6956$ & \ensuremath{\cdots} & \ensuremath{\cdots} & $\cdots$ \\
$4P_1$ &  & \ensuremath{\cdots} & $6727$ & \ensuremath{\cdots} & $6946$ & \ensuremath{\cdots} & \ensuremath{\cdots} & $\cdots$ \\
$4P'_1$ &  & \ensuremath{\cdots} & $6734$ & \ensuremath{\cdots} & $6959$ & \ensuremath{\cdots} & \ensuremath{\cdots} & $\cdots$ \\
$\theta_{4P}$ &  & \ensuremath{\cdots} & $-42.2^{\circ}$ & \ensuremath{\cdots} & $30.4^{\circ}$ & \ensuremath{\cdots} & \ensuremath{\cdots} & $\cdots$ \\
\addlinespace[3pt]
$1^3D_1$ & $B_{sJ}(6114)^0$~\cite{LHCb:2020pet} & {$6114\pm3\pm5$} & $6096$ & $6142$ & $6182$ & $6209$ & $6057$ & $\cdots$ \\
$1^3D_3$ &  & \ensuremath{\cdots} & $6124$ & $6096$ & $6179$ & $6191$ & $6063$ & $\cdots$ \\
$1D_2$ &  & \ensuremath{\cdots} & $6114$ & $6087$ & $6169$ & $6189$ & $6064$ & $\cdots$ \\
$1D'_2$ &  & \ensuremath{\cdots} & $6119$ & $6159$ & $6196$ & $6218$ & $6059$ & $\cdots$ \\
$\theta_{1D}$ &  & \ensuremath{\cdots} & $47.4^{\circ}$ & $-39.8^{\circ}$ & $40.0^{\circ}$ & $38.1^{\circ}$ & $39.2^{\circ}$ & $\cdots$ \\
\addlinespace[3pt]
$2^3D_1$ &  & \ensuremath{\cdots} & $6402$ & $6527$ & $6542$ & $6629$ & $6390$ & $\cdots$ \\
$2^3D_3$ &  & \ensuremath{\cdots} & $6428$ & $6500$ & $6535$ & $6637$ & $6399$ & $\cdots$ \\
$2D_2$ &  & \ensuremath{\cdots} & $6418$ & $6492$ & $6526$ & $6625$ & $6397$ & $\cdots$ \\
$2D'_2$ &  & \ensuremath{\cdots} & $6420$ & $6542$ & $6553$ & $6651$ & $6395$ & $\cdots$ \\
$\theta_{2D}$ &  & \ensuremath{\cdots} & $49.7^{\circ}$ & $-39.8^{\circ}$ & $39.5^{\circ}$ & $41.1^{\circ}$ & $39.2^{\circ}$ & $\cdots$ \\
\addlinespace[3pt]
$3^3D_1$ &  & \ensuremath{\cdots} & $6650$ & \ensuremath{\cdots} & $6855$ & \ensuremath{\cdots} & \ensuremath{\cdots} & $\cdots$ \\
$3^3D_3$ &  & \ensuremath{\cdots} & $6674$ & \ensuremath{\cdots} & $6849$ & \ensuremath{\cdots} & \ensuremath{\cdots} & $\cdots$ \\
$3D_2$ &  & \ensuremath{\cdots} & $6664$ & \ensuremath{\cdots} & $6841$ & \ensuremath{\cdots} & \ensuremath{\cdots} & $\cdots$ \\
$3D'_2$ &  & \ensuremath{\cdots} & $6665$ & \ensuremath{\cdots} & $6864$ & \ensuremath{\cdots} & \ensuremath{\cdots} & $\cdots$ \\
$\theta_{3D}$ &  & \ensuremath{\cdots} & $-45.0^{\circ}$ & \ensuremath{\cdots} & $39.1^{\circ}$ & \ensuremath{\cdots} & \ensuremath{\cdots} & $\cdots$ \\
\addlinespace[3pt]
$4^3D_1$ &  & $\cdots$ & $6856$ & $\cdots$ & $\cdots$ & $\cdots$ & $\cdots$ & $\cdots$ \\
$4^3D_3$ &  & $\cdots$ & $6878$ & $\cdots$ & $\cdots$ & $\cdots$ & $\cdots$ & $\cdots$ \\
$4D_2$ &  & $\cdots$ & $6868$ & $\cdots$ & $\cdots$ & $\cdots$ & $\cdots$ & $\cdots$ \\
$4D'_2$ &  & $\cdots$ & $6870$ & $\cdots$ & $\cdots$ & $\cdots$ & $\cdots$ & $\cdots$ \\
$\theta_{4D}$ &  & \ensuremath{\cdots} & $-43.1^{\circ}$ & \ensuremath{\cdots} & \ensuremath{\cdots} & \ensuremath{\cdots} & \ensuremath{\cdots} & $\cdots$ \\
\addlinespace[3pt]
$1^3F_2$ &  & \ensuremath{\cdots} & $6350$ & $6412$ & $6454$ & $6501$ & $6273$ & $\cdots$ \\
$1^3F_4$ &  & \ensuremath{\cdots} & $6349$ & $6319$ & $6432$ & $6475$ & $6267$ & $\cdots$ \\
$1F_3$ &  & \ensuremath{\cdots} & $6342$ & $6313$ & $6425$ & $6468$ & $6277$ & $\cdots$ \\
$1F'_3$ &  & \ensuremath{\cdots} & $6362$ & $6422$ & $6462$ & $6515$ & $6265$ & $\cdots$ \\
$\theta_{1F}$ &  & \ensuremath{\cdots} & $48.7^{\circ}$ & $-41.1^{\circ}$ & $41.1^{\circ}$ & $40.1^{\circ}$ & $40.9^{\circ}$ & $\cdots$ \\
\addlinespace[3pt]
$2^3F_2$ &  & \ensuremath{\cdots} & $6600$ & \ensuremath{\cdots} & $6768$ & \ensuremath{\cdots} & \ensuremath{\cdots} & $\cdots$ \\
$2^3F_4$ &  & \ensuremath{\cdots} & $6604$ & \ensuremath{\cdots} & $6748$ & $6878$ & \ensuremath{\cdots} & $\cdots$ \\
$2F_3$ &  & \ensuremath{\cdots} & $6597$ & \ensuremath{\cdots} & $6742$ & $6880$ & \ensuremath{\cdots} & $\cdots$ \\
$2F'_3$ &  & \ensuremath{\cdots} & $6610$ & \ensuremath{\cdots} & $6775$ & \ensuremath{\cdots} & \ensuremath{\cdots} & $\cdots$ \\
$\theta_{2F}$ &  & \ensuremath{\cdots} & $48.9^{\circ}$ & \ensuremath{\cdots} & $41.0^{\circ}$ & \ensuremath{\cdots} & \ensuremath{\cdots} & $\cdots$ \\
\end{longtable}
\par
\endgroup
\twocolumngrid

For $B(1{}^1S_0)$ and $B^*(1{}^3S_1)$, we obtained masses of $5278$ and $5334~\mathrm{MeV}$, respectively, which lie $1.63~\mathrm{MeV}$ below and $9.25~\mathrm{MeV}$ above the experimental values. The $B_s$ and $B_s^*$ masses of $5362$ and $5420~\mathrm{MeV}$ are $4.91~\mathrm{MeV}$ below and $4.2~\mathrm{MeV}$ above the experimental values, respectively~\cite{ParticleDataGroup:2026mpi}. The calculated $1S$ hyperfine splittings for $B$ and $B_s$ systems are $56$ and $58~\mathrm{MeV}$, respectively, approximately $10.9$ and $9.11~\mathrm{MeV}$ larger than the values of $45.1$ and $48.9~\mathrm{MeV}$ obtained from the experimental masses as shown in Tables \ref{tab:B} and \ref{tab:Bs}. In contrast, the spin-weighted average masses, defined as $\overline{M}_{1S}=[M(1{}^1S_0)+3M(1{}^3S_1)]/4$, are approximately $6.53$ and $1.92~\mathrm{MeV}$ above the corresponding experimental averages for the $B$ and $B_s$ systems, respectively. The spin-averaged ground-state masses are relatively close to experiment.

For the $1P$ states, the $M[B(1{}^3P_2)]=5754~\mathrm{MeV}$ and $M[B_s(1{}^3P_2)]=5843~\mathrm{MeV}$ are $16.7$ and $3.08~\mathrm{MeV}$ above the tabulated $B_2^*(5747)$ and $B_{s2}^*(5840)$ masses, respectively, providing a good description of the masses of these two narrow states. In addition, the PDG lists masses of $5726.1\pm1.2$ and $5828.73\pm0.20~\mathrm{MeV}$ for $B_1(5721)^0$ and $B_{s1}(5830)^0$, respectively~\cite{ParticleDataGroup:2026mpi}. The predicted masses $M[B(1P_1^\prime)]=5736~\mathrm{MeV}$ and $M[B_s(1P_1^\prime)]=5823~\mathrm{MeV}$ differ from these experimental central values by $+9.9$ and $-5.73~\mathrm{MeV}$. 

The calculated $1P$ mixing angles are $-36.2^\circ$ for $B$ and $-35.1^\circ$ for $B_s$, whose magnitudes are close to the angle Arctan$[\sqrt{\frac{L}{L+1}}]$ given by the heavy-quark limit, after allowing for basis-phase conventions \cite{Asghar:2019qjl,bokade2025bc}. In the heavy-quark limit, the narrow $1^+$ and $2^+$ states are usually associated with the $P$-wave doublet with light-degree-of-freedom angular momentum $j_\ell=3/2$. The studies in Refs.~\cite{DiPierro:2001dwf,Sun:2014wea} provide the relevant theoretical background through spectroscopy and decay analyses. The masses obtained in this work are compatible with this interpretation.

The prediction $M[B(1{}^3P_0)]=5629~\mathrm{MeV}$ is $54$, $127$, $120$, $75$, and $70~\mathrm{MeV}$ below the corresponding NR, GI, RQM, PM, and BS predictions, respectively~\cite{Lu:2016bbk,PhysRevD.94.054025,Ebert:2009ua,Asghar:2018tha,Li:2024dyi}. The predicted mass of the lower axial-vector state, $M[B(1P_1)]=5684~\mathrm{MeV}$, is also substantially lower than the corresponding tabulated predictions. In contrast, the fitted $1{}^3P_2$ mass and the predicted mass of the higher axial-vector state $B(1P_1^\prime)$ lie relatively close to those of the corresponding observed narrow states.

In the $B_s$ system, our predicted mass of $1{}^3P_0$, $5746~\mathrm{MeV}$, is $85$ and $87~\mathrm{MeV}$ lower than GI and RQM, respectively, but only $10$ and $24~\mathrm{MeV}$ lower than the NR and PM predictions. Thus, whether a prediction is substantially lower depends on the reference model; MGI does not produce the same downward shift relative to all models. The lowest $0^+$ and $1^+$ bottom-strange states may also be affected by dynamics near the $BK$ and $B^*K$ thresholds. 
Lattice QCD studies that include scattering channels also highlight such effects~\cite{Lang:2015hza}.

The prediction $M[B(3{}^1S_0)]=6206~\mathrm{MeV}$ is $163$, $129$, $173$, $82$, and $110~\mathrm{MeV}$ below the corresponding NR, GI, RQM, PM, and BS predictions, respectively. The predicted  $B(5{}^1S_0)$ mass $6709~\mathrm{MeV}$ is $288$ and $420~\mathrm{MeV}$ lower than the GI and RQM predictions, respectively. The prediction $M[B_s(5{}^1S_0)]=6808~\mathrm{MeV}$ is likewise  $255$ and $423~\mathrm{MeV}$ lower than the GI and RQM predictions, respectively. This trend is qualitatively consistent with the lowering of highly excited states by a screened confining potential~\cite{Li:2009ad,Li:2023wgq}. 

Nor do orbital excitations exhibit a uniform downward shift. For example, the calculated $B(1{}^3D_3)$ mass of $6035~\mathrm{MeV}$ is only $4~\mathrm{MeV}$ larger than the tabulated PM result, whereas the $B_s(1{}^3D_3)$ mass of $6124~\mathrm{MeV}$ is $55~\mathrm{MeV}$ smaller than the GI value but $61~\mathrm{MeV}$ larger than the PM value.

For the reported $B_J$ structure whose spectroscopic assignment remains uncertain, the PDG gives a mass of $5971\pm5~\mathrm{MeV}$ for $B_J(5970)^0$~\cite{ParticleDataGroup:2026mpi}. The prediction  $M[B(1{}^3D_1)]=5990~\mathrm{MeV}$ lies $19~\mathrm{MeV}$ above this value, whereas $M[B(2{}^3S_1)]=5866~\mathrm{MeV}$ is $105~\mathrm{MeV}$ lower. On the basis of mass comparisons alone, the $1{}^3D_1$ assignment is therefore closer to the observed mass than the $2{}^3S_1$ assignment in the present calculation. 
Reference~\cite{Xu:2014mka} investigates the $2{}^3S_1$ interpretation within an effective Lagrangian framework, and the mass and strong-decay analysis in Ref.~\cite{Sun:2014wea} also supports this radial-excitation assignment. In contrast, Ref.~\cite{Gandhi:2022nnk} proposes the $1{}^3D_1$ assignment based on a strong-decay analysis. The candidate classification considered here is therefore compatible with the latter interpretation. Because both $2{}^3S_1$ and $1{}^3D_1$ have $J^P=1^-$, decays into $B\pi$ and $B^*\pi$ are allowed for both states. Observation in a single allowed channel cannot distinguish between them, and further tests require their full decay patterns.

LHCb has observed a signal denoted as $B_{s0}^*(5700)^0$
in the $B_s^0\pi^0$ invariant-mass spectrum.
Its measured mass is $5698.9\pm1.5\pm0.6~\mathrm{MeV}$,
and its global significance exceeds
$7\sigma$~\cite{LHCb:2026knl}.
If this structure is interpreted as the lowest-lying
$J^P=0^+$ bottom-strange state, our calculated
$B_s(1\,{}^3P_0)$ mass is $5746~\mathrm{MeV}$.
This result is closer to the experimental central value
than the NR, GI, RQM, PM, and BS results listed in
Table~\ref{tab:Bs}, but still exceeds it by
$47.1~\mathrm{MeV}$.
For comparison, the coupled-channel calculation in
Ref.~\cite{Hao:2026ppm} yields a scalar mass of
$5727~\mathrm{MeV}$ after including $BK$ coupling,
$19~\mathrm{MeV}$ below our result.

A complementary interpretation identifies this scalar
state as the chiral partner of the ground-state
$B_s^0$~\cite{Nowak:2026zhc}.
The interpolation estimate in that study, including
$1/m_Q$ corrections, gives a mass splitting of approximately
$332~\mathrm{MeV}$ between the positive- and negative-parity states.
Combined with the $B_s^0$ ground-state mass adopted there,
this implies a scalar mass of approximately
$5699~\mathrm{MeV}$, close to the observed value. 
Near-threshold chiral dynamics provides another description:
earlier studies constrained by lattice QCD predicted the lowest
scalar bottom-strange state below the $BK$ threshold, with masses
of approximately
$5707$--$5724~\mathrm{MeV}$~\cite{Albaladejo:2016lbb,Albaladejo:2016ztm};
a lattice QCD calculation including the $BK$ scattering channel
likewise obtained a bound-state mass of
$5711\pm13\pm19~\mathrm{MeV}$~\cite{Lang:2015hza}. 

More recently, Ref.~\cite{Albaladejo:2026yfz} analyzed the LHCb
$B_s^0\pi^0$ spectrum using lattice-constrained coupled-channel
chiral amplitudes.
After adjusting one amplitude parameter, that study obtained
a pole mass of $5699.6^{+1.6}_{-1.7}~\mathrm{MeV}$, together
with a sizable $BK$ component and an isospin-violating decay
width of tens of keV.
This pole mass is approximately $46~\mathrm{MeV}$ smaller
than the present $B_s(1{}^3P_0)$ result of $5746~\mathrm{MeV}$.
These results suggest that near-threshold chiral dynamics
may provide a concrete mechanism for understanding the
discrepancy between the present result and the observed mass.

Additional information on bottom-strange spectroscopy is provided by the structures observed by LHCb in the $B^+K^-$ invariant-mass spectrum and reported in 2021. Under a two-state hypothesis in which both states decay directly to $B^+K^-$, the fitted masses are $6063.5\pm1.2\pm0.8$ and $6114\pm3\pm5~\mathrm{MeV}$~\cite{LHCb:2020pet}.

Following the scenarios discussed in Refs.~\cite{Garg:2023hvp,Gandhi:2022nnk}, we consider the lower peak as a $B_s(2{}^3S_1)$ candidate and the higher peak as a $B_s(1{}^3D_1)$ candidate. The corresponding calculated masses are $5961$ and $6096~\mathrm{MeV}$, respectively, which fall short of the experimental central values by $103$ and $18~\mathrm{MeV}$. Thus, the present calculation provides a closer mass reference for the $1{}^3D_1$ interpretation of the higher peak, but does not reproduce the $2{}^3S_1$ mass of the lower peak equally well. 
Whether these discrepancies are related to $S$--$D$ mixing or coupled-channel corrections requires an explicit dynamical calculation.

For higher excitations, the predicted $B(3{}^3S_1)$ and $B_s(3{}^3S_1)$ masses are $6225$ and $6322~\mathrm{MeV}$, respectively. The corresponding $3P$ multiplets span $6366$--$6414$ and $6468$--$6512~\mathrm{MeV}$. In addition, the $B(2{}^3D_3)$ and $B_s(2{}^3D_3)$ masses are $6333$ and $6428~\mathrm{MeV}$, respectively, while the $B(2{}^3F_4)$ and $B_s(2{}^3F_4)$ masses are $6511$ and $6604~\mathrm{MeV}$, respectively.

\subsection{The $B_c$ meson spectrum}

Using the parameters in Table~\ref{tab2}, we obtained the $B_c$ meson mass spectrum. Table~\ref{tab:Bc} presents the masses of the $S$-, $P$-, $D$-, and $F$-wave states, together with available experimental data and other theoretical predictions.
\subsubsection{S-wave $B_c$ mesons}

We first compare our calculated masses with the PDG masses of $B_c^+$ and $B_c(2S)^+$~\cite{ParticleDataGroup:2026mpi}. For $B_c^+$, the calculated mass of $M=6267~\mathrm{MeV}$ lies approximately $7.47~\mathrm{MeV}$ below the central value of the PDG result, $6274.47\pm0.32~\mathrm{MeV}$. For $B_c(2S)^+$, the calculated mass of $6856~\mathrm{MeV}$ is $15.2~\mathrm{MeV}$ below the central value of the PDG result, $6871.2\pm1.0~\mathrm{MeV}$.

Recently, ATLAS reported the first observation of a state decaying to $B_c^+\gamma$, with a measured central mass of $6339.0~\mathrm{MeV}$ and a signal significance exceeding $8\sigma$~\cite{ATLAS:2026ubk}. The observed properties are compatible with the $B_c^{*+}(1\,{}^3S_1)$ interpretation. The predicted mass of $6339~\mathrm{MeV}$ for the lowest $J^P=1^-$ state coincides with this central value at the precision displayed in Table~\ref{tab:Bc}. Our predicted splitting of $M[B_c^{*+}(1\,{}^3S_1)]-M[B_c^{+}(1\,{}^1S_0)]=72~\mathrm{MeV}$ exceeds the ATLAS central value of $64.5~\mathrm{MeV}$ by $7.5~\mathrm{MeV}$~\cite{ATLAS:2026ubk}.

Compared with the NR, GI, RQM, PM, and BS predictions listed in Table~\ref{tab:Bc}~\cite{Li:2019tbn,Godfrey:2004ya,Ebert:2002pp,Eichten:2019gig,Wang:2022cxy}, the results for low-lying states generally are in similar mass ranges.  Our $M[B_c(1{}^3S_1)]=6339~\mathrm{MeV}$ differs from each of these reference predictions by no more than $13~\mathrm{MeV}$.

\par
\onecolumngrid
\begingroup
\footnotesize
\setlength{\tabcolsep}{2.2pt}
\renewcommand{\arraystretch}{1.0}
\setlength{\LTleft}{0pt}
\setlength{\LTright}{0pt}
\setlength{\LTcapwidth}{\textwidth}
\setlength{\LTpre}{10pt}
\setlength{\LTpost}{4pt}
\setlength{\parindent}{0pt}
\begin{longtable}{@{\extracolsep{\fill}}l*{8}{c}@{}}
\caption[Mass spectrum of the $B_c$ meson family. Masses are given in $\mathrm{MeV}$; mixing angles follow the conventions specified in the corresponding references.]{Mass spectrum of the $B_c$ meson family. Masses are given in $\mathrm{MeV}$; mixing angles follow the conventions specified in the corresponding references. 
Experimental entries adopted in this work; see the PDG compilation~\cite{ParticleDataGroup:2026mpi}. For states included in the fit, the experimental masses and uncertainties reproduce Table~\ref{tab1}; other entries retain their cited experimental references.  NR: nonrelativistic potential model~\cite{Li:2019tbn}; GI: Godfrey--Isgur model~\cite{Godfrey:2004ya}; RQM: relativistic quasipotential model~\cite{Ebert:2002pp}; PM: potential model~\cite{Eichten:2019gig}; BS: Salpeter-equation model~\cite{Wang:2022cxy}.  In the ``This work'' column, primed states denote the higher-mass members of the mixed pairs.  An ellipsis indicates that no value is quoted in the present table.
}
\label{tab:Bc}\\
\toprule[1pt]
\toprule[1pt]
States & Meson & Expt. & This work
& NR
& GI
& RQM
& PM
& BS \\
\midrule[0.6pt]
\endfirsthead
\caption[]{Mass spectrum of the $B_c$ meson family (continued).}\\
\toprule[1pt]
\toprule[1pt]
States & Meson & Expt. & This work
& NR
& GI
& RQM
& PM
& BS \\
\midrule[0.6pt]
\endhead
\midrule
\multicolumn{9}{r}{\footnotesize Continued on next page}\\
\endfoot
\bottomrule[1pt]
\bottomrule[1pt]
\endlastfoot
$1^1S_0$ & $B_c^+$ & $6274.47\pm0.3191$ & $6267$ & $6271$ & $6271$ & $6270$ & $6275$ & $6277$ \\
$2^1S_0$ & $B_c(2S)^+$ & $6871.2\pm1$ & $6856$ & $6871$ & $6855$ & $6835$ & $6867$ & $6867$ \\
$3^1S_0$ &  & \ensuremath{\cdots} & $7221$ & $7239$ & $7250$ & $7193$ & $7254$ & $7228$ \\
$4^1S_0$ &  & \ensuremath{\cdots} & $7495$ & $7540$ & \ensuremath{\cdots} & \ensuremath{\cdots} & $7573$ & \ensuremath{\cdots} \\
$5^1S_0$ &  & \ensuremath{\cdots} & $7718$ & $7805$ & \ensuremath{\cdots} & \ensuremath{\cdots} & \ensuremath{\cdots} & \ensuremath{\cdots} \\
\addlinespace[3pt]

$1^3S_1$ & $B_c^{*+}$~\cite{ATLAS:2026ubk} & \smash{{$6339.0\pm1.4^{+1.0}_{-1.4}\pm0.3$}} & $6339$ & $6326$ & $6338$ & $6332$ & $6329$ & $6332$ \\
$2^3S_1$ &  & \ensuremath{\cdots} & $6891$ & $6890$ & $6887$ & $6881$ & $6898$ & $6911$ \\
$3^3S_1$ &  & \ensuremath{\cdots} & $7244$ & $7252$ & $7272$ & $7235$ & $7280$ & $7272$ \\
$4^3S_1$ &  & \ensuremath{\cdots} & $7513$ & $7550$ & \ensuremath{\cdots} & \ensuremath{\cdots} & $7596$ & \ensuremath{\cdots} \\
$5^3S_1$ &  & \ensuremath{\cdots} & $7732$ & $7813$ & \ensuremath{\cdots} & \ensuremath{\cdots} & \ensuremath{\cdots} & \ensuremath{\cdots} \\
\addlinespace[3pt]

$1^3P_0$ &  & \ensuremath{\cdots} & $6699$ & $6714$ & $6706$ & $6699$ & $6693$ & $6705$ \\
$1^3P_2$ &  & \ensuremath{\cdots} & $6775$ & $6787$ & $6768$ & $6762$ & $6751$ & $6762$ \\
$1P_1$ &  & \ensuremath{\cdots} & $6744$ & $6757$ & $6741$ & $6734$ & $6731$ & $6739$ \\
$1P'_1$ &  & \ensuremath{\cdots} & $6755$ & $6776$ & $6750$ & $6749$ & $6739$ & $6748$ \\
$\theta_{1P}$ &  & \ensuremath{\cdots} & $-25.9^{\circ}$ & $35.5^{\circ}$ & $22.4^{\circ}$ & $20.5^{\circ}$ & $18.7^{\circ}$ & $32.2^{\circ}$ \\
\addlinespace[3pt]

$2^3P_0$ &  & \ensuremath{\cdots} & $7098$ & $7107$ & $7122$ & $7091$ & $7105$ & $7112$ \\
$2^3P_2$ &  & \ensuremath{\cdots} & $7151$ & $7160$ & $7164$ & $7156$ & $7155$ & $7163$ \\
$2P_1$ &  & \ensuremath{\cdots} & $7129$ & $7134$ & $7145$ & $7126$ & $7136$ & $7144$ \\
$2P'_1$ &  & \ensuremath{\cdots} & $7136$ & $7150$ & $7150$ & $7145$ & $7144$ & $7149$ \\
$\theta_{2P}$ &  & \ensuremath{\cdots} & $-26.2^{\circ}$ & $38.0^{\circ}$ & $18.9^{\circ}$ & $23.2^{\circ}$ & $21.2^{\circ}$ & $30.9^{\circ}$ \\
\addlinespace[3pt]

$3^3P_0$ &  & \ensuremath{\cdots} & $7393$ & $7420$ & \ensuremath{\cdots} & \ensuremath{\cdots} & $7437$ & $7408$ \\
$3^3P_2$ &  & \ensuremath{\cdots} & $7434$ & $7464$ & \ensuremath{\cdots} & \ensuremath{\cdots} & $7483$ & $7456$ \\
$3P_1$ &  & \ensuremath{\cdots} & $7416$ & $7441$ & \ensuremath{\cdots} & \ensuremath{\cdots} & $7465$ & $7440$ \\
$3P'_1$ &  & \ensuremath{\cdots} & $7422$ & $7458$ & \ensuremath{\cdots} & \ensuremath{\cdots} & $7474$ & $7442$ \\
$\theta_{3P}$ &  & \ensuremath{\cdots} & $-26.9^{\circ}$ & $39.7^{\circ}$ & \ensuremath{\cdots} & \ensuremath{\cdots} & \ensuremath{\cdots} & $29.9^{\circ}$ \\
\addlinespace[3pt]

$4^3P_0$ &  & \ensuremath{\cdots} & $7630$ & $7693$ & \ensuremath{\cdots} & \ensuremath{\cdots} & \ensuremath{\cdots} & \ensuremath{\cdots} \\
$4^3P_2$ &  & \ensuremath{\cdots} & $7663$ & $7732$ & \ensuremath{\cdots} & \ensuremath{\cdots} & \ensuremath{\cdots} & \ensuremath{\cdots} \\
$4P_1$ &  & \ensuremath{\cdots} & $7649$ & $7710$ & \ensuremath{\cdots} & \ensuremath{\cdots} & \ensuremath{\cdots} & \ensuremath{\cdots} \\
$4P'_1$ &  & \ensuremath{\cdots} & $7653$ & $7727$ & \ensuremath{\cdots} & \ensuremath{\cdots} & \ensuremath{\cdots} & \ensuremath{\cdots} \\
$\theta_{4P}$ &  & \ensuremath{\cdots} & $-27.4^{\circ}$ & $39.7^{\circ}$ & \ensuremath{\cdots} & \ensuremath{\cdots} & \ensuremath{\cdots} & \ensuremath{\cdots} \\
\addlinespace[3pt]

$1^3D_1$ &  & \ensuremath{\cdots} & $7024$ & $7020$ & $7028$ & $7072$ & $7007$ & $7014$ \\
$1^3D_3$ &  & \ensuremath{\cdots} & $7046$ & $7030$ & $7045$ & $7081$ & $7011$ & $7035$ \\
$1D_2$ &  & \ensuremath{\cdots} & $7036$ & $7024$ & $7041$ & $7077$ & $7006$ & $7025$ \\
$1D'_2$ &  & \ensuremath{\cdots} & $7038$ & $7032$ & $7036$ & $7079$ & $7016$ & $7029$ \\
$\theta_{1D}$ &  & \ensuremath{\cdots} & $37.5^{\circ}$ & $45.0^{\circ}$ & $44.5^{\circ}$ & $-35.9^{\circ}$ & $-49.2^{\circ}$ & $31.8^{\circ}$ \\
\addlinespace[3pt]

$2^3D_1$ &  & \ensuremath{\cdots} & $7327$ & $7336$ & \ensuremath{\cdots} & \ensuremath{\cdots} & $7347$ & $7335$ \\
$2^3D_3$ &  & \ensuremath{\cdots} & $7347$ & $7348$ & \ensuremath{\cdots} & \ensuremath{\cdots} & $7354$ & $7355$ \\
$2D_2$ &  & \ensuremath{\cdots} & $7339$ & $7343$ & \ensuremath{\cdots} & \ensuremath{\cdots} & $7349$ & $7345$ \\
$2D'_2$ &  & \ensuremath{\cdots} & $7340$ & $7347$ & \ensuremath{\cdots} & \ensuremath{\cdots} & $7356$ & $7349$ \\
$\theta_{2D}$ &  & \ensuremath{\cdots} & $18.0^{\circ}$ & $45.0^{\circ}$ & \ensuremath{\cdots} & \ensuremath{\cdots} & $-40.3^{\circ}$ & $32.6^{\circ}$ \\
\addlinespace[3pt]

$3^3D_1$ &  & \ensuremath{\cdots} & $7571$ & $7611$ & \ensuremath{\cdots} & \ensuremath{\cdots} & \ensuremath{\cdots} & \ensuremath{\cdots} \\
$3^3D_3$ &  & \ensuremath{\cdots} & $7589$ & $7625$ & \ensuremath{\cdots} & \ensuremath{\cdots} & \ensuremath{\cdots} & \ensuremath{\cdots} \\
$3D_2$ &  & \ensuremath{\cdots} & $7581$ & $7620$ & \ensuremath{\cdots} & \ensuremath{\cdots} & \ensuremath{\cdots} & \ensuremath{\cdots} \\
$3D'_2$ &  & \ensuremath{\cdots} & $7582$ & $7623$ & \ensuremath{\cdots} & \ensuremath{\cdots} & \ensuremath{\cdots} & \ensuremath{\cdots} \\
$\theta_{3D}$ &  & \ensuremath{\cdots} & $-11.6^{\circ}$ & $45.0^{\circ}$ & \ensuremath{\cdots} & \ensuremath{\cdots} & \ensuremath{\cdots} & \ensuremath{\cdots} \\
\addlinespace[3pt]

$1^3F_2$ &  & \ensuremath{\cdots} & $7252$ & $7235$ & $7269$ & \ensuremath{\cdots} & $7234$ & $7239$ \\
$1^3F_4$ &  & \ensuremath{\cdots} & $7258$ & $7227$ & $7271$ & \ensuremath{\cdots} & $7224$ & \ensuremath{\cdots} \\
$1F_3$ &  & \ensuremath{\cdots} & $7253$ & $7224$ & $7276$ & \ensuremath{\cdots} & $7222$ & \ensuremath{\cdots} \\
$1F'_3$ &  & \ensuremath{\cdots} & $7259$ & $7240$ & $7266$ & \ensuremath{\cdots} & $7238$ & \ensuremath{\cdots} \\
$\theta_{1F}$ &  & \ensuremath{\cdots} & $48.2^{\circ}$ & $41.4^{\circ}$ & $41.4^{\circ}$ & \ensuremath{\cdots} & $-49.5^{\circ}$ & \ensuremath{\cdots} \\
\addlinespace[3pt]

$2^3F_2$ &  & \ensuremath{\cdots} & $7506$ & $7518$ & \ensuremath{\cdots} & \ensuremath{\cdots} & \ensuremath{\cdots} & $7508$ \\
$2^3F_4$ &  & \ensuremath{\cdots} & $7512$ & $7514$ & \ensuremath{\cdots} & \ensuremath{\cdots} & \ensuremath{\cdots} & \ensuremath{\cdots} \\
$2F_3$ &  & \ensuremath{\cdots} & $7508$ & $7508$ & \ensuremath{\cdots} & \ensuremath{\cdots} & \ensuremath{\cdots} & \ensuremath{\cdots} \\
$2F'_3$ &  & \ensuremath{\cdots} & $7512$ & $7525$ & \ensuremath{\cdots} & \ensuremath{\cdots} & \ensuremath{\cdots} & \ensuremath{\cdots} \\
$\theta_{2F}$ &  & \ensuremath{\cdots} & $48.1^{\circ}$ & $43.4^{\circ}$ & \ensuremath{\cdots} & \ensuremath{\cdots} & \ensuremath{\cdots} & \ensuremath{\cdots} \\
\end{longtable}
\par
\endgroup
\twocolumngrid
The predicted $B_c(2{}^3S_1)$ mass of $6891~\mathrm{MeV}$ is $20~\mathrm{MeV}$ below the BS value of $6911~\mathrm{MeV}$, but $4$ and $10~\mathrm{MeV}$ above the GI and RQM results, respectively. 
For higher excitations awaiting experimental tests, the predicted $B_c(3{}^1S_0)$ and $B_c(3{}^3S_1)$ masses are $7221$ and $7244~\mathrm{MeV}$, respectively.

For the $4S$ and $5S$ states, the predicted masses are lower than the corresponding reference values. The predicted $B_c(4{}^1S_0)$ and $B_c(4{}^3S_1)$ masses of $7495$ and $7513~\mathrm{MeV}$ are $78$ and $83~\mathrm{MeV}$ lower than the corresponding PM predictions, respectively.
For comparison, the screened Cornell potential study in Ref.~\cite{Li:2022bre} gives masses of $7551$ and $7561~\mathrm{MeV}$ for the two $4S$ states, both higher than the present results.  Since these calculations also differ in their parameters and other model ingredients, the numerical differences cannot be attributed solely to screening.
The $B_c(5{}^1S_0)$ and $B_c(5{}^3S_1)$ masses of $7718$ and $7732~\mathrm{MeV}$ are $87$ and $81~\mathrm{MeV}$ below the corresponding NR predictions, respectively. This trend is qualitatively consistent with the tendency of a screened confining potential to lower highly excited states~\cite{Li:2023wgq}.

\subsubsection{P-wave, D-wave, and  F-wave $B_c$ mesons}
In 2025, LHCb reported a peaking structure compatible with orbitally excited $B_c$ states in the $B_c^+\gamma$ invariant-mass spectrum. The structure was described using a minimal effective two-peak model, with fitted peak positions of $6704.8~\mathrm{MeV}$ and $6752.4~\mathrm{MeV}$~\cite{LHCb:2025uce,LHCb:2025ubr}. The PDG lists this observation under $B_c(1P)^\pm$, without assigning definite $J^P$ quantum numbers to the individual fitted peaks~\cite{ParticleDataGroup:2026mpi}.

In the present model, the lowest $P$-wave multiplet contains four physical states, namely $1\,{}^3P_0$, $1P_1$, $1P_1^\prime$, and $1\,{}^3P_2$, with the two axial-vector states formed from mixtures of $1\,{}^1P_1$ and $1\,{}^3P_1$. Our corresponding predicted masses are $6699~\mathrm{MeV}$, $6744~\mathrm{MeV}$, $6755~\mathrm{MeV}$, and $6775~\mathrm{MeV}$, respectively. Within this multiplet, the prediction $M[B_c(1{}^3P_0)]=6699~\mathrm{MeV}$ is close to the GI and BS predictions.

In leading-order electric-dipole (E1) transitions, the \mbox{$1\,{}^3P_0$} and \mbox{$1\,{}^3P_2$} states cannot decay directly into \mbox{$B_c^+\gamma$}, whereas both axial-vector mixed states, \mbox{$1P_1$} and \mbox{$1P_1^\prime$}, can decay into this final state through their spin-singlet components. For a direct-decay comparison, the predicted $B_c(1P_1)$ mass of $6744~\mathrm{MeV}$ is $39.2~\mathrm{MeV}$ above the lower fitted peak but $8.4~\mathrm{MeV}$ below the higher one. The predicted $B_c(1P_1^\prime)$ mass of $6755~\mathrm{MeV}$ lies $2.6~\mathrm{MeV}$ above the higher fitted peak. Thus, both axial-vector masses are near the higher peak, and their predicted separation of $11~\mathrm{MeV}$ is smaller than the $47.6~\mathrm{MeV}$ separation of the two effective fitted peaks.

Furthermore, all four $1P$ states can also contribute to the reconstructed $B_c^+\gamma$ spectrum through $B_c(1P)^+\to B_c^{*+}\gamma$, followed by $B_c^{*+}\to B_c^+\gamma_{\mathrm{s}}$, when the soft photon $\gamma_{\mathrm{s}}$ is not reconstructed. Such cascade contributions appear at reconstructed masses lower than the corresponding parent-state masses~\cite{LHCb:2025ubr}. These mass comparisons and transition patterns suggest that the higher-mass structure may receive direct contributions from the two axial-vector states, while the lower-mass structure may contain cascade contributions. Establishing the composition of the observed structures requires further studies of the radiative widths and reconstructed line shapes.

A complementary approach is provided by the study of the isospin-violating decays $B_c(1P)^+\to B_c^{(*)+}\pi^0$ in Ref.~\cite{Wang:2025vue}. That study adopts $6704.8$ and $6752.4~\mathrm{MeV}$ as the $1{}^3P_0$ and $1{}^3P_2$ mass inputs, respectively, rather than determining these quantum numbers from a fit to the photon spectrum. Relative to these inputs, the scalar and tensor masses differ by $-5.8$ and $+22.6~\mathrm{MeV}$.

Reference~\cite{Eichten:2019gig} emphasizes the importance of studying $B_c$ states near open-flavor thresholds and their decays. As shown in Table~\ref{tab:Bc}, the calculated $2P$ multiplet spans $7098$--$7151~\mathrm{MeV}$, close to the lowest $BD$ threshold at approximately $7144~\mathrm{MeV}$. The $2{}^3P_0$ state is below this threshold, whereas $2{}^3P_2$ is above it; both axial-vector states remain below the lowest $B^*D$ threshold at approximately $7189~\mathrm{MeV}$. The calculated $2P$ mixing angle is $\theta_{2P}=-26.2^\circ$.
The $3P$ multiplet spans $7393$--$7434~\mathrm{MeV}$, and its axial-vector mixing angle is $\theta_{3P}=-26.9^\circ$.

For the $D$- and $F$-wave states, $M[B_c(1{}^3D_3)]=7046~\mathrm{MeV}$ is $35~\mathrm{MeV}$ above the PM prediction, and $M[B_c(1{}^3F_4)]=7258~\mathrm{MeV}$ lies $31$ and $34~\mathrm{MeV}$ above the NR and PM predictions, respectively.
In addition, the $B_c(2{}^3D_3)$ and $B_c(2{}^3F_4)$ masses are $7347$ and $7512~\mathrm{MeV}$, respectively.
Additional $B_c$ mass predictions are listed in Table~\ref{tab:Bc}.

\subsection{The charmonium  spectrum}

Table~\ref{tab:cc} presents the charmonium mass spectrum calculated in this work, together with available experimental data and other theoretical predictions.

\par
\onecolumngrid
\begingroup
\footnotesize
\setlength{\tabcolsep}{2.2pt}
\renewcommand{\arraystretch}{1.0}
\setlength{\LTleft}{0pt}
\setlength{\LTright}{0pt}
\setlength{\LTcapwidth}{\textwidth}
\setlength{\LTpre}{10pt}
\setlength{\LTpost}{4pt}
\setlength{\parindent}{0pt}
\begin{longtable}{@{\extracolsep{\fill}}l*{9}{c}@{}}
\caption[Mass spectrum of the $c\bar{c}$ system. Masses are given in $ \mathrm{MeV}$.]{Mass spectrum of the $c\bar{c}$ system. Masses are given in $ \mathrm{MeV}$. 
Experimental entries adopted in this work; see the PDG compilation~\cite{ParticleDataGroup:2026mpi}. For states included in the fit, the experimental masses and uncertainties reproduce Table~\ref{tab1}; other entries retain their cited experimental references. NR: nonrelativistic potential model~\cite{Barnes:2005pb}; GI: Godfrey--Isgur model~\cite{Barnes:2005pb}; RQM: relativistic quasipotential model~\cite{Ebert:2002pp}; PM: screened potential model~\cite{Li:2009zu}; BS: covariant Bethe--Salpeter model~\cite{Fischer:2014cfa}; MGI: screened relativistic quark model~\cite{Wang:2019mhs}. An ellipsis indicates that no value is quoted in the present table.}

\label{tab:cc}\\
\toprule[1pt]
\toprule[1pt]
States & Meson & Expt. & This work & NR & GI & RQM & PM & BS & MGI \\
\midrule[0.6pt]
\endfirsthead
\caption[]{Mass spectrum of the $c\bar c$ system (continued).}\\
\toprule[1pt]
\toprule[1pt]
States & Meson & Expt. & This work & NR & GI & RQM & PM & BS & MGI \\
\midrule[0.6pt]
\endhead
\midrule
\multicolumn{10}{r}{\footnotesize Continued on next page}\\
\endfoot
\bottomrule[1pt]
\bottomrule[1pt]
\endlastfoot
$1^{1}S_{0}$ & $\eta_c(1S)$ & $2984.09\pm0.3$ & $2979$ & $2982$ & $2975$ & $2979$ & $2979$ & $2925$  & {$2981$} \\
$2^{1}S_{0}$ & $\eta_c(2S)$ & $3637.8\pm0.6$ & $3621$ & $3630$ & $3623$ & $3588$ & $3623$ & $\cdots$  & {$3642$} \\
$3^{1}S_{0}$ & $\eta_c(3945)$~\cite{LHCb:2024vfz} & {$3945^{+28}_{-17}{}^{+37}_{-28}\,{}$} & $4014$ & $4043$ & $4064$ & $3991$ & $3991$ & $\cdots$  & {$4013$} \\
$4^{1}S_{0}$ &  & $\cdots$ & $4307$ & $4384$ & $4425$ & $\cdots$ & $4250$ & $\cdots$  & {$4260$} \\
$5^{1}S_{0}$ &  & $\cdots$ & $4542$ & $\cdots$ & $\cdots$ & $\cdots$ & $4446$ & $\cdots$  & {$4433$} \\
$1^{3}S_{1}$ & $J/\psi$ & $3096.9\pm0.006$ & $3105$ & $3090$ & $3098$ & $3096$ & $3097$ & $3113$  & {$3096$} \\
$2^{3}S_{1}$ & $\psi(2S)$ & $3686.097\pm0.01$ & $3675$ & $3672$ & $3676$ & $3686$ & $3673$ & $\cdots$  & {$3683$} \\
$3^{3}S_{1}$ & $\psi(4040)$ & $4039.6\pm4.3$ & $4049$ & $4072$ & $4100$ & $4088$ & $4022$ & $\cdots$  & {$4035$} \\
$4^{3}S_{1}$ & {$\psi(4230)$}& {$4221.7\pm2.5$}& $4332$ & $4406$ & $4450$ & $\cdots$ & $4273$ & $\cdots$  & {$4274$} \\
$5^{3}S_{1}$ & $Y(4500)$~\cite{BESIII:2022joj} & {$4484.7\pm13.3\pm24.1\,{}$} & $4562$ & $\cdots$ & $\cdots$ & $\cdots$ & $4463$ & $\cdots$  & {$4443$} \\
\addlinespace[3pt]

$1^{3}P_{0}$ & $\chi_{c0}(1P)$ & $3415.5\pm0.19$ & $3424$ & $3424$ & $3445$ & $3424$ & $3433$ & $3323$  & {$3464$} \\
$1^{3}P_{2}$ & $\chi_{c2}(1P)$ & $3556.17\pm0.07$ & $3555$ & $3556$ & $3550$ & $3556$ & $3554$ & $3550$  & {$3571$} \\
$1^{1}P_{1}$ & $h_c(1P)$ & $3525.37\pm0.14$ & $3519$ & $3516$ & $3517$ & $3526$ & $3519$ & $3433$  & {$3538$} \\
$1^{3}P_{1}$ & $\chi_{c1}(1P)$ & $3510.67\pm0.05$ & $3511$ & $3505$ & $3510$ & $3510$ & $3510$ & $3489$  & {$3530$} \\
$2^{3}P_{0}$ & $\chi_{c0}(3915)$ & $3922.1\pm1.8$ & $3866$ & $3852$ & $3916$ & $3854$ & $3842$ & $\cdots$  & {$3896$} \\
$2^{3}P_{2}$ & $\chi_{c2}(3930)$ & $3922.5\pm1$ & $3950$ & $3972$ & $3979$ & $3972$ & $3937$ & $\cdots$  & {$3952$} \\
$2^{1}P_{1}$ & $h_c(4000)$~\cite{LHCb:2024vfz} & {$4000^{+17}_{-14}{}^{+29}_{-22}\,{}$} & $3924$ & $3934$ & $3956$ & $3945$ & $3908$ & $\cdots$  & {$3933$} \\
$2^{3}P_{1}$ & $\chi_{c1}(4010)$~\cite{LHCb:2024vfz} & {$4012.5^{+3.6}_{-3.9}{}^{+4.1}_{-3.7}\,{}$} & $3920$ & $3925$ & $3953$ & $3929$ & $3901$ & $\cdots$  & {$3929$} \\
$3^{3}P_{0}$ &  & $\cdots$ & $4187$ & $4202$ & $4292$ & $\cdots$ & $4131$ & $\cdots$  & {$4177$} \\
$3^{3}P_{2}$ &  & $\cdots$ & $4250$ & $4317$ & $4337$ & $\cdots$ & $4208$ & $\cdots$  & {$4213$} \\
$3^{1}P_{1}$ & $h_c(4300)$~\cite{LHCb:2024vfz} & {$4307.3^{+6.4}_{-6.6}{}^{+3.3}_{-4.1}\,{}$} & $4229$ & $4279$ & $4318$ & $\cdots$ & $4184$ & $\cdots$  & {$4200$} \\
$3^{3}P_{1}$ &  & $\cdots$ & $4226$ & $4271$ & $4317$ & $\cdots$ & $4178$ & $\cdots$  & {$4197$} \\
$4^{3}P_{0}$ &  & $\cdots$ & $4442$ & $\cdots$ & $\cdots$ & $\cdots$ & $\cdots$ & $\cdots$  & {$4374$} \\
$4^{3}P_{2}$ &  & $\cdots$ & $4491$ & $\cdots$ & $\cdots$ & $\cdots$ & $\cdots$ & $\cdots$  & {$4398$} \\
$4^{1}P_{1}$ &  & $\cdots$ & $4474$ & $\cdots$ & $\cdots$ & $\cdots$ & $\cdots$ & $\cdots$  & {$4389$} \\
$4^{3}P_{1}$ &  & $\cdots$ & $4472$ & $\cdots$ & $\cdots$ & $\cdots$ & $\cdots$ & $\cdots$  & {$4387$} \\
\addlinespace[3pt]

$1^{3}D_{1}$ & $\psi(3770)$ & $3778.1\pm0.7$ & $3801$ & $3785$ & $3819$ & $3798$ & $3787$ & $\cdots$  & {$3830$} \\
$1^{3}D_{3}$ & $\psi_3(3842)$ & {$3842.71\pm0.20$} & $3839$ & $3806$ & $3849$ & $3815$ & $3799$ & $3896$  & {$3859$} \\
$1^{1}D_{2}$ &  & $\cdots$ & $3826$ & $3799$ & $3837$ & $3811$ & $3796$ & $3806$  & {$3848$} \\
$1^{3}D_{2}$ & $\psi_2(3823)$ & $3823.51\pm0.34$ & $3826$ & $3800$ & $3838$ & $3813$ & $3798$ & $3739$  & {$3848$} \\
$2^{3}D_{1}$ & {$\psi(4160)$}& $4191\pm5$ & $4125$ & $4142$ & $4194$ & $\cdots$ & $4089$ & $\cdots$  & {$4125$} \\
$2^{3}D_{3}$ &  & $\cdots$ & $4157$ & $4167$ & $4217$ & $\cdots$ & $4103$ & $\cdots$  & {$4144$} \\
$2^{1}D_{2}$ &  & $\cdots$ & $4146$ & $4158$ & $4208$ & $\cdots$ & $4099$ & $\cdots$  & {$4137$} \\
$2^{3}D_{2}$ &  & $\cdots$ & $4146$ & $4158$ & $4208$ & $\cdots$ & $4100$ & $\cdots$  & {$4137$} \\
$3^{3}D_{1}$ &  & $\cdots$ & $4385$ & $\cdots$ & $\cdots$ & $\cdots$ & $4317$ & $\cdots$  & {$4334$} \\
$3^{3}D_{3}$ &  & $\cdots$ & $4413$ & $\cdots$ & $\cdots$ & $\cdots$ & $4331$ & $\cdots$  & {$4348$} \\
$3^{1}D_{2}$ &  & $\cdots$ & $4402$ & $\cdots$ & $\cdots$ & $\cdots$ & $4326$ & $\cdots$  & {$4343$} \\
$3^{3}D_{2}$ &  & $\cdots$ & $4402$ & $\cdots$ & $\cdots$ & $\cdots$ & $4327$ & $\cdots$  & {$4343$} \\
\addlinespace[3pt]

{$4^{3}D_{1}$} &  & $\cdots$ & $4600$ & $\cdots$ & $\cdots$ & $\cdots$ & $\cdots$ & $\cdots$  & {$4484$} \\
{$4^{3}D_{3}$} &  & $\cdots$ & $4624$ & $\cdots$ & $\cdots$ & $\cdots$ & $\cdots$ & $\cdots$  & {$4494$} \\
{$4^{1}D_{2}$} &  & $\cdots$ & $4615$ & $\cdots$ & $\cdots$ & $\cdots$ & $\cdots$ & $\cdots$  & {$4490$} \\
{$4^{3}D_{2}$} &  & $\cdots$ & $4615$ & $\cdots$ & $\cdots$ & $\cdots$ & $\cdots$ & $\cdots$  & {$4490$} \\
\addlinespace[3pt]

$1^{3}F_{2}$ &  & $\cdots$ & $4053$ & $4029$ & $4092$ & $\cdots$ & $\cdots$ & $\cdots$  & {$4070$} \\
$1^{3}F_{4}$ &  & $\cdots$ & $4064$ & $4021$ & $4095$ & $\cdots$ & $\cdots$ & $\cdots$  & {$4076$} \\
$1^{1}F_{3}$ &  & $\cdots$ & $4060$ & $4026$ & $4094$ & $\cdots$ & $\cdots$ & $4037$  & {$4074$} \\
$1^{3}F_{3}$ &  & $\cdots$ & $4062$ & $4029$ & $4097$ & $\cdots$ & $\cdots$ & $3999$  & {$4075$} \\
$2^{3}F_{2}$ &  & $\cdots$ & $4321$ & $4351$ & $4422$ & $\cdots$ & $\cdots$ & $\cdots$  & {$4293$} \\
$2^{3}F_{4}$ &  & $\cdots$ & $4333$ & $4348$ & $4425$ & $\cdots$ & $\cdots$ & $\cdots$  & {$4298$} \\
$2^{1}F_{3}$ &  & $\cdots$ & $4329$ & $4350$ & $4424$ & $\cdots$ & $\cdots$ & $\cdots$  & {$4296$} \\
$2^{3}F_{3}$ &  & $\cdots$ & $4329$ & $4352$ & $4426$ & $\cdots$ & $\cdots$ & $\cdots$  & {$4297$} \\
\end{longtable}
\par
\endgroup
\twocolumngrid

\subsubsection{Fitted states}
For the fitted states, our calculated masses for $J/\psi(1S)$ and $\psi(2S)$ are $3105$ and $3675~\mathrm{MeV}$, respectively, with signed differences of $+8.1$ and $-11.1~\mathrm{MeV}$ from the corresponding PDG central values~\cite{ParticleDataGroup:2026mpi}. The results for $\chi_{c1}(1P)$ and $\chi_{c2}(1P)$ are $3511$ and $3555~\mathrm{MeV}$, respectively, with signed differences of $+0.33$ and $-1.17~\mathrm{MeV}$ from the experimental values of $3510.67$ and $3556.17~\mathrm{MeV}$~\cite{ParticleDataGroup:2026mpi}. The calculated masses of $\eta_c(1S)$, $\eta_c(2S)$, and $h_c(1P)$ lie approximately $5.09$, $16.8$, and $6.37~\mathrm{MeV}$ lower than experiment, respectively, whereas the $\chi_{c0}(1P)$ mass is $8.5~\mathrm{MeV}$ higher~\cite{ParticleDataGroup:2026mpi}. The $1S$ and $2S$ hyperfine splittings are $126$ and $54~\mathrm{MeV}$, respectively, whereas the corresponding values obtained from the tabulated experimental masses are approximately $113$ and $48.3~\mathrm{MeV}$.
The present calculation thus provides a qualitative description of the low-lying states.

The fitted $\psi(4040)$ is assigned to $3{}^3S_1$, with a calculated mass of $4049~\mathrm{MeV}$, compared with the fit input $4039.6\pm4.3~\mathrm{MeV}$; the $3{}^3S_1$ assignment is also discussed in Ref.~\cite{Wang:2019mhs}. The fitted $\chi_{c0}(3915)$ and $\chi_{c2}(3930)$ are assigned to $2{}^3P_0$ and $2{}^3P_2$, with masses of $3866$ and $3950~\mathrm{MeV}$, respectively, compared with inputs of $3922.1\pm1.8$ and $3922.5\pm1.0~\mathrm{MeV}$. The fitted $\psi(4160)$ is assigned to $2{}^3D_1$, with a calculated mass of $4125~\mathrm{MeV}$ compared with the input $4191\pm5~\mathrm{MeV}$.

The fitted $\psi(3770)$ and $\psi_2(3823)$ are assigned to $1{}^3D_1$ and $1{}^3D_2$. Their calculated masses are $3801$ and $3826~\mathrm{MeV}$, with signed differences of approximately $+22.9$ and $+2.49~\mathrm{MeV}$ from the corresponding experimental central values~\cite{ParticleDataGroup:2026mpi}.

\subsubsection{Additional experimental comparisons}

The unfitted $1{}^3D_3$ level has a calculated mass of $3839~\mathrm{MeV}$, approximately $3.71~\mathrm{MeV}$ below the mass of $\psi_3(3842)$~\cite{ParticleDataGroup:2026mpi}. The spectroscopic classification of $D$-wave charmonia also requires tests of decay channels and widths.

The $4{}^3S_1$ mass of 4332 MeV provides a comparison with the $\psi(4230)$ assignment proposed in Ref.~\cite{Wang:2019mhs}.  
Intermodel differences become more pronounced for some higher radial excitations. The $4^3S_1$ mass is $4332~\mathrm{MeV}$, which is $74$ and $118~\mathrm{MeV}$ lower than the NR and GI predictions, respectively, but only $59~\mathrm{MeV}$ above the screened-potential PM result~\cite{Barnes:2005pb,Li:2009zu}. 

The mass gap between the $3{}^3S_1$ and $4{}^3S_1$ levels is
$283~\mathrm{MeV}$ in the present calculation, compared with
$350~\mathrm{MeV}$ in the GI model. This smaller gap is qualitatively
consistent with the trend reviewed in Ref.~\cite{Bai:2026atm}:
screening flattens the confining potential at large distances and
reduces the spacing between higher radial excitations.

The modified rovibrational model in Ref.~\cite{Fang:2022bft} provides another comparison baseline, and its mass relation is not a screened-potential Hamiltonian.  In this previous work, $3{}^1S_0$ and $3{}^3S_1$ masses are $4003$ and $4037~\mathrm{MeV}$, differing from the present results by $11$ and $12~\mathrm{MeV}$, respectively. 
However, their $4{}^3S_1$ mass of $4231~\mathrm{MeV}$ is $101~\mathrm{MeV}$ smaller than the present value, whereas their $2{}^3P_1$ mass of $4013~\mathrm{MeV}$ is $93~\mathrm{MeV}$ larger.

In an amplitude analysis of $B^+\to D^{*\pm}D^\mp K^+$ decays, LHCb reported several charmonium or charmoniumlike structures, including $\eta_c(3945)$, $h_c(4000)$, $\chi_{c1}(4010)$, and $h_c(4300)$, with $J^{PC}=0^{-+}$, $1^{+-}$, $1^{++}$, and $1^{+-}$, respectively~\cite{LHCb:2024vfz}. Under the tentative $3\,{}^1S_0$, $2\,{}^1P_1$, $2\,{}^3P_1$, and $3\,{}^1P_1$ assignments, respectively, the predicted masses differ from the measured central values by approximately $+69$, $-76$, $-92.5$, and $-78.3~\mathrm{MeV}$. The relative agreement is state dependent. The $3\,{}^1S_0$ prediction is closer to the central mass of $\eta_c(3945)$ than the NR and GI predictions, but farther from it than the RQM and PM predictions. For the proposed assignments of $h_c(4000)$, $\chi_{c1}(4010)$, and $h_c(4300)$, the NR and GI predictions have smaller absolute mass discrepancies than the corresponding results of this work.

BESIII reported a mass of $4484.7\pm13.3\pm24.1~\mathrm{MeV}$ for the $Y(4500)$ structure in 2022~\cite{BESIII:2022joj}. A subsequent study also reported the vector structure $Y(4710)$, with a mass of approximately $4708~\mathrm{MeV}$~\cite{BESIII:2023wqy}. Using a nonrelativistic screened potential together with a decay analysis, Ref.~\cite{Cui:2025low} proposes the  $2{}^3P_0$,  $3{}^3P_0$, $2{}^3D_1$, $3{}^3D_1$, and $4{}^3D_1$ interpretations for $\chi_{c0}(3915)$, $X(4160)$, $\psi(4160)$, $Y(4500)$, and $Y(4710)$, respectively. The present $2{}^3P_0$  and $3{}^3P_0$ mass of 3866 MeV and $4187~\mathrm{MeV}$ are close to their predictions of approximately 3846 MeV and $4186~\mathrm{MeV}$, and are therefore compatible at the mass level with their candidate classifications of $\chi_{c0}(3915)$ and  $X(4160)$. For the vector states, the two studies differ more substantially: the present $2{}^3D_1$ and $3{}^3D_1$ mass of $4125~\mathrm{MeV}$ and $4385~\mathrm{MeV}$ are smaller than their result of approximately $4143~\mathrm{MeV}$ and $4423~\mathrm{MeV}$ and are also smaller than the measured central mass of $\psi(4160)$ and $Y(4500)$. The present $4{}^3D_1$ mass of $4600~\mathrm{MeV}$ is likewise smaller than the approximately $4659~\mathrm{MeV}$ result in Ref.~\cite{Cui:2025low} and $116~\mathrm{MeV}$ above the mass 4484 MeV obtained by Ref. \cite{Wang:2019mhs}. 

Our $5{}^3S_1$ mass of $4562~\mathrm{MeV}$ is $77.3~\mathrm{MeV}$ higher than the BESIII central value of $4484.7~\mathrm{MeV}$ for $Y(4500)$. We can only tentatively consider $5{}^3S_1$ as a candidate interpretation at the level of the present mass spectrum. 
As reviewed in Ref.~\cite{Wang:2025dur}, BESIII observed an enhancement
near $4.47~\mathrm{GeV}$ in the $e^+e^-\to\pi^+D^{*0}D^{*-}$ cross
section, with a fitted mass consistent with that of the $Y(4500)$
observed in the $K^+K^-J/\psi$ channel. A comparison of amplitude
analyses of these open-charm and hidden-charm channels could provide
a further test of the tentative $5{}^3S_1$ assignment beyond the mass
comparison alone. 
Ref.~\cite{Wang:2019mhs} discusses $\psi(4415)$ through $5S$--$4D$ mixing and predicts a partner state with a mass of approximately $4489$--$4529~\mathrm{MeV}$. The present calculation does not include $S$--$D$ mixing, so it provides no prediction for the corresponding mixing angle. The partner-state prediction in Ref.~\cite{Wang:2019mhs} predates the BESIII report of $Y(4500)$ and provides a mixed-state candidate framework in the same mass region, which can be compared with future mixing calculations.

\subsubsection{Predictions for unobserved states}

Our $4\,{}^1S_0$ mass of $4307~\mathrm{MeV}$ is $118~\mathrm{MeV}$ below the GI prediction and $57~\mathrm{MeV}$ above the PM prediction. 

For the $3P$ multiplet, the $3\,{}^3P_0$, $3\,{}^3P_1$, and $3^3P_2$ masses are $4187$, $4226$, and $4250~\mathrm{MeV}$, respectively, approximately $87$--$105~\mathrm{MeV}$ lower than the corresponding GI predictions. This trend is qualitatively consistent with the lowering of highly excited states by a screened confining potential~\cite{Li:2009ad,Li:2023wgq}.

In addition to the experimental candidates discussed above, the present calculation predicts a $5{}^1S_0$ mass of $4542~\mathrm{MeV}$. The $4P$ and $3D$ multiplets span $4442$--$4491$ and $4385$--$4413~\mathrm{MeV}$, respectively, while the $1F$ and $2F$ multiplets span $4053$--$4064$ and $4321$--$4333~\mathrm{MeV}$, respectively. The other masses of the higher excited states can be found in Table~\ref{tab:cc}.

\subsection{The bottomonium  spectrum}

Table~\ref{tab:bb} presents the bottomonium mass spectrum calculated in this work, together with available experimental data and other theoretical predictions.

\par
\onecolumngrid
\begingroup
\footnotesize
\setlength{\tabcolsep}{2.2pt}
\renewcommand{\arraystretch}{1.0}
\setlength{\LTleft}{0pt}
\setlength{\LTright}{0pt}
\setlength{\LTcapwidth}{\textwidth}
\setlength{\LTpre}{10pt}
\setlength{\LTpost}{4pt}
\setlength{\parindent}{0pt}
\begin{longtable}{@{\extracolsep{\fill}}l*{9}{c}@{}}
\caption[Mass spectrum of the $b\bar{b}$ system. Masses are given in $\mathrm{MeV}$.]{Mass spectrum of the $b\bar{b}$ system. Masses are given in $\mathrm{MeV}$. Experimental entries adopted in this work; see the PDG compilation~\cite{ParticleDataGroup:2026mpi}. For states included in the fit, the experimental masses and uncertainties reproduce Table~\ref{tab1}; other entries retain their cited experimental references. NR: nonrelativistic constituent quark model~\cite{Segovia:2016xqb}; GI: Godfrey--Isgur model~\cite{Godfrey:2015dia}; RQM: relativistic quasipotential model~\cite{Ebert:2002pp}; PM: screened potential model~\cite{Li:2009nr}; BS: covariant Bethe--Salpeter model~\cite{Fischer:2014cfa};  MGI: screened relativistic quark model~\cite{Wang:2018rjg}. An ellipsis indicates that no value is quoted in the present table.}

\label{tab:bb}\\
\toprule[1pt]
\toprule[1pt]
States & Meson & Expt. & This work & NR & GI & RQM & PM & BS  & MGI \\
\midrule[0.6pt]
\endfirsthead
\caption[]{Mass spectrum of the $b\bar b$ system (continued).}\\
\toprule[1pt]
\toprule[1pt]
States & Meson & Expt. & This work & NR & GI & RQM & PM & BS  & MGI \\
\midrule[0.6pt]
\endhead
\midrule
\multicolumn{10}{r}{\footnotesize Continued on next page}\\
\endfoot
\bottomrule[1pt]
\bottomrule[1pt]
\endlastfoot
$1^{1}S_{0}$ & $\eta_b(1S)$ & $9398.7\pm2$ & $9397.7$ & $9455.0$ & $9402.0$ & $9400.0$ & $9389.0$ & $9414.0$  & {$9398.0$} \\
$2^{1}S_{0}$ & $\eta_b(2S)$ & $9999\pm4.2157$ & $9988.5$ & $9990.0$ & $9976.0$ & $9993.0$ & $9987.0$ & $\cdots$  & {$9989.0$} \\
$3^{1}S_{0}$ &  & $\cdots$ & $10335$ & $10330$ & $10336$ & $10328$ & $10330$ & $\cdots$  & {$10336$} \\
$4^{1}S_{0}$ &  & $\cdots$ & $10595$ & $\cdots$ & $10623$ & $\cdots$ & $10595$ & $\cdots$  & {$10597$} \\
$5^{1}S_{0}$ &  & $\cdots$ & $10805$ & $\cdots$ & $10869$ & $\cdots$ & $10817$ & $\cdots$  & {$10810$} \\
{$6^{1}S_{0}$} &  & $\cdots$ & $10985$ & $\cdots$ & {$11097$} & $\cdots$ & {$11011$} & $\cdots$  & {$10991$} \\
\addlinespace[3pt]

$1^{3}S_{1}$ & $\Upsilon(1S)$ & $9460.4\pm0.1$ & $9466.5$ & $9502.0$ & $9465.0$ & $9460.0$ & $9460.0$ & $9490.0$  & {$9463.0$} \\
$2^{3}S_{1}$ & $\Upsilon(2S)$ & $10023.4\pm0.5$ & $10019$ & $10015$ & $10003$ & $10023$ & $10016$ & $\cdots$  & {$10017$} \\
$3^{3}S_{1}$ & $\Upsilon(3S)$ & $10355.1\pm0.5$ & $10356$ & $10349$ & $10354$ & $10355$ & $10351$ & $\cdots$  & {$10356$} \\
$4^{3}S_{1}$ & $\Upsilon(4S)$ & $10579.4\pm1.2$ & $10610$ & $10607$ & $10635$ & $\cdots$ & $10611$ & $\cdots$  & {$10612$} \\
$5^{3}S_{1}$ & $\Upsilon(10860)$ & $10885.2\pm2.1$ & $10817$ & $10818$ & $10878$ & $\cdots$ & $10831$ & $\cdots$  & {$10822$} \\
{$6^{3}S_{1}$} & {$\Upsilon(11020)$}  & $11000\pm4$ & $10994$ & {$10995$} & {$11102$} & $\cdots$ & {$11023$} & $\cdots$  & {$11001$} \\
\addlinespace[3pt]

$1^{3}P_{0}$ & $\chi_{b0}(1P)$ & $9859.44\pm0.522$ & $9854.9$ & $9855.0$ & $9847.0$ & $9863.0$ & $9865.0$ & $9815.0$  & {$9858.0$} \\
$1^{3}P_{2}$ & $\chi_{b2}(1P)$ & $9912.21\pm0.4046$ & $9910.9$ & $9886.0$ & $9897.0$ & $9913.0$ & $9918.0$ & $9906.0$  & {$9910.0$} \\
$1^{1}P_{1}$ & $h_b(1P)$ & $9899.3\pm0.8$ & $9895.0$ & $9879.0$ & $9882.0$ & $9901.0$ & $9903.0$ & $9806.0$  & {$9894.0$} \\
$1^{3}P_{1}$ & $\chi_{b1}(1P)$ & $9892.78\pm0.4046$ & $9889.1$ & $9874.0$ & $9876.0$ & $9892.0$ & $9897.0$ & $9842.0$  & {$9889.0$} \\
\addlinespace[3pt]

$2^{3}P_{0}$ & $\chi_{b0}(2P)$ & $10232.5\pm0.6403$ & $10230$ & $10221$ & $10226$ & $10234$ & $10226$ & $\cdots$  & {$10235$} \\
$2^{3}P_{2}$ & $\chi_{b2}(2P)$ & $10268.65\pm0.5463$ & $10269$ & $10246$ & $10261$ & $10268$ & $10269$ & $\cdots$  & {$10269$} \\
$2^{1}P_{1}$ & $h_b(2P)$ & $10259.8\pm1.2083$ & $10258$ & $10240$ & $10250$ & $10261$ & $10256$ & $\cdots$  & {$10259$} \\
$2^{3}P_{1}$ & $\chi_{b1}(2P)$ & $10255.46\pm0.5463$ & $10254$ & $10236$ & $10246$ & $10255$ & $10251$ & $\cdots$  & {$10255$} \\
\addlinespace[3pt]

$3^{3}P_{0}$ &  & $\cdots$ & $10506$ & $10500$ & $10522$ & $\cdots$ & $10502$ & $\cdots$  & {$10513$} \\
$3^{3}P_{2}$ & $\chi_{b2}(3P)$ & $10524.02\pm0.7783$ & $10537$ & $10521$ & $10550$ & $\cdots$ & $10540$ & $\cdots$  & {$10539$} \\
$3^{1}P_{1}$ &  & $\cdots$ & $10528$ & $10516$ & $10541$ & $\cdots$ & $10529$ & $\cdots$  & {$10530$} \\
$3^{3}P_{1}$ & $\chi_{b1}(3P)$ & $10513.42\pm0.6701$ & $10525$ & $10513$ & $10538$ & $\cdots$ & $10524$ & $\cdots$  & {$10527$} \\
\addlinespace[3pt]

$4^{3}P_{0}$ &  & $\cdots$ & $10729$ & $\cdots$ & $10775$ & $\cdots$ & $10732$ & $\cdots$  & {$10736$} \\
$4^{3}P_{2}$ &  & $\cdots$ & $10754$ & $10744$ & $10798$ & $\cdots$ & $10767$ & $\cdots$  & {$10758$} \\
$4^{1}P_{1}$ &  & $\cdots$ & $10746$ & $\cdots$ & $10790$ & $\cdots$ & $10757$ & $\cdots$  & {$10751$} \\
$4^{3}P_{1}$ &  & $\cdots$ & $10744$ & $\cdots$ & $10788$ & $\cdots$ & $10753$ & $\cdots$  & {$10749$} \\
\addlinespace[3pt]

$1^{3}D_{1}$ &  & $\cdots$ & $10152$ & $10117$ & $10138$ & $10153$ & $10145$ & $\cdots$  & {$10153$} \\
$1^{3}D_{3}$ &  & $\cdots$ & $10170$ & $10127$ & $10155$ & $10162$ & $10156$ & $10232$  & {$10170$} \\
$1^{1}D_{2}$ &  & $\cdots$ & $10163$ & $10123$ & $10148$ & $10158$ & $10152$ & $10194$  & {$10163$} \\
$1^{3}D_{2}$ & $\Upsilon_2(1D)$ & {$10163.7\pm1.4$} & $10162$ & $10122$ & $10147$ & $10158$ & $10151$ & $10145$  & {$10162$} \\
\addlinespace[3pt]

$2^{3}D_{1}$ &  & $\cdots$ & $10440$ & $10414$ & $10441$ & $\cdots$ & $10432$ & $\cdots$  & {$10442$} \\
$2^{3}D_{3}$ &  & $\cdots$ & $10455$ & $10422$ & $10455$ & $\cdots$ & $10442$ & $\cdots$  & {$10456$} \\
$2^{1}D_{2}$ &  & $\cdots$ & $10449$ & $10419$ & $10450$ & $\cdots$ & $10439$ & $\cdots$  & {$10450$} \\
$2^{3}D_{2}$ &  & $\cdots$ & $10448$ & $10418$ & $10449$ & $\cdots$ & $10438$ & $\cdots$  & {$10450$} \\
\addlinespace[3pt]

$3^{3}D_{1}$ &  & $\cdots$ & $10670$ & $10653$ & $10698$ & $\cdots$ & $10670$ & $\cdots$  & {$10675$} \\
$3^{3}D_{3}$ &  & $\cdots$ & $10683$ & $10660$ & $10711$ & $\cdots$ & $10680$ & $\cdots$  & {$10686$} \\
$3^{1}D_{2}$ &  & $\cdots$ & $10678$ & $10658$ & $10706$ & $\cdots$ & $10677$ & $\cdots$  & {$10681$} \\
$3^{3}D_{2}$ &  & $\cdots$ & $10677$ & $10657$ & $10705$ & $\cdots$ & $10676$ & $\cdots$  & {$10681$} \\
\addlinespace[3pt]

{$4^{3}D_{1}$} &  & $\cdots$ & $10863$ & {$10853$} & {$10928$} & $\cdots$ & {$10877$} & $\cdots$  & {$10871$} \\
{$4^{3}D_{3}$} &  & $\cdots$ & $10875$ & {$10860$} & {$10939$} & $\cdots$ & {$10886$} & $\cdots$  & {$10880$} \\
{$4^{1}D_{2}$} &  & $\cdots$ & $10870$ & $\cdots$ & {$10935$} & $\cdots$ & {$10883$} & $\cdots$  & {$10876$} \\
{$4^{3}D_{2}$} &  & $\cdots$ & $10870$ & $\cdots$ & {$10934$} & $\cdots$ & {$10882$} & $\cdots$  & {$10876$} \\
\addlinespace[3pt]

{$5^{3}D_{1}$} &  & $\cdots$ & $11031$ & {$11023$} & $\cdots$ & $\cdots$ & {$11060$} & $\cdots$  & {$11041$} \\
{$5^{3}D_{3}$} &  & $\cdots$ & $11041$ & $\cdots$ & $\cdots$ & $\cdots$ & {$11069$} & $\cdots$  & {$11049$} \\
{$5^{1}D_{2}$} &  & $\cdots$ & $11037$ & $\cdots$ & $\cdots$ & $\cdots$ & {$11066$} & $\cdots$  & {$11046$} \\
{$5^{3}D_{2}$} &  & $\cdots$ & $11036$ & $\cdots$ & $\cdots$ & $\cdots$ & {$11065$} & $\cdots$  & {$11045$} \\
\addlinespace[3pt]

$1^{3}F_{2}$ &  & $\cdots$ & $10360$ & $10315$ & $10350$ & $\cdots$ & $\cdots$ & $\cdots$  & {$10362$} \\
$1^{3}F_{4}$ &  & $\cdots$ & $10369$ & $\cdots$ & $10358$ & $\cdots$ & $\cdots$ & $\cdots$  & {$10369$} \\
$1^{1}F_{3}$ &  & $\cdots$ & $10365$ & $10322$ & $10355$ & $\cdots$ & $\cdots$ & $10319$  & {$10366$} \\
$1^{3}F_{3}$ &  & $\cdots$ & $10365$ & $10321$ & $10355$ & $\cdots$ & $\cdots$ & $10302$  & {$10366$} \\
\addlinespace[3pt]

$2^{3}F_{2}$ &  & $\cdots$ & $10602$ & $10569$ & $10615$ & $\cdots$ & $\cdots$ & $\cdots$  & {$10605$} \\
$2^{3}F_{4}$ &  & $\cdots$ & $10609$ & $\cdots$ & $10622$ & $\cdots$ & $\cdots$ & $\cdots$  & {$10612$} \\
$2^{1}F_{3}$ &  & $\cdots$ & $10606$ & $10573$ & $10619$ & $\cdots$ & $\cdots$ & $\cdots$  & {$10609$} \\
$2^{3}F_{3}$ &  & $\cdots$ & $10606$ & $10573$ & $10619$ & $\cdots$ & $\cdots$ & $\cdots$  & {$10609$} \\
\end{longtable}
\par
\endgroup
\twocolumngrid

\subsubsection{Fitted  states}
Among the 18 bottomonium states included in the fit, our calculated $\Upsilon(1S)$ mass is $9466.5~\mathrm{MeV}$, $6.1~\mathrm{MeV}$ above the adopted experimental central value. The $\eta_b(1S)$ mass of $9397.7~\mathrm{MeV}$ lies $1.0~\mathrm{MeV}$ lower than the experimental value of $9398.7~\mathrm{MeV}$. The calculated $M[\eta_b(2S)]=9988.5~\mathrm{MeV}$ and $M[\Upsilon(2S)]=10019~\mathrm{MeV}$ are $10.5$ and $4.4~\mathrm{MeV}$ below the corresponding experimental values of $9999.0$ and $10023.4~\mathrm{MeV}$, respectively. For the $1P$ multiplet, the calculated masses for $h_b(1P)$, $\chi_{b0}(1P)$, $\chi_{b1}(1P)$, and $\chi_{b2}(1P)$ are $9895.0$, $9854.9$, $9889.1$, and $9910.9~\mathrm{MeV}$, respectively, approximately $4.3$, $4.54$, $3.68$, and $1.31~\mathrm{MeV}$ below the adopted experimental central values of $9899.3$, $9859.44$, $9892.78$, and $9912.21~\mathrm{MeV}$~\cite{ParticleDataGroup:2026mpi}. The absolute mass deviations for these eight low-lying fitted states range from $1.0$ to $10.5~\mathrm{MeV}$.

The experimental masses and additional comparisons provide a test of the model's description across fitted and unfitted levels. The  fitted $3\,{}^3S_1$ and $4\,{}^3S_1$ masses are $10356$ and $10610~\mathrm{MeV}$, respectively, lying $0.9$ and $30.6~\mathrm{MeV}$ above the corresponding experimental central masses of $\Upsilon(3S)$ and $\Upsilon(4S)$ quoted by the PDG~\cite{ParticleDataGroup:2026mpi}.
 As a comparison within the same class of models, the screened GI calculation in Ref.~\cite{Wang:2018rjg} gives $10356$ and $10612~\mathrm{MeV}$, which equal the present $3S$ result and exceed the present $4S$ result by $2~\mathrm{MeV}$, respectively. 
Our fitted $3\,{}^3S_1$ mass is $7$, $2$, $1$, and $5~\mathrm{MeV}$ above the corresponding NR, GI, RQM, and PM predictions, respectively~\cite{Segovia:2016xqb,Godfrey:2015dia,Ebert:2002pp,Li:2009nr}. 

As the radial excitation increases further, the fitted $4\,{}^3S_1$ mass of $10610~\mathrm{MeV}$ lies $25~\mathrm{MeV}$ below the GI prediction, whereas  the fit result  $M[5\,{}^3S_1]=10817~\mathrm{MeV}$ is $61~\mathrm{MeV}$ below it. Meanwhile, our fitted $5\,{}^3S_1$ mass is also $14~\mathrm{MeV}$ below the screened-potential PM result.

For the two higher vector resonances, the $5{}^3S_1$ and $6{}^3S_1$ assignments adopted in Table~\ref{tab:bb} are also used in the screened GI study of Ref.~\cite{Wang:2018rjg}. Our masses of $10817$ and $10994~\mathrm{MeV}$ are $68.2$ and $6~\mathrm{MeV}$ below the adopted experimental central values of $\Upsilon(10860)$ and $\Upsilon(11020)$, respectively. Reference~\cite{Wang:2018rjg} obtains $10822$ and $11001~\mathrm{MeV}$ for these two states. Relative to the experimental values~\cite{ParticleDataGroup:2026mpi}, its $5{}^3S_1$ result is also low, by $63.2~\mathrm{MeV}$, whereas its $6{}^3S_1$ mass differs by only $1.0~\mathrm{MeV}$.

In our calculation, the masses of  $\chi_{b0}(2P)$, $\chi_{b1}(2P)$, $\chi_{b2}(2P)$, and $h_b(2P)$ in Table~\ref{tab:bb} have signed differences of $-2.5$, $-1.46$, $+0.35$, and $-1.8~\mathrm{MeV}$ from the experimental central values, respectively.  Its internal splittings are also described closely: the calculated $M[\chi_{b1}(2P)]-M[\chi_{b0}(2P)]$ and $M[\chi_{b2}(2P)]-M[\chi_{b1}(2P)]$ mass differences are $24$ and $15~\mathrm{MeV}$, compared with $23.0$ and $13.2~\mathrm{MeV}$ from the adopted experimental central values. Reference~\cite{Wang:2018rjg} gives $10235$, $10255$, $10269$, and $10259~\mathrm{MeV}$ for the same four states, respectively.

For the observed $\chi_{b1}(3P)$ and $\chi_{b2}(3P)$ states, the present masses are $10525$ and $10537~\mathrm{MeV}$, which exceed the experimental central values by $11.6$ and $13.0~\mathrm{MeV}$, respectively. Nevertheless, their calculated separation of $12~\mathrm{MeV}$ is close to the experimental central-value difference of $10.6~\mathrm{MeV}$. The screened GI calculation of Ref.~\cite{Wang:2018rjg} gives $10527$ and $10539~\mathrm{MeV}$, corresponding to a similar splitting of $12~\mathrm{MeV}$ but slightly larger deviations in the absolute masses.

\subsubsection{Additional experimental comparisons}

The corresponding $5\,{}^1S_0$ mass of $10805~\mathrm{MeV}$ is $64~\mathrm{MeV}$  lower than  the GI prediction.  For orbital excitations, the prediction  $M[3\,{}^3D_1]=10670~\mathrm{MeV}$ is $28~\mathrm{MeV}$ below the GI prediction and $17~\mathrm{MeV}$ above the NR prediction and equal to the PM prediction. The signs and magnitudes of these intermodel mass differences therefore depend on both the state and the reference model.

Recent studies of $\Upsilon(10753)$ provide new experimental constraints. In a joint Belle and Belle II analysis in 2026, a simultaneous fit to the $e^+e^-\to\omega\chi_{b1,b2}(1P)$ cross sections yielded a mass of $10756.1\pm3.4\pm2.7~\mathrm{MeV}$~\cite{Belle:2025syc}. The predicted   $3\,{}^3D_1$ mass lies $86.1~\mathrm{MeV}$ below this central value, whereas the $5\,{}^3S_1$ mass  is $60.9~\mathrm{MeV}$ above it. 
 
The present unmixed spectrum is therefore insufficient to determine a unique assignment. Ref.~\cite{Li:2022leg} investigates transitions from this structure to $\Upsilon(1\,{}^3D_J)\eta$ under a $4S$--$3D$ mixing interpretation, providing a way to test its composition through decay information. The $4\,{}^3S_1$ mass of $10610~\mathrm{MeV}$ is also smaller than the mass of this structure, motivating further investigation of its mixing with $3D$. 

The present $S$- and $D$-wave masses are unmixed predictions; no $S$--$D$ mixing angle was calculated. The study in Ref.~\cite{Bicudo:2022ihz} uses lattice-QCD static potentials as input to investigate coupled channels involving bottomonium and open-bottom meson pairs. It obtains an $S$-wave-sector candidate at $10773^{+1}_{-2}~\mathrm{MeV}$ and a $D$-wave-sector candidate at $10799\pm2~\mathrm{MeV}$ related to $\Upsilon(10753)$. These values exceed the experimental central mass quoted above by $16.9$ and $42.9~\mathrm{MeV}$, respectively. Comparison with the present unmixed states indicates that explicit channel coupling provides a different interpretation worth testing.

The experimentally established $\Upsilon_2(1D)$ offers a further test of the orbital spectrum. Identifying it with $1{}^3D_2$, our result of $10162~\mathrm{MeV}$ is $1.7~\mathrm{MeV}$ below the adopted experimental mass of $10163.7\pm1.4~\mathrm{MeV}$. Reference~\cite{Wang:2018rjg} obtains $10162~\mathrm{MeV}$ and identifies the radiative transition to $\chi_{b1}(1P)\gamma$ as its dominant decay mode, illustrating the additional information available from radiative cascades. 

\subsubsection{Predictions for unobserved states}

For higher excitations, the predicted $\eta_b(3S)$ and $\eta_b(4S)$ masses are $10335$ and $10595~\mathrm{MeV}$, respectively. The $2D$ and $1F$ multiplets span $10440$--$10455$ and $10360$--$10369~\mathrm{MeV}$, respectively. These states provide mass references for future searches through radiative cascades and hadronic transitions.

The $1\,{}^3F_2$ mass of $10360~\mathrm{MeV}$ is $45$ and $10~\mathrm{MeV}$ above the NR and GI predictions, respectively, whereas $M[1\,{}^3F_3]=10365~\mathrm{MeV}$ is $63~\mathrm{MeV}$ above the BS prediction~\cite{Fischer:2014cfa}. These intermodel differences may reflect variations in model parameters and dynamical treatments.

The other masses of the higher excitations of bottomonium are shown in Table~\ref{tab:bb}.

\subsection{The toponium  spectrum}

In the toponium calculation, we fixed the top-quark mass at $m_t=172.6~\mathrm{GeV}$, close to the direct-measurement average, $172.56\pm0.31~\mathrm{GeV}$, in the PDG 2025 update~\cite{ParticleDataGroup:2024cfk}.

Table~\ref{tab:tt} presents the mass eigenvalues obtained by extrapolating the MGI model to the $t\bar t$ system in the stable-top-quark approximation, together with results from other theoretical calculations.

\par
\onecolumngrid
\begingroup
\footnotesize
\setlength{\tabcolsep}{2.2pt}
\renewcommand{\arraystretch}{1.0}
\setlength{\LTleft}{0pt}
\setlength{\LTright}{0pt}
\setlength{\LTcapwidth}{\textwidth}
\setlength{\LTpre}{10pt}
\setlength{\LTpost}{4pt}
\setlength{\parindent}{0pt}
\begin{longtable}{@{\extracolsep{\fill}}l*{6}{c}@{}}
\caption[Potential-model mass spectrum of the $t\bar{t}$ system. Masses are given in $\mathrm{GeV}$. The results of this work are obtained in the stable-top-quark approximation and should not be identified directly with individually resolvable narrow resonances.]{ Potential-model mass spectrum of the $t\bar{t}$ system. Masses are given in $\mathrm{GeV}$. The results of this work are obtained in the stable-top-quark approximation and should not be identified directly with individually resolvable narrow resonances. 
NR: nonrelativistic potential model~\cite{Akbar:2024brg}; GIC: Godfrey--Isgur--Capstick model~\cite{Luo:2025psq}; PM: screened potential model~\cite{Jiang:2024fyw}; BS: Salpeter model with a Cornell potential~\cite{Wang:2024hzd}. 
}
\label{tab:tt}\\
\toprule[1pt]
\toprule[1pt]
States & This work & NR & GIC & PM & BS \\
\midrule[0.6pt]
\endfirsthead
\caption[]{Potential-model spectrum of the $t\bar t$ system (continued).}\\
\toprule[1pt]
\toprule[1pt]
States & This work & NR & GIC & PM & BS \\
\midrule[0.6pt]
\endhead
\midrule
\multicolumn{7}{r}{\footnotesize Continued on next page}\\
\endfoot
\toprule[1pt]\toprule[1pt]
\endlastfoot
$1^{1}S_{0}$   & $342.424$ & $342.867$ & $342.618$ & $341.267$ & $343.620$ \\
$2^{1}S_{0}$   & $343.666$ & $344.405$ & $\cdots$ & $344.179$ & $344.590$ \\
$3^{1}S_{0}$  & $344.083$ & $344.742$ & $\cdots$ & $344.759$ & $344.930$ \\
$4^{1}S_{0}$   & $344.327$ & $344.910$ & $\cdots$ & $345.000$ & $\cdots$ \\
$5^{1}S_{0}$  & $344.502$ & $345.026$ & $\cdots$ & $345.153$ & $\cdots$ \\
\addlinespace[3pt]

$1^{3}S_{1}$  & $342.434$ & $342.914$ & $342.628$ & $341.650$ & $343.620$ \\
$2^{3}S_{1}$   & $343.668$ & $344.411$ & $\cdots$ & $344.227$ & $344.590$ \\
$3^{3}S_{1}$  & $344.084$ & $344.745$ & $\cdots$ & $344.759$ & $344.930$ \\
$4^{3}S_{1}$  & $344.328$ & $344.911$ & $\cdots$ & $345.000$ & $\cdots$ \\
$5^{3}S_{1}$   & $344.503$ & $345.027$ & $\cdots$ & $345.153$ & $\cdots$ \\
\addlinespace[3pt]

$1^{3}P_{0}$  & $343.457$ & $344.396$ & $\cdots$ & $344.181$ & $344.390$ \\
$1^{3}P_{2}$  & $343.469$ & $344.401$ & $\cdots$ & $344.215$ & $344.390$ \\
$1^{1}P_{1}$  & $343.464$ & $344.400$ & $\cdots$ & $344.206$ & $344.390$ \\
$1^{3}P_{1}$   & $343.463$ & $344.399$ & $\cdots$ & $344.200$ & $344.390$ \\
\addlinespace[3pt]

$2^{3}P_{0}$   & $343.981$ & $344.733$ & $\cdots$ & $344.742$ & $344.830$ \\
$2^{3}P_{2}$  & $343.985$ & $344.735$ & $\cdots$ & $344.754$ & $344.830$ \\
$2^{1}P_{1}$  & $343.984$ & $344.734$ & $\cdots$ & $344.751$ & $344.830$ \\
$2^{3}P_{1}$  & $343.983$ & $344.734$ & $\cdots$ & $344.749$ & $344.830$ \\
\addlinespace[3pt]

$3^{3}P_{0}$  & $344.257$ & $344.901$ & $\cdots$ & $344.987$ & $345.060$ \\
$3^{3}P_{2}$  & $344.259$ & $344.902$ & $\cdots$ & $344.994$ & $345.060$ \\
$3^{1}P_{1}$  & $344.258$ & $344.901$ & $\cdots$ & $344.992$ & $345.060$ \\
$3^{3}P_{1}$  & $344.258$ & $344.901$ & $\cdots$ & $344.991$ & $345.060$ \\
\addlinespace[3pt]

$4^{3}P_{0}$  & $344.447$ & $345.017$ & $\cdots$ & $\cdots$ & $\cdots$ \\
$4^{3}P_{2}$  & $344.449$ & $345.018$ & $\cdots$ & $\cdots$ & $\cdots$ \\
$4^{1}P_{1}$ & $344.448$ & $345.018$ & $\cdots$ & $\cdots$ & $\cdots$ \\
$4^{3}P_{1}$  & $344.448$ & $345.018$ & $\cdots$ & $\cdots$ & $\cdots$ \\
\addlinespace[3pt]

$1^{3}D_{1}$  & $343.933$ & $344.716$ & $\cdots$ & $344.733$ & $344.720$ \\
$1^{3}D_{3}$  & $343.936$ & $344.717$ & $\cdots$ & $344.736$ & $344.720$ \\
$1^{1}D_{2}$  & $343.935$ & $344.717$ & $\cdots$ & $344.735$ & $344.720$ \\
$1^{3}D_{2}$  & $343.934$ & $344.717$ & $\cdots$ & $344.735$ & $344.720$ \\
\addlinespace[3pt]

$2^{3}D_{1}$   & $344.215$ & $344.884$ & $\cdots$ & $344.975$ & $344.990$ \\
$2^{3}D_{3}$  & $344.217$ & $344.884$ & $\cdots$ & $344.977$ & $344.990$ \\
$2^{1}D_{2}$  & $344.216$ & $344.884$ & $\cdots$ & $344.976$ & $344.990$ \\
$2^{3}D_{2}$  & $344.216$ & $344.884$ & $\cdots$ & $344.976$ & $344.990$ \\
\addlinespace[3pt]

$3^{3}D_{1}$  & $344.411$ & $345.001$ & $\cdots$ & $345.129$ & $345.160$ \\
$3^{3}D_{3}$  & $344.413$ & $345.002$ & $\cdots$ & $345.131$ & $345.160$ \\
$3^{1}D_{2}$  & $344.412$ & $345.002$ & $\cdots$ & $345.130$ & $345.160$ \\
$3^{3}D_{2}$  & $344.412$ & $345.001$ & $\cdots$ & $345.130$ & $345.160$ \\
\addlinespace[3pt]

\addlinespace[3pt]
$4^{3}D_{1}$   & $344.562$ & $345.096$ & $\cdots$ & $\cdots$ & $\cdots$ \\
$4^{3}D_{3}$  & $344.563$ & $345.096$ & $\cdots$ & $\cdots$ & $\cdots$ \\
$4^{1}D_{2}$   & $344.562$ & $345.096$ & $\cdots$ & $\cdots$ & $\cdots$ \\
$4^{3}D_{2}$   & $344.562$ & $345.096$ & $\cdots$ & $\cdots$ & $\cdots$ \\
\addlinespace[3pt]
$1^{3}F_{2}$  & $344.174$ & $344.862$ & $\cdots$ & $\cdots$ & $344.930$ \\
$1^{3}F_{4}$  & $344.175$ & $344.862$ & $\cdots$ & $\cdots$ & $344.930$ \\
$1^{1}F_{3}$  & $344.174$ &  {$344.930$} & $\cdots$ & $\cdots$ & $344.930$ \\
$1^{3}F_{3}$  & $344.174$ & $344.862$ & $\cdots$ & $\cdots$ & $344.930$ \\
\addlinespace[3pt]

$2^{3}F_{2}$   & $344.375$ & $344.980$ & $\cdots$ & $\cdots$ & $345.120$ \\
$2^{3}F_{4}$   & $344.376$ & $344.981$ & $\cdots$ & $\cdots$ & $345.120$ \\
$2^{1}F_{3}$  & $344.376$ & $344.980$ & $\cdots$ & $\cdots$ & $345.120$ \\
$2^{3}F_{3}$  & $344.376$ & $344.980$ & $\cdots$ & $\cdots$ & $345.120$ \\
\end{longtable}
\par
\endgroup
\twocolumngrid

Unlike conventional charmonium and bottomonium systems, the $t\bar t$ system requires explicit treatment of the finite top-quark lifetime in its threshold dynamics. The mass eigenvalues obtained in the stable-top-quark approximation cannot be identified directly with individually resolvable experimental resonances~\cite{CMS:2025kzt,ATLAS:2026dbe}. Therefore, the tabulated masses are used as references for potential-model extrapolation in this subsection.

Our calculated masses of the lowest $1\,{}^1S_0$ and $1\,{}^3S_1$ states are $342.424$ and $342.434~\mathrm{GeV}$, respectively, with a hyperfine splitting of $10~\mathrm{MeV}$.

This splitting is close to the $10~\mathrm{MeV}$ GIC extrapolation result in Ref.~\cite{Luo:2025psq}. Both calculations therefore predict an approximately spin-degenerate ground-state doublet. However, the present pseudoscalar mass is $194~\mathrm{MeV}$ lower than that result.

The present ground-state masses are also lower than those of the Cornell-potential Salpeter calculation in Ref.~\cite{Wang:2024hzd}, but the spin structure should be compared separately. That study adopts $m_t=172.7~\mathrm{GeV}$ and reports degeneracy within multiplets with the same radial and orbital quantum numbers in the approximation used.

The corresponding $2S$ and $3S$ hyperfine splittings decrease to $2.0$ and $1.0~\mathrm{MeV}$, respectively. In the vector channel, the successive $2S-1S$, $3S-2S$, $4S-3S$, and $5S-4S$ radial spacings are $1.23\times10^3$, $416$, $244$, and $175~\mathrm{MeV}$, respectively. The calculated $1P$ masses lie between $343.457$ and $343.469~\mathrm{GeV}$, corresponding to a total mass spread of $12~\mathrm{MeV}$. The calculated spin splittings are therefore small compared with the radial level spacings, indicating approximate spin degeneracy within the corresponding multiplets in this model extrapolation.

Different models yield substantially different predictions for the absolute masses. The calculated $1\,{}^1S_0$ mass is   $443$, $194$, and $1.20\times10^3~\mathrm{MeV}$ below the NR, GIC, and BS predictions, respectively, but $1.16\times10^3~\mathrm{MeV}$ above the PM prediction~\cite{Akbar:2024brg,Luo:2025psq,Wang:2024hzd,Jiang:2024fyw}. For $1\,{}^3S_1$, the corresponding differences are $-480$, $-194$, $-1.19\times10^3$, and $+784~\mathrm{MeV}$, respectively. For $2\,{}^1S_0$, the present result is $739$, $513$, and $924~\mathrm{MeV}$ below the NR, PM, and BS predictions, respectively. These differences reflect not only the interaction models but also differences in the adopted top-quark masses and other model parameters.

The calculated $1S$ hyperfine splitting of $10~\mathrm{MeV}$ is close to the $10~\mathrm{MeV}$ value of the GIC calculation, but is considerably smaller than the NR and PM results of $47$ and $383~\mathrm{MeV}$, respectively. This comparison illustrates the strong model dependence of the predicted short-distance spin-dependent structure.

Recent experimental studies near the $t\bar t$ threshold provide a reference for the above extrapolation. The 2025 CMS dilepton analysis reported a significant excess over fixed-order perturbative QCD predictions, with spin-correlation observables consistent with a color-singlet pseudoscalar $S$-wave quasibound-state interpretation~\cite{CMS:2025kzt}. An ATLAS analysis published in 2026 also reported a threshold enhancement with a significance exceeding $8\sigma$~\cite{ATLAS:2026dbe}. An August 2026 CMS analysis further reported a $6.1\sigma$ excess in the single-lepton-plus-jets final state, providing a complementary test in a different decay channel~\cite{CMS:2026ani}.

In conventional spectroscopic notation, the pseudoscalar color-singlet $t\bar t$ channel corresponds to the ${}^1S_0$ with $J^{PC}=0^{-+}$. Our calculated $1\,{}^1S_0$ mass of $342.424~\mathrm{GeV}$ lies approximately $0.576~\mathrm{GeV}$ below the $343~\mathrm{GeV}$ mass assumed in the CMS 2025 simplified signal model~\cite{CMS:2025kzt}. However, this value is a fixed input to the simulation, not an independently measured resonance mass. The present potential-model calculation does not include the finite top-quark width or continuum contributions to the threshold production process; a quantitative comparison requires an appropriate threshold-dynamics treatment.

In the stable-top-quark approximation, we further obtained $2S$ states at $343.666$--$343.668~\mathrm{GeV}$, a $1D$ multiplet at $343.933$--$343.936~\mathrm{GeV}$, and a $1F$ multiplet at $344.174$--$344.175~\mathrm{GeV}$; the  mass of $5S$ states are at $344.502$--$344.503~\mathrm{GeV}$. These results provide model benchmarks for increasingly dense levels and small spin splittings in the extremely heavy-quark limit.

\section{Conclusion}\label{sec4}

Within the MGI framework, we introduce a screened confining potential and systematically calculate the mass spectra of light mesons, heavy-light mesons, and heavy quarkonia. The model parameters are determined through a simultaneous fit to 90 adopted meson masses in different flavor sectors, and the masses of radial and orbital excitations, together with the corresponding mixing angles, are subsequently obtained. 
This result provides a good overall description of the experimental values across multiple systems.

Some low-lying light-meson states are well described, while sizeable deviations remain for the pion, scalar states, and several orbital excitations and spin splittings. The light-meson results provide mass-based references for spectroscopic assignments, including assigning $\pi(1800)$ and $\rho(1450)$ to the $3\,{}^1S_0$ and $2\,{}^3S_1$ states, respectively; regarding $\eta_2(1645)$ and $\eta_2(2030)$ as candidates for the nonstrange $1\,{}^1D_2$ and $2\,{}^1D_2$ states, respectively; and interpreting $K_1(1861)$ and $K_1(1911)$ as a mixed $2P$ doublet. For $K_2(2250)$, the higher-lying $2D$ state obtained in this work supports its interpretation as a candidate for the first radially excited $D$-wave state. Meanwhile, this work provides predictions for the masses and mixing angles of the higher $3D$, $4D$, and $3F$ doublets. Compared with some existing screened models, the overall masses of these highly excited doublets obtained in this work are higher, and the candidate $2G$ state for $K_4(2500)$ also lies above the experimental central value, which may be attributable to differences in the parameter fitting and the long-range confining interaction. On the other hand, the masses of some highly radially excited states are lower than the GI-model results, which is qualitatively consistent with the tendency of the screened potential to lower highly excited energy levels.

For the recently reported $D_{s1}(2933)^+$, with a statistical significance exceeding $10\sigma$, our heavy-light meson results are compatible with its candidate interpretation as a mixed $2P$ state. Meanwhile, this work also provides possible candidate interpretations of $D_{s0}(2590)$ as a $2\,{}^1S_0$ state and $D_2(2740)$ as a mixed $1D$ state. For $D_3^*(2750)^0$, the possible assignment proposed in this work is $1\,{}^3D_3$. 
In the unquenched relativistic quark model employed in this study, the puzzle surrounding the low mass of the $D_{s0}^*(2317)$ is partially alleviated.

For $B_J(5970)$, its mass lies closer to the $1\,{}^3D_1$ state obtained in this work. Under the assumption of a direct $BK$ decay, $B_{sJ}(6114)$ may also retain this candidate assignment, whereas the $2\,{}^3S_1$ interpretation of $B_{sJ}(6063)$ still exhibits a mass discrepancy, which requires further detailed discussion incorporating $S$--$D$ mixing or coupled-channel corrections. For the newly observed $B_{s0}^*(5700)^0$ structure, the predicted scalar mass of $5746~\mathrm{MeV}$ lies $47~\mathrm{MeV}$ above the observed peak,  the origin of this discrepancy is not resolved by the present calculation.

The ground-state mass of $B_c^*$ predicted in this work is close to the newly observed value. For the $B_c(1P)$ structure, the $E1$ selection rules support the two mixed axial-vector states as candidates in the higher-mass region, whereas the candidate structure in the lower-mass region is taken to comprise multiple cascade components.

The $X(4160)$ and $Y(4500)$ in the charmonium sector can be compared with the $3\,{}^3P_0$ and $5\,{}^3S_1$ levels, respectively. A firm assignment of $\Upsilon(10753)$ requires an analysis of $S$--$D$ mixing and coupling to open-bottom channels. The toponium predictions obtained under the stable-top-quark approximation exhibit small spin splittings.

The masses of higher radial and orbital excitations and the applicable mixing angles presented in this work also provide cross-flavor references for future experimental searches. Further studies combining explicit coupled-channel calculations with decay calculations can test these candidate assignments and predictions for highly excited states.

\section{ACKNOWLEDGMENTS}
This work is supported  by the National Natural Science Foundation of China under Grants No. 11965016 and No. 12047501, and by the Natural Science Foundation of Qinghai Province under Grant No. 2022-ZJ-939Q. Zhe-Qi Huang and Lei Huang contributed equally to this work.
\section{DATA AVAILABILITY}
The data that support the findings of this article are not publicly available. The data are available from the authors upon reasonable request.
\bibliography{hepref}
\end{document}